\documentclass[article,nojss]{jss}

\let\maxwidth\textwidth

\usepackage{amsfonts,amstext,amsmath,amssymb,amsthm}
\usepackage{accents}
\usepackage{rotating}
\usepackage{nicefrac}
\usepackage{subfigure}
\usepackage{floatrow}
\usepackage{orcidlink}
\usepackage{multirow}
\usepackage{subfigure}

\newcommand{\AYcite}[2]{\cite{#2}}

\newtheorem{assumption}{Assumption}

\usepackage[flushleft]{threeparttable}
\usepackage{empheq, cases}

\newcommand{\indep}{\perp \!\!\! \perp}

\usepackage{booktabs}

\newcommand{\rZ}{Z}
\newcommand{\rY}{Y}
\newcommand{\rX}{\mX}

\newcommand{\ry}{y}
\newcommand{\rx}{\xvec}

\newcommand{\Xspace}{\mathcal{X}}

\newcommand{\pZ}{F}

\newcommand{\dZ}{f}

\newcommand{\h}{h}

\newcommand{\eparm}{\vartheta}

\newcommand{\ie}{\textit{i.e.}~}

\renewcommand{\Prob}{\mathbb{P}}
\newcommand{\Ex}{\mathbb{E}}
\newcommand{\RR}{\mathbb{R}}
\newcommand{\NN}{\mathbb{N}}

\usepackage{dsfont}
\newcommand{\I}{\mathds{1}}

 \DeclareMathOperator{\logit}{logit}
 \DeclareMathOperator{\cloglog}{cloglog}
 \DeclareMathOperator{\loglog}{loglog}

 \DeclareMathOperator*{\argmin}{{arg\,min}}

 \DeclareMathOperator{\ND}{N}

 \DeclareMathOperator{\BD}{B}
 \DeclareMathOperator{\MD}{M}

 \DeclareMathOperator{\WD}{W}

\def \xvec {\text{\boldmath$x$}}    \def \mX {\text{\boldmath$X$}}

\newcommand{\ubar}[1]{\underaccent{\bar}{#1}}

\author{Susanne Dandl~\orcidlink{0000-0003-4324-4163} \\ LMU M\"unchen, MCML \\
\And Andeas Bender~\orcidlink{0000-0001-5628-8611} \\ LMU M\"unchen, MCML \\
\And Torsten Hothorn~\orcidlink{0000-0001-8301-0471} \\ Universit\"at Z\"urich}
\Plainauthor{Dandl, Bender, Hothorn}

\title{Heterogeneous Treatment Effect Estimation for Observational Data using Model-based Forests}
\Shorttitle{Forest-based HTE Estimation}

\Abstract{
The estimation of heterogeneous treatment effects (HTEs) has attracted
considerable interest in many disciplines, most prominently in medicine and
economics.  Contemporary research has so far primarily focused on continuous and
binary responses where HTEs are traditionally estimated by a linear
model, which allows the estimation of constant or heterogeneous effects even
under certain model misspecifications.  More complex models
for survival, count, or ordinal outcomes require stricter assumptions to
reliably estimate the treatment effect. Most importantly, the noncollapsibility issue
necessitates the joint estimation of treatment and prognostic effects. 
Model-based forests allow simultaneous estimation of covariate-dependent
treatment and prognostic effects, but only for
randomized trials.  In this paper, we propose modifications to model-based
forests to address the confounding issue in
observational data.  In particular, we evaluate an orthogonalization strategy originally
proposed by Robinson (1988, Econometrica) in the context of
model-based forests targeting HTE estimation in generalized linear models
and transformation models.  We found that this strategy reduces confounding effects in a simulated study with various outcome distributions. 
We demonstrate the practical aspects of HTE estimation
for survival and ordinal outcomes by an assessment of the potentially
heterogeneous effect of Riluzole on the progress of Amyotrophic Lateral
Sclerosis.
 }

\Keywords{Heterogeneous treatment effects, personalized medicine, random forest, observational data, censored survival data, generalized linear model, transformation model}

\Address{
  Susanne Dandl, Andreas Bender \\
  Institut f\"ur Statistik, LMU M\"unchen, Germany \\
  Munich Center for Machine Learning (MCML), Germany \\

  Torsten Hothorn\\
  Institut f\"ur Epidemiologie, Biostatistik und Pr\"avention, Universit\"at Z\"urich \\
  Hirschengraben 84, CH-8001 Z\"urich, Switzerland \\
  E-mail: \email{Torsten.Hothorn@R-project.org}\\
}

\begin{document}

\section{Introduction}
 
Over the past years, there has been emerging interest in methods to estimate heterogeneous treatment effects (HTEs) in various application
fields. In healthcare, HTE estimation can be understood as a core principle
driving personalized medicine.
As opposed to average treatment effects, which assume a constant
effect of a treatment on an outcome for the whole population, HTEs account
for the heterogeneity in the effect for subgroups or individuals based on
their characteristics.
Most research on HTE estimation has mainly focused on continuous and binary response variables. 
These methods have typically built upon Rubin's potential outcomes framework, a
statistical approach to formulating and inferring causal effects in various
designs \citep{Rubin_1974,Rubin_2005}.

Traditionally, statistical models were used to estimate the treatment effect, but machine learning methods have been more and more adapted for these tasks over the past decade. Machine learning models rely on weaker assumptions and can automatically learn complex relationships such as higher order interaction effects, resulting in greater predictive performance in a variety of applications.
In the case of continuous or binary responses, prominent methods to estimate HTEs are based on random forests \citep{Foster_Taylor_Ruberg_2011,Lu_Sadiq_Feaster_Ishwaran_2018,Athey_Tibshirani_Wager_2019,Powers_Qian_Jung_2018,Su_Pena_Liu_2018,Li_Levine_Fan_2022}, Bayesian additive regression trees (BART) \citep{Hill_2011,Hu_2020}, or neural networks \citep{Shalit_Johansson_Sontag_2017,Curth_Lee_Schaar_2021,Chapfuwa_Assaad_Zeng_2021}. 
\AYcite{Kuenzel et al.}{Kuenzel_Sekhon_2019} proposed general frameworks -- T-learners, S-learners, U-learners, and X-learners
-- that base treatment effect estimates on arbitrary machine learning models.
\AYcite{Chernozhukov et al.}{Chernozhukov_Chetverikov_Demirer_2018} coined the term double/debiased machine learning models, which uses machine learning models for nuisance
parameter estimations. The approach still relies on parametric models for estimating treatment effects, but \AYcite{Nie and Wager}{Nie_Wager_2020} derived so-called R-learners that allow for arbitrary (nonparametric or semiparametric) models.

Beyond continuous or binary responses, research on machine learning methods for HTE estimation have primarily focused on (right-censored) survival data.
Methods have been proposed based on Bayesian additive regression trees (BART) \citep{Henderson_Louis_2018}, random forest-type methods \citep{Cui_Kosorok_Wager_Zhu_2021,Tabib_Larocque_2020}, or deep learning approaches \citep{Curth_Lee_Schaar_2021,Chapfuwa_Assaad_Zeng_2021}.
Theoretically, any machine learning model for survival analysis -- such as random survival forests \citep{Ishwaran_Kogalur_2008} or a Cox regression-based deep neural network (deepSurv) \citep{Katzman_2018} -- can estimate HTEs \citep{Hu_Ji_Li_2021}. These models can estimate survival or hazard functions in both treatment groups separately; HTEs are then defined as the difference in derived properties of the two functions, e.g., as differences in the median survival time. 
However, \AYcite{Hu et al.}{Hu_Ji_Li_2021} found that methods specifically designed for HTE estimation, like the adapted BART \citep{Henderson_Louis_2018}, produce more reliable estimates.

In general, for a continuous or binary outcome $Y$ conditional on treatment $w$ and covariates $\rx$, the
conditional average treatment effect $\tau(\rx)$ (CATE) can be estimated from the
model $\Ex(Y \mid W = w, \rX = \rx) = \mu(\rx) + \tau(\rx) w$ even if the
model is misspecified, e.g., when the prognostic effect $\mu(\rx)$
cannot be fully estimated due to missing covariate information.
Beyond mean regression, stricter assumptions are necessary
both for randomized and for observational studies to estimate HTEs. For example, under a true Cox
model with survivor function $\exp(-\exp(\h(t) + \mu(\rx) + \tau w))$
with log-cumulative baseline hazard $\h(t)$ at time $t$ and log-hazard ratio $\tau$, the prognostic effect
$\mu(\rx)$ must be specified correctly, even in a randomized trial.
Estimated marginal log-hazard ratios $\hat{\tau}$ -- i.e., when the model is fitted under the constraint
$\mu(\rx) \equiv 0$ -- are shrunken towards zero if this constraint is
unrealistic \citep{AalenCookRoysland2015}. Naturally, this problem carries over to
heterogeneous log-hazard ratios $\tau(\rx)$.

Consequently, HTE estimation in more complex
models requires the simultaneous estimation of both the prognostic part
$\mu(\rx)$ and the predictive HTE $\tau(\rx)$.
Model-based forests have been demonstrated to allow estimation of $\mu(\rx)$ and $\tau(\rx)$ 
in randomized trials \citep{Seibold_Zeileis_Hothorn_2015,Seibold_Zeileis_Hothorn_2017,Korepanova_Seibold_Steffen_2019,Buri_Hothorn_2020,Fokkema_Smits_Zeileis_2017,Hothorn_Zeileis_2021}.
In a nutshell, model-based forests combine the parametric modeling framework with random forests to estimate individual treatment effects \citep{Seibold_Zeileis_Hothorn_2017}.
By using generalized linear models and transformation models, model-based forests can be adapted for survival data \citep{Seibold_Zeileis_Hothorn_2015,Seibold_Zeileis_Hothorn_2017,Korepanova_Seibold_Steffen_2019}, ordinal data \citep{Buri_Hothorn_2020}, or clustered data \citep{Fokkema_Smits_Zeileis_2017}.
A unique feature of model-based forests is the simultaneous estimation of
both treatment and prognostic effects in the same forest model.

In observational studies the treatment group assignment is not under control of the researcher and confounding effects could bias the estimation of HTEs. 
In this work, we propose and evaluate novel variants of model-based forests
for HTE estimation in observational studies. Adaptions of Robinson's
orthogonalization strategy for generalized linear models and transformation
models are discussed and implemented. We review key components
of model-based forests for HTE estimation in randomized trials in
Section~\ref{sec:method}.
In Section~\ref{subsec:ortho}, we start introducing the orthogonalization approach
by \AYcite{Robinson}{Robinson_1988}, which is instrumental for achieving
robustness to confounding effects in the non-randomized situation. We
motivate previous developments using linear models\citep{Dandl_Hothorn_Seibold_2022} and leverage adaptations to more complex
models discussed by \AYcite{Gao and Hastie}{Gao_Hastie_2022} to define novel
model-based forest variants suitable for HTE in the observational setting. 
These variants' performances are empirically assessed in a simulation study with a range of outcome distributions
in Section~\ref{sec:sim}.
Finally, in Section~\ref{sec:als} presenting a re-analysis of the patient-specific effect of Riluzole in patients with Amyotrophic Lateral Sclerosis (ALS), practical
aspects of model estimation and interpretation are discussed.
 
\section{Review of model-based forests for randomized trials}
\label{sec:method}

We are interested in estimating HTEs based on i.i.d.\ observations $(y, \rx, w)$, where $y$, $\rx$ and $w$ are realizations of the outcome $\rY$, covariates $\rX \in \Xspace$, and control vs. treatment indicator $W \in \{0, 1\}$.
$Y(0)$ and $Y(1)$ denote the potential outcomes under the two treatment conditions $W \in \{0, 1\}$.
Throughout this paper, we assume that $\rX$ includes all relevant variables to explain heterogeneity both in the treatment effect and the outcome $\rY$, and that the base model underlying model-based forests is correctly specified. 

We review model-based forests for HTE estimation
based on randomized trials as introduced by \AYcite{Seibold et al.}{Seibold_Zeileis_Hothorn_2017} and \AYcite{Korepanova et al.}{Korepanova_Seibold_Steffen_2019}. 
Within this section, we only consider settings where the treatment assignment is
randomized and, therefore, follows a binomial model $W \mid \rX = \rx \sim
\BD(1, \pi(\rx))$ with constant propensities $\pi(\rx) \equiv \pi$.
We omit discussion of the abstract framework underlying model-based forests
and instead discuss the important linear, generalized linear
\citep{Seibold_Zeileis_Hothorn_2017}, and 
transformation models \citep{Korepanova_Seibold_Steffen_2019} in detail.

\subsection{Linear model}
\label{sec:methodlin}

For a continuous outcome $\rY \in \RR$ with symmetric error distribution, a model-based forest
might be defined based on the model 
\begin{eqnarray}
	(\rY \mid \rX = \rx, W = w) & = & \mu(\rx) + \tau(\rx) w + \phi \rZ 
	\label{eq:mobnorm} 
\end{eqnarray}
where the residuals are given by the error term $\phi \rZ$ with $\Ex(\rZ| \rX, W) = 0$ and standard deviation $\phi > 0$ \citep{Dandl_Hothorn_Seibold_2022}.
We are mainly interested in estimating $\tau(\rx)$, the treatment effect that depends on \textit{predictive} variables in $\rx$.
With model-based forests, however, we also obtain an estimated value for the prognostic effect $\mu(\rx)$, which depends on \textit{prognostic} variables in $\rx$. A variable might be predictive and prognostic at the same time. We refer to these situations as ``overlays''.

Because we assume in this section that $\pi(\rx) \equiv \pi$ applies, $W \indep \rX$ holds. Consequently, $\tau(\rx)$ can be interpreted as a CATE
\begin{equation}
\tau(\rx) = \text{CATE}(\rx) = \Ex(\rY(1) - \rY(0) \mid \rX = \rx)
\label{eq:cate}
\end{equation}
on the absolute scale. 
To estimate $(\mu(\rx), \tau(\rx))^\top$ 
the $L_2$ loss
\begin{eqnarray}
	\ell(\mu(\rx), \tau(\rx)) = \nicefrac{1}{2}\left(\rY - \mu(\rx) - \tau(\rx)w\right)^2
\label{eq:ellmob}
\end{eqnarray}
is minimized 
w.r.t.~$\mu$ and $\tau$ using an ensemble of trees.
Inspired by recursive partitioning techniques \citep{Hothorn_Hornik_Zeileis_2006, Zeileis_Hothorn_Hornik_2008}, split variable and split point selection are separated.
The split variable is the variable that has the lowest $p$-value for the bivariate permutation tests for the $H_0$-hypothesis that $\mu$ and $\tau$ are constant and independent of any split variable.
The cut-point is the point of the chosen split variable at which the score functions
\begin{eqnarray*} \label{MOBscore}
s(\hat{\mu}, \hat{\tau}) := (\rY - \hat{\mu} - \hat{\tau} w) (1, w)^\top
\end{eqnarray*}
in the two resultant subgroups differ the most; details are available in 
Appendix 2 of \AYcite{Seibold et al.}{Seibold_Zeileis_Hothorn_2017}.

Once $B \in \NN$ trees were fitted to subsamples of the training data, predictions for the treatment effect for a new observation $\rx$ are obtained via local maximum likelihood aggregation \citep{Hothorn_Lausen_Benner_2004, Meinshausen_2006, Lin_Jeon_2006, Athey_Tibshirani_Wager_2019, Hothorn_Zeileis_2021}.
First, for the $i$-th training sample, the frequency $\alpha_i$ with which it falls in the same leaf as $\rx$ over all $B$ trees is measured. The obtained weighting vector $(\alpha_1, ..., \alpha_n)$ is used as an input for minimizing
\begin{equation}
(\hat{\mu}(\rx), \hat{\tau}(\rx))^\top = \argmin_{\mu, \tau} 
\sum_{i = 1}^{n} \alpha_i(\rx) \ell_i(\mu, \tau)
\label{eq:persmodMOB}
\end{equation}
where $\ell_i$ denotes the loss for the $i$-th sample.
Model-based forests easily allow adaptions if HTEs for an outcome variable $\rY$ that is not well represented by equation~\eqref{eq:mobnorm} should be estimated.
In this case, model-based forests can build on generalized linear models or transformation models in the recursive partitioning framework \citep{Zeileis_Hothorn_Hornik_2008}.
As detailed in the following sections, the loss function $\ell$ in equation~\eqref{eq:ellmob} changes from the squared error to the negative (partial) log-likelihood of some appropriate model.
 
\subsection{Generalized linear models} \label{sec:GLM}

When the conditional outcome distribution is better described through a
generalized linear model 
\begin{eqnarray*}
(\rY \mid \rX = \rx, W = w) \sim \text{ExpFam}(\theta(\mu(\rx) + \tau(\rx)w), \phi)
\label{eq:expfam}
\end{eqnarray*}
with parameter $\theta$ depending on the additive function $\mu(\rx) +
\tau(\rx)w$, the conditional mean 
\begin{eqnarray}
g(\Ex(\rY \mid \rX = \rx, W = w)) = \mu(\rx) + \tau(\rx)w =: \eta_w(\rx)
\label{eq:eta}
\end{eqnarray}
is linear on the scale of a link function $g$. Thus, the interpretation of
$\tau(\rx)$ as CATE \eqref{eq:cate} generally no longer holds. Instead,
the predictive effect is understood as the difference in natural parameters
(DINA \citep{Gao_Hastie_2022})
\begin{eqnarray}
\tau(\rx) = \text{DINA}(\rx) = \eta_1(\rx) - \eta_0(\rx).
\label{eq:DINA}
\end{eqnarray}
In contrast to the linear model case, HTEs $\tau(\rx)$ are now defined on relative
scales, such as odds ratios in binary logistic regression models or multiplicative mean effects in a Poisson or Gaussian model with a log-link.
The negative log-likelihood contribution of
some observation $(\rY, \rx, w)$ is
$$\ell(\mu, \tau, \phi) = -\log(\dZ(\rY \mid \theta(\mu(\rx) + \tau(\rx)w), \phi))$$
with $f$ as the conditional density of an exponential family distribution
\begin{equation*}
\dZ(Y \mid \theta(\mu(\rx) + \tau(\rx)w), \phi).
\label{eq:glm}
\end{equation*}

Model-based trees and forests
\citep{Zeileis_Hothorn_Hornik_2008,Seibold_Zeileis_Hothorn_2015,Seibold_Zeileis_Hothorn_2017}
jointly estimate the prognostic effect $\mu(\rx)$ and the predictive
effect $\tau(\rx)$. The procedure 
simultaneously minimizes the negative log-likelihood with respect
to $\mu(\rx)$ and $\tau(\rx)$. 
In each node of the model-based forest, $\mu$, $\tau$, and potentially
$\phi$ are estimated by minimizing
\begin{eqnarray} 
\ell(\mu, \tau, \phi) = -\log(\dZ(\rY \mid \theta(\mu + \tau w), \phi))
\label{eq:likeliglm}
\end{eqnarray}
and regressing the bivariate gradient
\begin{eqnarray*}
\left.\frac{\partial \ell(\mu, \tau, \phi)}{\partial (\mu, \tau)} \right|_{\hat{\mu}, \hat{\tau}, \hat{\phi}}
\end{eqnarray*}
on $\rx$. This means that one is not explicitly looking for changes in the
scale parameter $\phi$, but this could be implemented by looking at the
three-variate gradient
\begin{eqnarray*}
\left.\frac{\partial \ell(\mu, \tau, \phi)}{\partial (\mu, \tau, \phi)} \right|_{\hat{\mu}, \hat{\tau}, \hat{\phi}}
\end{eqnarray*}
for example, in a heteroscedastic normal linear model
\begin{eqnarray*}
(\rY \mid \rX = \rx, W = w) = \mu(\rx) + \tau(\rx)w + \phi(\rx) \rZ.
\end{eqnarray*}
After the tree fitting phase, a HTE is 
estimated with equation~\eqref{eq:persmodMOB} with $\ell(\mu, \tau, \phi)$ of equation~\eqref{eq:likeliglm} as
the corresponding loss function.

Thus, model-based forests can be directly applied to estimate HTEs on relative scales for binary outcomes (binary logistic or probit
regression, for example), counts (Poisson or quasi-Poisson regression), or
continuous outcomes where a multiplicative effect is of interest (normal
model with log-link).
 
\subsection{Transformation models}
\label{subsec:transform}
More complex responses like ordered categorical or time-to-event outcomes
are not covered by generalized linear models but can be analysed
using transformation models; corresponding model-based forests for survival
analysis have been introduced by \AYcite{Korepanova et al.}{Korepanova_Seibold_Steffen_2019}. For some at least ordered outcome $\rY$, we
write the conditional distribution function as
\begin{eqnarray}
\Prob(\rY \le \ry \mid \rX = \rx, W = w) = \pZ(\h(\ry) - \underbrace{(\mu(\rx) +
\tau(\rx)w)}_{=:\eta_w(\rx)}). \label{eq:trafo}
\end{eqnarray}
The transformation function $\h$ is monotone non-decreasing and the inverse
link function $\pZ$ governs the interpretability of $\tau$ as log-odds
ratios ($\pZ = \logit^{-1}$), log-hazard ratios ($\pZ = \cloglog^{-1}$), log-reverse time hazard ratios 
($\pZ = \loglog^{-1}$), or shift effects ($\pZ = \Phi$, the cumulative distribution function of the standard normal).
The shift term $\eta_w(\rx)$ differs between the two treatment groups $w \in \{0, 1\}$.
The distribution functions of the potential outcomes are $\pZ(\h(\ry) -
\mu(\rx))$ for $\rY(0)$ and $\pZ(\h(\ry) - \mu(\rx) - \tau(\rx))$ for
$\rY(1)$.
The negative log-likelihood of a discrete or interval-censored observation
$(\ubar{\ry}, \bar{\ry}]$ (where $\ubar{\ry}$ is the lower interval bound, $\bar{\ry}$ is the upper) is
\begin{eqnarray*}
\ell_\text{Trafo}(\h, \mu, \tau) & = & -\log(\Prob(\ubar{\ry} < \rY \le \bar{\ry} \mid \rX = \rx, W = w)) \\ \nonumber
& = & -\log(\pZ(\h(\bar{\ry}) - \mu(\rx) - \tau(\rx)w) - \pZ(\h(\ubar{\ry}) - \mu(\rx) - \tau(\rx)w)). \label{eq:int}
\end{eqnarray*}
For a continuous datum $\ry \in \RR$, we obtain
\begin{eqnarray*} 
	\ell_\text{Trafo}(\h, \mu, \tau) = -\{\log(\pZ^\prime(\h(\ry) - \mu(\rx) - \tau(\rx)w)) + \log(\h^\prime(\ry))\};
\end{eqnarray*}
details are given in \AYcite{Hothorn et al.}{Hothorn_Moest_Buehlmann_2018}.
Transformation forests apply the model-based recursive partitioning principle 
and estimate $\tau$ in each node along with the transformation function $\h$ (a ``nuisance''
parameter) by minimising $\ell_\text{Trafo}(\h, \mu \equiv 0, \tau)$ \citep{Hothorn_Zeileis_2021}.
Because $\h$ contains an intercept term, the parameter $\mu$ is not
identified. We thus estimate the model under the constraint $\mu \equiv 0$.
Variable and cut-points are selected using the bivariate gradient
\begin{eqnarray*}
\left.\frac{\partial \ell_\text{Trafo}(\h, \mu \equiv 0, \tau)}{\partial (\mu, \tau)}\right|_{\mu = 0, \hat{\tau}}
\end{eqnarray*}
This model family includes proportional odds logistic regression (for
ordered categorical, count or continuous outcomes), Box-Cox type models,
Cox proportional hazards model, Weibull proportional hazards
models for discrete and continuous outcomes, reverse time proportional
hazards models relying on Lehmann alternatives, and many more \citep{Hothorn_Moest_Buehlmann_2018}.
Forests for ordinal outcomes were evaluated by \AYcite{Buri and Hothorn}{Buri_Hothorn_2020}, and 
a general approach to ``transformation forests'' is described in \AYcite{Hothorn and Zeileis}{Hothorn_Zeileis_2021}.

Application of the ideas underlying model-based forests allows HTEs to be estimated for such outcomes under all types of random
censoring and truncation \citep{Korepanova_Seibold_Steffen_2019}.
For example, for Weibull distributed outcomes under right censoring, $h(y)
= \nu_1 + \nu_2
\log(y)$ is chosen for the conditional distribution function in equation~\eqref{eq:trafo} \citep{Hothorn_Moest_Buehlmann_2018}.

In this case, we define $Y$ as the event time, $C$ as the censoring time and $T = \min(Y, C)$ as the observed time.   
For identification of $\tau(\rx)$ under potential censoring, the following assumption must hold \citep{Cui_Kosorok_Wager_Zhu_2021}:
\begin{assumption}[Ignorable censoring] \label{asu4} 
	Censoring time $C$ is independent of survival time $Y$ conditional on treatment indicator $W$ and covariates $X$
$$ (Y(0), Y(1)) \indep C \mid \rX = \rx, W = w.$$
\end{assumption}

\noindent
An important special case represents the Cox proportional hazards model,
where the profile likelihood over the baseline hazard function defines the
partial log-likelihood $\ell_\text{PL}(\mu, \tau)$ with $\mu \equiv 0$. The 
scores with respect to the constant $\mu \equiv 0$ are known as martingale
residuals. Model-based forests for such models, and extensions to time-varying prognostic
and predictive effects, are discussed in \AYcite{Korepanova et al.}{Korepanova_Seibold_Steffen_2019}.
 
\subsection{Noncollapsibility}
\label{subsec:noncollaps}

As mentioned in the introduction, one problem with the Cox model is that misspecifications of prognostic effects $\mu(\rx)$ lead to biased estimates such that the estimated hazard ratios cannot be interpreted causally. 
This issue arises from the noncollapsiblity of the Cox model, the notion of which is characterized by the fact that in these models, the mean of the conditional effect estimates defined over covariates $\rX$ does not coincide with the marginal effect over $\rX$.
Because the noncollapsiblity of the Cox model arises from its nonadditivity of the hazard function, models such as the Weibull model do not suffer from this issue because they satisfy the additivity condition. Consequently, misspecifications of prognostic effects do not affect treatment effect estimates \citep{AalenCookRoysland2015}. 

The noncollapsibility issue is not limited to the Cox model but also affects members of the exponential family without identity or linear link functions. 
Without adjustments, effect estimates can only be interpreted causally if there is no treatment effect ($\tau \equiv 0$) or there are no prognostic covariates \citep{Daniel_Zhang_Farewell_2021}.

If this is not the case, specific methods are needed; ignoring the estimation of $\mu(\rx)$ at all and only focusing on $\tau(\rx)$ does not solve the problem.
Conditioning on available prognostic variables is a common solution and is already applied by model-based forests, because they estimate both the prognostic effect $\mu(\rx)$ and $\tau(\rx)$. 
The ensemble of trees used to estimate these effects provides a high degree of flexibility and might therefore retain some of the potential complexity in the underlying $\mu(\rx)$ to mitigate misspecification. 
Whether conditioning resolves the non-collapsibility issue depends heavily on the assumption that all prognostic variables are known which is often not the case in the real world \citep{AalenCookRoysland2015}. 

For members of the exponential family and the Cox model, \AYcite{Gao and Hastie}{Gao_Hastie_2022} derived a method to account for noncollapsibility in the context of observational data with confounding effects. 
While we consider the noncollapsibility issue beyond the scope of this work, we briefly review the work of Gao and Hastie and discuss its applicability to model-based forests in Section~A of the Supplementary Material.

\section{Model-based forests for observational studies}
\label{subsec:ortho}

In the previous section, we described model-based forests in the randomized setting under the assumption that $\pi(\rx) = \pi$.
In observational studies in which the treatment group assignment is not under the control of the researcher, the propensity score (and therefore, the probability of being in the treatment group) often depends on covariates $\rx$
\begin{equation}
	\pi(\rx) := \Prob(W = 1 \mid \rX = \rx) = \Ex(W \mid \rX = \rx).
	\label{eq:robinsonpi}
\end{equation}
In this case, confounding effects could bias the estimation of treatment effects $\tau(\rx)$, and stricter assumptions are necessary in order to interpret $\tau(\rx)$ causally \citep{Rosenbaum_Rubin_1983}. 
\begin{assumption}[Ignorability/Unconfoundedness] \label{asu2}  The treatment assignment is independent of the potential outcomes conditional on covariates $\rx$ $$(Y(0), Y(1)) \indep W \mid \rX = \rx.$$
\end{assumption} 
\begin{assumption}[Positivity] \label{asu3} The propensity score $\pi(\rx)$ must be bounded away from 0 and 1 
	$$ 0 < \pi(\rx) = \Prob(W = 1 \mid \rX = \rx) = \Ex(W \mid \rX = \rx) < 1. $$
\end{assumption}
\noindent Assumption~\ref{asu2} could be violated by an unmeasured confounder, while Assumption~\ref{asu3} could be violated if all observations in a certain group (defined via $\rx$) are in the treatment group.

\AYcite{Dandl et al.}{Dandl_Hothorn_Seibold_2022} showed for mean regression models that model-based forests are not robust to confounding effects and need further adaptions to estimate causal effects in case of observational data.
One strategy for dealing with confounding effects is the orthogonalization strategy originally introduced by \AYcite{Robinson}{Robinson_1988},
which has received considerable attention in recent years \citep{Chernozhukov_Chetverikov_Demirer_2018,Athey_Tibshirani_Wager_2019,Nie_Wager_2020}.
The reformulation of the linear model
\begin{eqnarray}
(\rY \mid \rX = \rx) & = & \mu(\rx) + \tau(\rx) W + \phi \rZ \label{eq:lm}
\end{eqnarray}
to 
\begin{eqnarray}
(\rY \mid \rX = \rx)  &=&  m(\rx) - m(\rx) + \mu(\rx) + \tau(\rx) W + \phi \rZ \nonumber \\
&=&  m(\rx) + \tau(\rx)(W - \pi(\rx)) + \phi \rZ
\label{eq:lm2}
\end{eqnarray}
given the conditional mean function
\begin{equation}
m(\rx) := \Ex(\rY \mid \rX = \rx) =  \mu(\rx) + \tau(\rx) \pi(\rx), 
\label{eq:elm}
\end{equation}
motivates this approach \citep{Dandl_Hothorn_Seibold_2022}. 

Overall, the orthogonalization strategy consists of two steps: 
First, nuisance parameters $m(\rx) = \Ex(\rY \mid \rX = \rx)$ and $\pi(\rx) = \Prob(W = 1 \mid \rX = \rx)$ are estimated.
Originally, \AYcite{Robinson}{Robinson_1988} used kernel estimators, but any machine learning method could be employed \citep{Chernozhukov_Chetverikov_Demirer_2018,Nie_Wager_2020}.
Regressing $Y - \hat{m}(\rx)$ on $W - \hat{\pi}(\rx)$ then yields unbiased estimates for $\tau(\rx)$.
Subtracting $\hat{m}(\rx)$ and $\hat{\pi}(\rx)$ from $Y$ and $W$, respectively, partially eliminates the association between
$\rX$ and $Y$ and between $\rX$ and $W$, respectively.
The orthogonalization strategy has the distinct advantage over other methods against confounding
-- such as inverse propensity weighting and matching -- that it is stable for extreme propensity scores and forgoes stratification \citep{Gao_Hastie_2022}.

\AYcite{Robinson}{Robinson_1988} and \AYcite{Chernozhukov et al.}{Chernozhukov_Chetverikov_Demirer_2018} use parametric models 
to estimate treatment effects based on residualized $W$ and $Y$, but these models could be replaced by non-parametric or local parametric models \citep{Nie_Wager_2020,Wager_Athey_2018}
-- such as model-based forests.  
For mean regression, \AYcite{Dandl et al.}{Dandl_Hothorn_Seibold_2022} adapted the orthogonalization strategy to model-based forests. Their approach closely follows causal forests, which were the first to combine the orthogonalization strategy with tree-based estimators for $\tau(\rx)$. 

\AYcite{Gao and Hastie}{Gao_Hastie_2022} proposed extensions of Robinson's strategy to members of the exponential family and the Cox model, where
DINA~\eqref{eq:DINA} is of interest.
\AYcite{Gao and Hastie}{Gao_Hastie_2022} assume $\tau(\rx) = \rx^\top \boldsymbol{\beta}$ and use
parametric models to estimate $\tau(\rx)$, but they conclude that
non-parametric or local parametric models could be applied instead. We review model-based forests in combination with
linear models for observational data in the next section and summarize 
the idea by \AYcite{Gao and Hastie}{Gao_Hastie_2022} in Section~\ref{subsec:gao}.
On this basis, we assess how the orthogonalization strategy could be employed in model-based forests beyond mean regression with generalized linear models and transformation models as base models.

\subsection{Review of Dandl et al. (2022)}
\label{subsec:dandl}

As noted above, \AYcite{Athey et al.}{Athey_Tibshirani_Wager_2019} were the first to combine the orthogonalization strategy of Robinson with tree-based estimators to estimate $\tau(\rx)$. 
First, the marginal model $m(\rx) = \Ex(\rY \mid \rX = \rx)$ and propensity score $\pi(\rx) = \Ex(W \mid
\rX = \rx)$ are estimated by regression forests. Afterwards, causal forests estimate individual treatment effects $\tau(\rx)$ in the model
\begin{equation}
(\rY \mid \rX = \rx, W = w) = \hat{m}(\rx) + \tau(\rx)(w - \hat{\pi}(\rx)) + \phi \rZ
\label{eq:cf}
\end{equation}
using the ``locally centered'' outcomes $\rY -  \hat{m}(\rx)$ and treatment indicators $W - \hat{\pi}(\rx)$.

Equation~\eqref{eq:cf} shows that causal forests and model-based forests share common foundations for mean regression.
The main difference is that the splitting scheme of model-based forests allows splitting according to heterogeneity in both treatment and prognostic effects, whereas causal forests only split with respect to heterogeneity in treatment effects (in equation~\eqref{eq:lm2}, $\mu(\rx)$ cancels out).   

\AYcite{Dandl et al.}{Dandl_Hothorn_Seibold_2022} identified which elements of both approaches lead to improved performance in randomized trials and observational studies by defining and evaluating blended versions of model-based forests and causal forests:
\begin{itemize}
	\item[(1)] $\text{mob}(\hat{W}, \hat{Y})$, which applies model-based forests to the model
$$\Ex(\rY \mid \rX = \rx, W = w) = \hat{m}(\rx) + \tilde{\mu}(\rx) + \tau(\rx)(w - \hat{\pi}(\rx)),$$
\ie after centering the treatment indicator $w$ and the outcome $\rY$.
        Both parameters $\tilde{\mu}$ and $\tau$ are estimated simultaneously.
	\item[(2)] $\text{mob}(\hat{W})$, which applies model-based forests to the model
$$\Ex(\rY \mid \rX = \rx, W = w) = \mu(\rx) + \tau(\rx)(w - \hat{\pi}(\rx)),$$
\ie after only centering the treatment indicator $w$ but \textit{not} outcome $\rY$.
        Both $\mu$ and $\tau$ are estimated.
	\item[(3)] cfmob, a method that applies model-based forests to the model 	$$\Ex(\rY \mid \rX = \rx, W = w) = \hat{m}(\rx) + \tau(\rx)(w - \hat{\pi}(\rx)),$$
\ie after only centering the treatment indicator $w$ and splitting only according to
        $\hat{\tau}$. That is, only the parameters $\tau$ are estimated in this variant.
\end{itemize}
Their blended approaches competed with the original implementations of (uncentered) model-based forests and causal forests in an extensive simulation study.
In case of confounding, the authors identified local centering of treatment indicator $w$  and simultaneous estimation of both predictive \textit{and} prognostic effects of the treatment indication ($\text{mob}(\hat{W})$) as the key driver for good performance.  
Additionally, centering $Y$ ($\text{mob}(\hat{W}, \hat{Y})$) is recommended, since it further improved performances in some cases. 
Splitting only according to $\hat{\tau}$ but not $\hat{\mu}$ (cfmob) resulted in lower performance.  
Even for settings with confounding, the performance of cfmob was inferior to that of uncentered model-based forests. 

\subsection{Review of Gao and Hastie (2022)}
\label{subsec:gao}

\AYcite{Robinson}{Robinson_1988} derived the orthogonalization strategy only for semi-parametric additive models with $Y \in
\RR$. \AYcite{Gao and Hastie}{Gao_Hastie_2022} 
extended the idea to a broader class of distributions including the exponential family and Cox' model.

Local centering of the treatment indicator works analogously to mean regression. 
First, propensity scores $\pi(\rx) = \Prob(W \mid \rX = \rx)$ are estimated. 
The effects of the covariates $\rX$ on the treatment assignment are then regressed out by subtracting $\hat{\pi}(\rx)$ from $W$. 

Orthogonalization of $Y$ is not straightforward due to the link function that relates the linear predictor $\eta_w(\rx)$ in equation~\eqref{eq:eta} to the outcome $Y$. 
To understand how Gao and Hastie derived $m(\rx)$ to center $Y$, we consider equation~\eqref{eq:lm} as a model of the exponential family with identity link function $g$. 
Now we can rewrite equation~\eqref{eq:elm} to 

\begin{eqnarray*}
	g(\Ex(Y \mid \rX = \rx)) & = & \Ex_W(g(\Ex(Y \mid \rX = \rx, W = w))) \\ \nonumber
	& = & \pi(\rx) \underbrace{(\mu(\rx) + \tau(\rx))}_{= \eta_1(\rx)} + (1-\pi(\rx)) \underbrace{\mu(\rx)}_{= \eta_0(\rx)} \\
	& = & \mu(\rx) + \pi(\rx) \tau(\rx) = m(\rx).
	\label{eq:robinsonm_gauss}
\end{eqnarray*}
Similarly, we derive $g(\Ex(Y \mid \rX = \rx))$ for all other distributions of the exponential family by
\begin{equation}
	m(\rx) = \pi(\rx) \eta_1(\rx) + (1- \pi(\rx)) \eta_0(\rx).
	\label{eq:robinsonm}
\end{equation}
We can regard the estimated $m(\rx)$ as an offset in the linear predictor
$$\hat{m}(\rx) + \tau(\rx) (W - \hat{\pi}(\rx)).$$ 
Note that equation~\eqref{eq:robinsonm_gauss} states that (only) for the Gaussian distribution we can directly estimate $m(\rx) = \Ex(\rY \mid \rX = \rx)$ without estimating $\eta_0(\rx)$ and $\eta_1(\rx)$. 
We can also derive $\hat{m}(\rx)$ for transformation models based on the definition of $\eta_0$ and $\eta_1$ in equation~\eqref{eq:trafo}.
As mentioned in Section~\ref{subsec:noncollaps}, compared to the difference in conditional means, the difference in natural parameters additionally suffers from the noncollapsibility issue \citep{Greenland_Pearl_Robins_1999}. \AYcite{Gao and Hastie}{Gao_Hastie_2022} also extend the Robinson strategy to tackle not only the confounding but also the noncollapsibility issue for members of the exponential family (without a linear or log link function, otherwise confounding is not an issue) and the Cox model. While the noncollapsibility issue is beyond the scope of this work, we 
briefly summarize and discuss the work of Gao and Hastie in Section~A of the Supplementary Material.

\subsection{Novel model-based forests for observational data}
\label{subsec:hypotheses}

As stated above, our main goal is to assess how the orthogonalization strategy proposed for continuous outcomes could be extended to models beyond mean regression, specifically generalized linear models and transformation models.
Based on \AYcite{Dandl et al.}{Dandl_Hothorn_Seibold_2022} and \AYcite{Gao and Hastie}{Gao_Hastie_2022} we propose
two different
versions of model-based forests, which should be more robust against confounding.
Following \AYcite{Dandl et al.}{Dandl_Hothorn_Seibold_2022}, we formulate research questions for these versions, which we aim to answer empirically in Section~\ref{sec:sim}.
An overview of all proposed versions is given in Table~\ref{tab:strategies}.

The first version of model-based forests directly applies Robinson's orthogonalization strategy: 
First, we estimate propensities $\pi(\rx)$ as well as $\eta_0(\rx)$ and $\eta_1(\rx)$ to derive $\hat{m}(\rx)$.
Then, we update the linear predictor of equation~\eqref{eq:eta} by centering $W$ by $\hat{\pi}(\rx)$ and by adding the offset $\hat{m}(\rx)$. 
For generalized linear models, we obtain
\begin{equation*}
g(\Ex(\rY \mid \rX = \rx, W = w)) = \hat{m}(\rx)  + \tilde{\mu}(\rx) + \tau(\rx)(w - \hat{\pi}(\rx)) 
\label{eq:robinsonglm}
\end{equation*}
and for the conditional distribution function of equation~\eqref{eq:trafo} in case of transformation models
\begin{equation*}
F[h(y) - \{\hat{m}(\rx) + \tilde{\mu}(\rx) + \tau(\rx) (w - \hat{\pi}(\rx)\}].
\label{eq:robinsontf}
\end{equation*}     
Based on the updated models, both prognostic and predictive effects
$\tilde{\mu}(\rx)$ and $\tau(\rx)$ are simultaneously estimated by model-based forests.

In the simulation study and practical example in Sections~\ref{sec:sim}~and~\ref{sec:als}, we use regression forests to estimate $\pi(\rx)$ and gradient boosting machines (with tailored loss functions) to estimate $\eta_0$ and $\eta_1$. 
In the following, we denote this version of model-based forests as 
\textit{Robinson} in recognition of \AYcite{Robinson}{Robinson_1988} while model-based forests without centering $W$ and without offset $\hat{m}(\rx)$ are called \textit{Naive}. 

\paragraph*{RQ 1} 
\textit{To what extent does centering $W$ by $\hat{\pi}(\rx)$ and including $\hat{m}(\rx)$ as an offset affect the performance of model-based forests in the presence of confounding?\\} 

Similar to Dandl et al.\ -- who saw an improvement in performance when only centering $W$ (compared to the naive model-based forests) -- we define an approach called \textit{Robinson}$_{\hat{W}}$ that applies model-based forests to models with linear predictors
$$ 
\mu(\rx) + \tau(\rx)(w-\hat{\pi}(\rx)).
$$
\paragraph*{RQ 2}
	\textit{Do centered treatment indicator model-based forests perform better than uncentered model-based forests in the presence of confounding?\\}
\paragraph*{RQ 3}
\textit{Are model-based forests with centered treatment indicators relevantly outperformed by model-based forests with $\hat{m}(\rx)$ as an additional offset in the presence of confounding?\\}

\begin{table}[ht]
	\renewcommand{\arraystretch}{1.2}
	\begin{center}
		\caption{Overview of proposed model-based forest versions.}
		\begin{tabular}{llll} \hline
			Method & Linear Predictor & Definitions  \\ \hline
			Naive & $\mu(\rx) + \tau(\rx) \, \, w$ &  \\ \hline
			Robinson$_{\hat{W}}$ & $\mu(\rx) + \tau(\rx) (w - \hat{\pi}(\rx))$ & $\pi(\rx) = \Prob(W = 1|\rX = \rx)$ \\
			Robinson & $\hat{m}(\rx) + \tilde{\mu}(\rx) + \tau(\rx) (w - \hat{\pi}(\rx))$ & $m(\rx) = \pi(\rx) \eta_1(\rx) - (1- \pi(\rx)) \eta_0(\rx)$  \\ \hline
		\end{tabular}
\label{tab:strategies}
	\end{center}
\end{table}
 
\section{Empirical evaluation}
\label{sec:sim}

We evaluated the performance of our proposed model-based forest versions
(Table~\ref{tab:strategies}) in a simulation study.
The study includes different outcome types, different
predictive and prognostic effects, and a varying number of observations and
covariates.
Model-based forests were fitted with the \pkg{model4you} \proglang{R} add-on
package \citep{Seibold_Zeileis_Hothorn_2019}.
Similar to \AYcite{Dandl et al.}{Dandl_Hothorn_Seibold_2022}, we base our
study settings on the four setups (A, B, C and D) of \AYcite{Nie and Wager}{Nie_Wager_2020}.
In addition, in Section~B of the Supplementary Material, we show the results for simulation settings first proposed by \AYcite{Wager and Athey}{Wager_Athey_2018} and later reused by \AYcite{Athey et al.}{Athey_Tibshirani_Wager_2019}.

\subsection{Data generating process}
\label{subsec:studynie}

Given $P = \{10, 20\}$, for Setup A, we sampled $\rX \sim U([0, 1]^P)$. For all
other setups, we used $\rX \sim N(0, \I_{P\times P})$.
The treatment indicator was binomially distributed with $W \mid \rX = \rx \sim \BD(1, \pi(\rx))$.
The propensity function $\pi(\rx)$ differed for the four considered setups:
\begin{eqnarray*}
\pi(\rx) = \left\{
  \begin{array}{l}
  	\pi_{A}(x_1, x_2) = \max\{0.1, \min\{\sin(\pi x_1 x_2), 1-0.1\}\} \\
  	\pi_{B} \equiv 0.5 \\
    \pi_{C}(x_2, x_3) = 1/(1 + \exp(x_2+x_3)) \\
    \pi_{D}(x_1, x_2) = 1/(1 + \exp(-x_1) + \exp(-x_2)).
  \end{array} \right.
\end{eqnarray*}
$\pi(\rx) \equiv 0.5$ in Setup B implies a randomized study.
The treatment effect function $\tau(\cdot)$ and the prognostic effect
function $\mu(\cdot)$ also differed between the setups
\begin{eqnarray*}
	\tau(\rx) = \left\{
	\begin{array}{l}
		\tau_{A}(x_1, x_2) = (x_1 + x_2)/2 \\
		\tau_{B}(x_1, x_2) = x_1 + \log(1 + \exp(x_2)) \\
		\tau_{C}\equiv 1 \\
		\tau_{D}(x_1, x_2, x_3, x_4, x_5) = \max\{x_1 + x_2 + x_3, 0\} - \max\{x_4 + x_5,
		0\}.
	\end{array} \right.
\end{eqnarray*}
\begin{eqnarray*}
\mu(\rx) = \left\{
  \begin{array}{l} \mu_A(x_1, x_2, x_3, x_4, x_5) = \sin(\pi x_1 x_2)+ 2(x_3 - 0.5)^2 + x_4 + 0.5 x_5 \\
                   \mu_B(x_1, x_2, x_3) = \max\{x_1 + x_2, x_3, 0\} + \max\{x_4 + x_5, 0\}\\
                   \mu_C(x_1, x_2, x_3) = 2\log (1 + \exp(x_1 + x_2 + x_3))\\
                   \mu_D(x_1, x_2, x_3, x_4, x_5) = (\max\{x_1 + x_2 + x_3, 0\} + \max\{x_4 + x_5,
                   0\})/2.
  \end{array} \right.
\end{eqnarray*}
Setup A has extensive confounding that must be eliminated before estimating an easily predictable treatment effect function $\tau(\rx)$.
Setup B needs no confounding adjustment for reliable estimation of $\tau$.
Although Setup C contains strong confounding, the propensity score function is easier to estimate than the prognostic effect, while the treatment effect is constant.
In Setup D, the treatment and control arms are unrelated, and therefore, learning the conditional expected outcomes of both arms jointly is not beneficial
\citep{Nie_Wager_2020,Dandl_Hothorn_Seibold_2022}.

We studied four different simulation models
\begin{subnumcases}{\hspace{-0.5cm}(\rY \mid \rX = \rx, W = w) \sim}
 \ND(\mu(\rx) + \tau(\rx)(w - 0.5), 1)  \label{m1} \\
 \BD(1, \text{expit}(\mu(\rx) + \tau(\rx)(w - 0.5)))  \label{m2}  \\
 \MD \text{with} \log(O(\ry_k \mid \rx, w)) = \eparm_k - \mu(\rx) - \tau(\rx)(w - 0.5) \label{m3} \\
 \WD \text{with} \log(H(\ry \mid \rx, w)) = 2 \log(y) - \mu(\rx) - \tau(\rx)(w - 0.5)
  \label{m4}
\end{subnumcases}
Model (\ref{m1}) is a normal linear regression model, model (\ref{m2}) is a binary
logistic regression model,
model (\ref{m3}) is a 4-nomial model with log-odds function
$\eparm_k - \mu(\rx) - \tau(\rx)(w - 0.5)$ with threshold parameters
$\eparm_k = \text{logit}(k / 4)$ for $k = 1, 2, 3$, and
model (\ref{m4}) is a Weibull model with
log-cumulative hazard function $2 \log(y) - \mu(\rx) - \tau(\rx)(w - 0.5)$.
We added $50\,\%$ random right-censoring to the Weibull-generated data. 
Additionally, we applied a Cox proportional hazards model to the Weibull data to 
determine if the performance of model-based forests degrades 
when the forests do not take the true underlying model as their base model. 

Due to $w - 0.5$ in all scenarios, half of the (negative) predictive effect $\tau(\rx)$ was
added to the prognostic effect.
We refer to the implied scenario -- where one variable which is both prognostic (impact in $\mu(\rx)$) and predictive (impact in $\tau(\rx)$) exists -- as overlay.
Apart from Setup C in which the treatment effect is constant and independent of any covariate,
overlay was present for all scenarios.

Like \AYcite{Dandl et al.}{Dandl_Hothorn_Seibold_2022}, we compared all study settings and outcome types
for a varying number of samples $N \in \{800, 1600\}$
and dimensions $P \in \{10, 20\}$.
All model-based forests were grown with the same hyperparameter options specified in 
Section~\ref{sec:comp}.
We used random forests as implemented in the \pkg{grf} package to estimate
$\pi(\rx)$ for centering $W$ \citep{pkg:grf}.
To estimate $\eta_0(\rx)$ and $\eta_1(\rx)$ to derive $\hat{m}(\rx)$,
we relied on different tree-based estimators depending on the outcome type.
For normally distributed outcomes (models \eqref{m1}),
we used \pkg{grf} regression forests \citep{pkg:grf}.
For all other outcomes, we relied on gradient boosting machines (with adapted loss functions) as implemented
in \pkg{mboost} and \pkg{gbm} \citep{pkg:mboost,pkg:gbm}.
The employed distribution varied depending on the outcome type.

\begin{figure}

\includegraphics[width=\maxwidth]{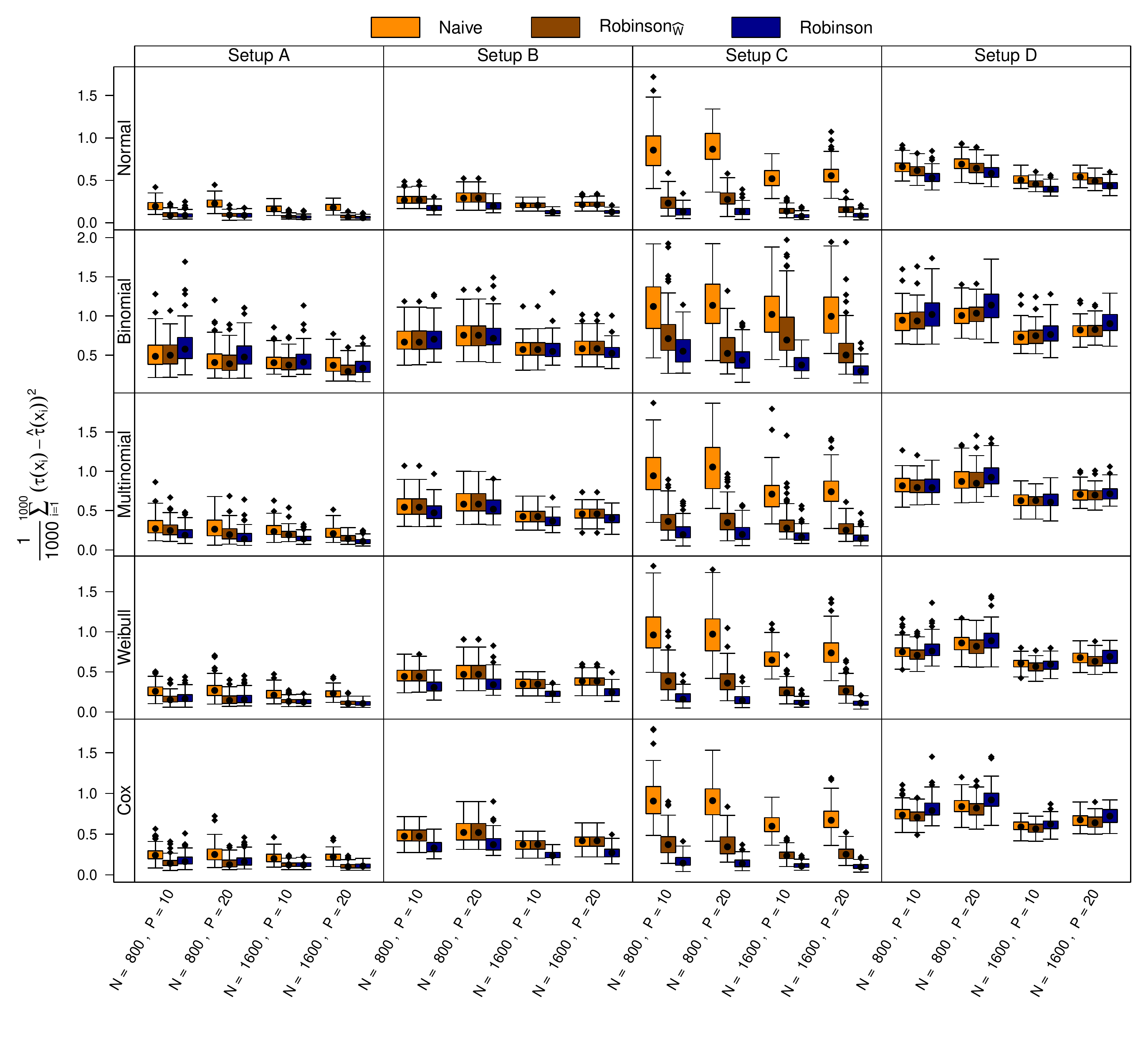} 

\caption{Model-based forest results for the empirical study (Section~\ref{sec:sim}), Cox means a Cox model applied to the
Weibull data. For the Weibull and Cox model, treatment effects $\tau(\rx)$
are estimated as conditional log hazard ratios.
Direct comparison of model-based forests without centering (\textit{Naive}),
model-based forests with local centering according to \AYcite{Robinson}{Robinson_1988} of $Y$ and $W$ (originally proposed) (\textit{Robinson}) or
only of $W$ (\textit{Robinson}$_{\widehat{W}}$).
\label{fig:otherB-nie}}
\end{figure}

In accordance with \AYcite{Dandl et al.}{Dandl_Hothorn_Seibold_2022}, we evaluated the models with respect to the mean squared error $\Ex_\mX\{(\hat{\tau}(\mX) - \tau(\mX))^2\}$ on a test sample of
size $1000$.
The results are shown in Figure~\ref{fig:otherB-nie}
and were statistically analyzed by means of a normal linear mixed model with a log-link.
The model explained the estimated mean squared error for
$\hat{\tau}(\rx)$ by a four-way interaction of the data generating process,
sample size $N$, dimension $P$, and random forest variant.
We estimated the mean squared error ratios between different model-based forest versions according to
the two research questions stated in Section~\ref{subsec:hypotheses}.
The corresponding tables are given in Tables~\ref{tab:lmeradaptive-nie}~to~\ref{tab:lmeradaptive-nie3}.

\begin{table}
\caption{Results of \textbf{RQ 1} for the experimental setups in Section~\ref{sec:sim}.  Comparison of mean
squared errors for $\hat{\tau}(\rx)$ in the different scenarios.  Estimates
and simultaneous $95$ \% confidence intervals were obtained from a normal
linear mixed model with log-link.  Cells printed in bold font correspond to
a superior reference of the \textit{Naive} model-based forests, 
and cells printed in italics indicate an inferior reference.
\label{tab:lmeradaptive-nie}}
\addtolength{\tabcolsep}{-2pt}
\tiny
\begin{tabular}{llllrrrrr}
\hline
&& && \multicolumn{5}{c}{Mean squared error ratio for \textbf{RQ 1}: Robinson vs. Naive}\\
\cline{5-5}\cline{6-6}\cline{7-7}\cline{8-8}\cline{9-9}
DGP&N&P && \multicolumn{1}{c}{Normal}&\multicolumn{1}{c}{Binomial}&\multicolumn{1}{c}{Multinomial}&\multicolumn{1}{c}{Weibull}&\multicolumn{1}{c}{Cox}\\
\hline

Setup A&800    &10      && \textit{0.465 (0.421, 0.512)} & \textbf{1.173 (1.045, 1.316)} & \textit{0.690 (0.629, 0.758)} & \textit{0.672 (0.609, 0.742)} & \textit{0.712 (0.650, 0.781)}\\
       &       &20      && \textit{0.396 (0.359, 0.438)} & \textbf{1.161 (1.014, 1.330)} & \textit{0.600 (0.540, 0.666)} & \textit{0.605 (0.547, 0.669)} & \textit{0.654 (0.596, 0.718)}\\
       &1600   &10      && \textit{0.414 (0.362, 0.474)} &           1.042 (0.892, 1.216) & \textit{0.582 (0.512, 0.662)} & \textit{0.580 (0.508, 0.663)} & \textit{0.589 (0.519, 0.669)}\\
       &       &20      && \textit{0.341 (0.295, 0.395)} &           0.898 (0.751, 1.075) & \textit{0.503 (0.428, 0.591)} & \textit{0.471 (0.405, 0.548)} & \textit{0.495 (0.430, 0.570)}\\
Setup B&800    &10      && \textit{0.643 (0.607, 0.681)} &           1.021 (0.929, 1.121) & \textit{0.868 (0.830, 0.907)} & \textit{0.692 (0.653, 0.733)} & \textit{0.703 (0.668, 0.739)}\\
       &       &20      && \textit{0.658 (0.625, 0.693)} &           0.977 (0.894, 1.067) & \textit{0.906 (0.870, 0.943)} & \textit{0.716 (0.681, 0.754)} & \textit{0.731 (0.699, 0.763)}\\
       &1600   &10      && \textit{0.603 (0.557, 0.653)} &           0.981 (0.873, 1.101) & \textit{0.852 (0.803, 0.903)} & \textit{0.657 (0.607, 0.710)} & \textit{0.656 (0.612, 0.702)}\\
       &       &20      && \textit{0.588 (0.544, 0.636)} &           0.912 (0.811, 1.026) & \textit{0.869 (0.822, 0.917)} & \textit{0.648 (0.603, 0.697)} & \textit{0.653 (0.614, 0.695)}\\
Setup C&800    &10      && \textit{0.153 (0.144, 0.163)} & \textit{0.474 (0.432, 0.520)} & \textit{0.250 (0.233, 0.268)} & \textit{0.174 (0.160, 0.189)} & \textit{0.176 (0.162, 0.191)}\\
       &       &20      && \textit{0.156 (0.147, 0.166)} & \textit{0.359 (0.322, 0.400)} & \textit{0.219 (0.204, 0.235)} & \textit{0.157 (0.142, 0.172)} & \textit{0.161 (0.146, 0.177)}\\
       &1600   &10      && \textit{0.154 (0.139, 0.171)} & \textit{0.361 (0.316, 0.412)} & \textit{0.260 (0.238, 0.284)} & \textit{0.181 (0.160, 0.206)} & \textit{0.187 (0.165, 0.212)}\\
       &       &20      && \textit{0.157 (0.142, 0.173)} & \textit{0.300 (0.256, 0.351)} & \textit{0.215 (0.195, 0.238)} & \textit{0.147 (0.129, 0.169)} & \textit{0.152 (0.133, 0.174)}\\
Setup D&800    &10      && \textit{0.818 (0.801, 0.835)} & \textbf{1.109 (1.037, 1.187)} &           0.996 (0.968, 1.026) & \textbf{1.036 (1.008, 1.065)} & \textbf{1.085 (1.057, 1.113)}\\
       &       &20      && \textit{0.851 (0.835, 0.867)} & \textbf{1.126 (1.058, 1.199)} & \textbf{1.054 (1.028, 1.082)} & \textbf{1.055 (1.029, 1.081)} & \textbf{1.099 (1.075, 1.124)}\\
       &1600   &10      && \textit{0.783 (0.762, 0.805)} &           1.075 (0.985, 1.175) &           0.994 (0.957, 1.032) &           0.968 (0.934, 1.004) &           1.029 (0.995, 1.063)\\
       &       &20      && \textit{0.803 (0.783, 0.824)} & \textbf{1.131 (1.046, 1.223)} &           1.016 (0.983, 1.051) &           1.021 (0.989, 1.053) & \textbf{1.076 (1.046, 1.108)}\\
\hline
\end{tabular}

\end{table}

\begin{table}
\caption{Results of \textbf{RQ 2} for the experimental setups in Section~\ref{sec:sim}.  Comparison of mean
squared errors for $\hat{\tau}(\rx)$ in the different scenarios.  Estimates
and simultaneous $95$ \% confidence intervals were obtained from a normal
linear mixed model with log-link.  Cells printed in bold font correspond to
a superior reference of the \textit{Naive} model-based forests, 
and cells printed in italics indicate an inferior reference.
\label{tab:lmeradaptive-nie2}}
\addtolength{\tabcolsep}{-2pt}
\centering
\tiny
\begin{tabular}{llllrrrrr}
\hline
&& && \multicolumn{5}{c}{Mean squared error ratio for \textbf{RQ 2}: Robinson$_{\hat{W}}$ vs. Naive}\\
\cline{5-5}\cline{6-6}\cline{7-7}\cline{8-8}\cline{9-9}
DGP&N&P && \multicolumn{1}{c}{Normal}&\multicolumn{1}{c}{Binomial}&\multicolumn{1}{c}{Multinomial}&\multicolumn{1}{c}{Weibull}&\multicolumn{1}{c}{Cox}\\
\hline

Setup A&800    &10      &&           1.029 (0.910, 1.164) & \textit{0.820 (0.729, 0.922)} & \textbf{1.259 (1.142, 1.388)} &           0.924 (0.820, 1.042) & \textit{0.844 (0.752, 0.947)}\\
       &       &20      &&           1.060 (0.933, 1.204) & \textit{0.784 (0.679, 0.905)} & \textbf{1.282 (1.144, 1.437)} &           0.935 (0.825, 1.060) & \textit{0.835 (0.740, 0.942)}\\
       &1600   &10      &&           1.126 (0.953, 1.330) &           0.915 (0.781, 1.072) & \textbf{1.370 (1.194, 1.571)} &           1.067 (0.911, 1.250) &           1.015 (0.870, 1.184)\\
       &       &20      &&           1.163 (0.970, 1.395) &           0.887 (0.726, 1.084) & \textbf{1.302 (1.086, 1.561)} &           1.063 (0.881, 1.283) &           0.994 (0.831, 1.188)\\
Setup B&800    &10      && \textbf{1.555 (1.468, 1.647)} &           0.980 (0.892, 1.077) & \textbf{1.152 (1.102, 1.205)} & \textbf{1.445 (1.363, 1.531)} & \textbf{1.423 (1.353, 1.496)}\\
       &       &20      && \textbf{1.520 (1.444, 1.600)} &           1.024 (0.938, 1.119) & \textbf{1.104 (1.060, 1.150)} & \textbf{1.396 (1.327, 1.469)} & \textbf{1.368 (1.309, 1.430)}\\
       &1600   &10      && \textbf{1.658 (1.530, 1.796)} &           1.019 (0.907, 1.144) & \textbf{1.174 (1.107, 1.245)} & \textbf{1.524 (1.409, 1.648)} & \textbf{1.525 (1.424, 1.634)}\\
       &       &20      && \textbf{1.700 (1.574, 1.837)} &           1.097 (0.975, 1.233) & \textbf{1.151 (1.090, 1.216)} & \textbf{1.542 (1.435, 1.657)} & \textbf{1.532 (1.440, 1.629)}\\
Setup C&800    &10      && \textbf{1.871 (1.743, 2.009)} & \textbf{1.377 (1.243, 1.526)} & \textbf{1.577 (1.456, 1.708)} & \textbf{2.331 (2.128, 2.553)} & \textbf{2.388 (2.182, 2.614)}\\
       &       &20      && \textbf{2.081 (1.944, 2.226)} & \textbf{1.294 (1.138, 1.470)} & \textbf{1.718 (1.588, 1.859)} & \textbf{2.565 (2.318, 2.839)} & \textbf{2.611 (2.363, 2.886)}\\
       &1600   &10      && \textbf{1.774 (1.573, 2.001)} & \textbf{2.619 (2.288, 2.999)} & \textbf{1.759 (1.594, 1.942)} & \textbf{2.198 (1.920, 2.517)} & \textbf{2.141 (1.874, 2.446)}\\
       &       &20      && \textbf{1.817 (1.629, 2.026)} & \textbf{1.800 (1.512, 2.144)} & \textbf{1.675 (1.494, 1.877)} & \textbf{2.541 (2.203, 2.932)} & \textbf{2.566 (2.228, 2.956)}\\
Setup D&800    &10      && \textbf{1.136 (1.113, 1.161)} & \textit{0.910 (0.851, 0.974)} &           0.992 (0.964, 1.021) & \textit{0.916 (0.890, 0.942)} & \textit{0.883 (0.860, 0.906)}\\
       &       &20      && \textbf{1.098 (1.077, 1.120)} & \textit{0.898 (0.844, 0.956)} & \textit{0.942 (0.918, 0.966)} & \textit{0.909 (0.886, 0.932)} & \textit{0.881 (0.861, 0.901)}\\
       &1600   &10      && \textbf{1.147 (1.114, 1.180)} &           0.950 (0.871, 1.037) &           0.994 (0.958, 1.032) &           0.965 (0.929, 1.001) & \textit{0.923 (0.892, 0.954)}\\
       &       &20      && \textbf{1.126 (1.097, 1.157)} & \textit{0.890 (0.823, 0.961)} &           0.972 (0.940, 1.005) & \textit{0.922 (0.893, 0.952)} & \textit{0.888 (0.862, 0.914)}\\
\hline
\end{tabular}

\end{table}

\begin{table}
\caption{Results of \textbf{RQ 3} for the experimental setups in Section~\ref{sec:sim}.  Comparison of mean
squared errors for $\hat{\tau}(\rx)$ in the different scenarios.  Estimates
and simultaneous $95$ \% confidence intervals were obtained from a normal
linear mixed model with log-link.  Cells printed in bold font correspond to
a superior reference of \textit{Robinson}$_{\hat{W}}$, 
and cells printed in italics indicate an inferior reference.
\label{tab:lmeradaptive-nie3}}
\addtolength{\tabcolsep}{-2pt}
\centering
\tiny
\begin{tabular}{llllrrrrr}
\hline
&& && \multicolumn{5}{c}{Mean squared error ratio for \textbf{RQ 3}: Robinson vs. Robinson$_{\hat{W}}$}\\
\cline{5-5}\cline{6-6}\cline{7-7}\cline{8-8}\cline{9-9}
DGP&N&P && \multicolumn{1}{c}{Normal}&\multicolumn{1}{c}{Binomial}&\multicolumn{1}{c}{Multinomial}&\multicolumn{1}{c}{Weibull}&\multicolumn{1}{c}{Cox}\\
\hline

Setup A&800    &10      &&           0.972 (0.859, 1.099) & \textbf{1.220 (1.085, 1.373)} & \textit{0.794 (0.720, 0.876)} &           1.082 (0.959, 1.220) & \textbf{1.185 (1.056, 1.329)}\\
       &       &20      &&           0.944 (0.831, 1.072) & \textbf{1.276 (1.105, 1.472)} & \textit{0.780 (0.696, 0.874)} &           1.070 (0.944, 1.212) & \textbf{1.197 (1.061, 1.351)}\\
       &1600   &10      &&           0.888 (0.752, 1.049) &           1.093 (0.933, 1.281) & \textit{0.730 (0.637, 0.838)} &           0.937 (0.800, 1.098) &           0.985 (0.844, 1.149)\\
       &       &20      &&           0.860 (0.717, 1.030) &           1.127 (0.922, 1.378) & \textit{0.768 (0.641, 0.921)} &           0.941 (0.780, 1.135) &           1.006 (0.841, 1.203)\\
Setup B&800    &10      && \textit{0.643 (0.607, 0.681)} &           1.020 (0.929, 1.121) & \textit{0.868 (0.830, 0.907)} & \textit{0.692 (0.653, 0.733)} & \textit{0.703 (0.669, 0.739)}\\
       &       &20      && \textit{0.658 (0.625, 0.692)} &           0.976 (0.894, 1.067) & \textit{0.906 (0.869, 0.943)} & \textit{0.716 (0.681, 0.754)} & \textit{0.731 (0.699, 0.764)}\\
       &1600   &10      && \textit{0.603 (0.557, 0.654)} &           0.981 (0.874, 1.102) & \textit{0.852 (0.803, 0.903)} & \textit{0.656 (0.607, 0.710)} & \textit{0.656 (0.612, 0.702)}\\
       &       &20      && \textit{0.588 (0.544, 0.635)} &           0.912 (0.811, 1.026) & \textit{0.869 (0.822, 0.917)} & \textit{0.649 (0.603, 0.697)} & \textit{0.653 (0.614, 0.695)}\\
Setup C&800    &10      && \textit{0.534 (0.498, 0.574)} & \textit{0.726 (0.655, 0.804)} & \textit{0.634 (0.586, 0.687)} & \textit{0.429 (0.392, 0.470)} & \textit{0.419 (0.383, 0.458)}\\
       &       &20      && \textit{0.481 (0.449, 0.514)} & \textit{0.773 (0.680, 0.878)} & \textit{0.582 (0.538, 0.630)} & \textit{0.390 (0.352, 0.431)} & \textit{0.383 (0.346, 0.423)}\\
       &1600   &10      && \textit{0.564 (0.500, 0.636)} & \textit{0.382 (0.333, 0.437)} & \textit{0.569 (0.515, 0.628)} & \textit{0.455 (0.397, 0.521)} & \textit{0.467 (0.409, 0.534)}\\
       &       &20      && \textit{0.550 (0.494, 0.614)} & \textit{0.555 (0.467, 0.661)} & \textit{0.597 (0.533, 0.669)} & \textit{0.393 (0.341, 0.454)} & \textit{0.390 (0.338, 0.449)}\\
Setup D&800    &10      && \textit{0.880 (0.861, 0.899)} & \textbf{1.099 (1.027, 1.175)} &           1.008 (0.979, 1.037) & \textbf{1.092 (1.061, 1.123)} & \textbf{1.133 (1.104, 1.163)}\\
       &       &20      && \textit{0.911 (0.893, 0.929)} & \textbf{1.113 (1.046, 1.185)} & \textbf{1.062 (1.035, 1.089)} & \textbf{1.101 (1.073, 1.128)} & \textbf{1.136 (1.110, 1.162)}\\
       &1600   &10      && \textit{0.872 (0.848, 0.898)} &           1.052 (0.964, 1.148) &           1.006 (0.969, 1.044) &           1.037 (0.999, 1.076) & \textbf{1.084 (1.048, 1.121)}\\
       &       &20      && \textit{0.888 (0.865, 0.912)} & \textbf{1.124 (1.040, 1.215)} &           1.029 (0.995, 1.064) & \textbf{1.085 (1.050, 1.120)} & \textbf{1.126 (1.094, 1.160)}\\
\hline
\end{tabular}

\end{table}

\subsection{Results}

The results for the normal distribution coincide with the results obtained
by \AYcite{Dandl et al.}{Dandl_Hothorn_Seibold_2022} summarized in Section~\ref{subsec:dandl}. 
To some degree, they also hold for the other distributions.
The boxplots are not directly comparable between different
data generating processes because of different signal-to-noise ratios.
In general, a more informative outcome
(binary $<$ ordered $<$ right-censored $<$ exact normal), more
data (higher $N$), and less noise (lower $P$) leads to better results.
Using a Cox model compared to a Weibull model (last two rows of
Figure~\ref{fig:otherB-nie}) 
did not lead to a major decrease in performance, 
although knowledge of the true functional form of the transformation
function did not enter the Cox modeling process.
 
For Setup A, model-based forests without centering (\textit{Naive}) were unable to cope with complex confounding, but solely centering of the treatment indicator (\textit{Robinson}$_{\hat{W}}$) was valuable.
Additionally adding $\hat{m}(\rx)$ as an offset (\textit{Robinson}) did not further improve the results for the normal, binomial, and Weibull distributions, but an improvement was observed for the multinomial distribution.

For Setup B, the \textit{Robinson} forests performed slightly better in disentangling the more complicated prognostic and predictive effects compared 
to \textit{Naive} and \textit{Robinson}$_{\hat{W}}$ model-based forests.
An exception is the binomial model: without overlay, \textit{Robinson}$_{\hat{W}}$ forests performed similarly to \textit{Robinson} forests.

In Setup C, over all distributions, uncentered model-based forests (\textit{Naive}) failed to overcome the strong confounding effect and therefore did not provide accurate estimates for the treatment effect.
The performance was fundamentally improved by centering the treatment indicator (\textit{Robinson}$_{\hat{W}}$) and was further improved by additionally adding $\hat{m}(\rx)$ as an offset (\textit{Robinson}).

In Setup D -- with unrelated treatment and control arms -- all methods had a higher mean squared error than in the other setups,
as jointly modeling the expected conditional outcomes for both arms has no benefit. 
Apart from the normal distributions, \textit{Robinson} forests
were inferior to the \textit{Robinson}$_{\hat{W}}$ and \textit{Naive} model-based forests.

The empirical evidence of our simulation study can be summarized as follows:
If confounding was present, model-based forests performed better when centering
$W$ by $\hat{\pi}(\rx)$ (\textit{Robinson}$_{\hat{W}}$) compared to not centering $W$ (\textit{Naive}).
Adding $\hat{m}(\rx)$ as an offset (\textit{Robinson}) further improved the performance
-- especially in cases with very strong confounding.
 
\section{Effect of Riluzole on progression of ALS}
\label{sec:als}

Amyotrophic lateral sclerosis (ALS) is a progressive nervous system disease causing loss of muscle control.
The status of the disease as well as the rate of progression is commonly evaluated by the ALS functional rating scale (ALSFRS) \citep{Brooks_Sanjak_Ringel_1996,Cedarbaum_Stambler_Malta_1999}. Here, physical abilities such as speaking, handwriting, and walking are assessed and rated on a scale from 0 (inability) to 4 (normal ability).
In 1995, the FDA approved the first drug to manage and slow progression of ALS, named Riluzole. 
The largest database for study results on the effect of Riluzole offers the Pooled Resource Open-Access Clinical Trials (PROACT) database
-- initiated by the non-profit organization Prize4Life (\url{http://www.prize4life.org}).
The data comes from different randomized and observational studies not disclosed in the data.
Thus, the assumption of random treatment assignment is quite hard to justify in an analysis.
Patient characteristics and treatment group sizes might vary greatly between the centers, which affect both the probability of receiving treatment as well as the outcome.  
To account for these potential confounding effects, we compared the treatment effects estimated by the naive model-based forests to the ones estimated with local centering by Robinson. 
As in Section~\ref{sec:sim}, we use random forests to estimate the propensity scores to center $W$ and gradient boosting machines (with adapted loss functions) to estimate the values of the linear predictors $\eta_0(\rx)$ and $\eta_1(\rx)$ to center $Y$. Model-based forests, random forests, and gradient boosting machines rely on the hyperparameter values stated in Section~\ref{sec:comp}.
As for \AYcite{Seibold et al.}{Seibold_Zeileis_Hothorn_2017} and \AYcite{Korepanova et al.}{Korepanova_Seibold_Steffen_2019}, 16 phase II and phase III randomized trials and one observational study from the PROACT database serve as a training dataset.
We analyze the effect of Riluzole with respect to two outcome variables: survival time and the handwriting ability score approximately six months after treatment
-- an item of the ALSFRS. 
We omitted observations with missing outcome values.
As splitting variables, \AYcite{Seibold et al.}{Seibold_Zeileis_Hothorn_2017} used demographic, medical history, and family history data, which were informative in the sense that not more than half of their values were missing.

\begin{figure}[t]
	\centering
	\includegraphics[width=0.65\textwidth]{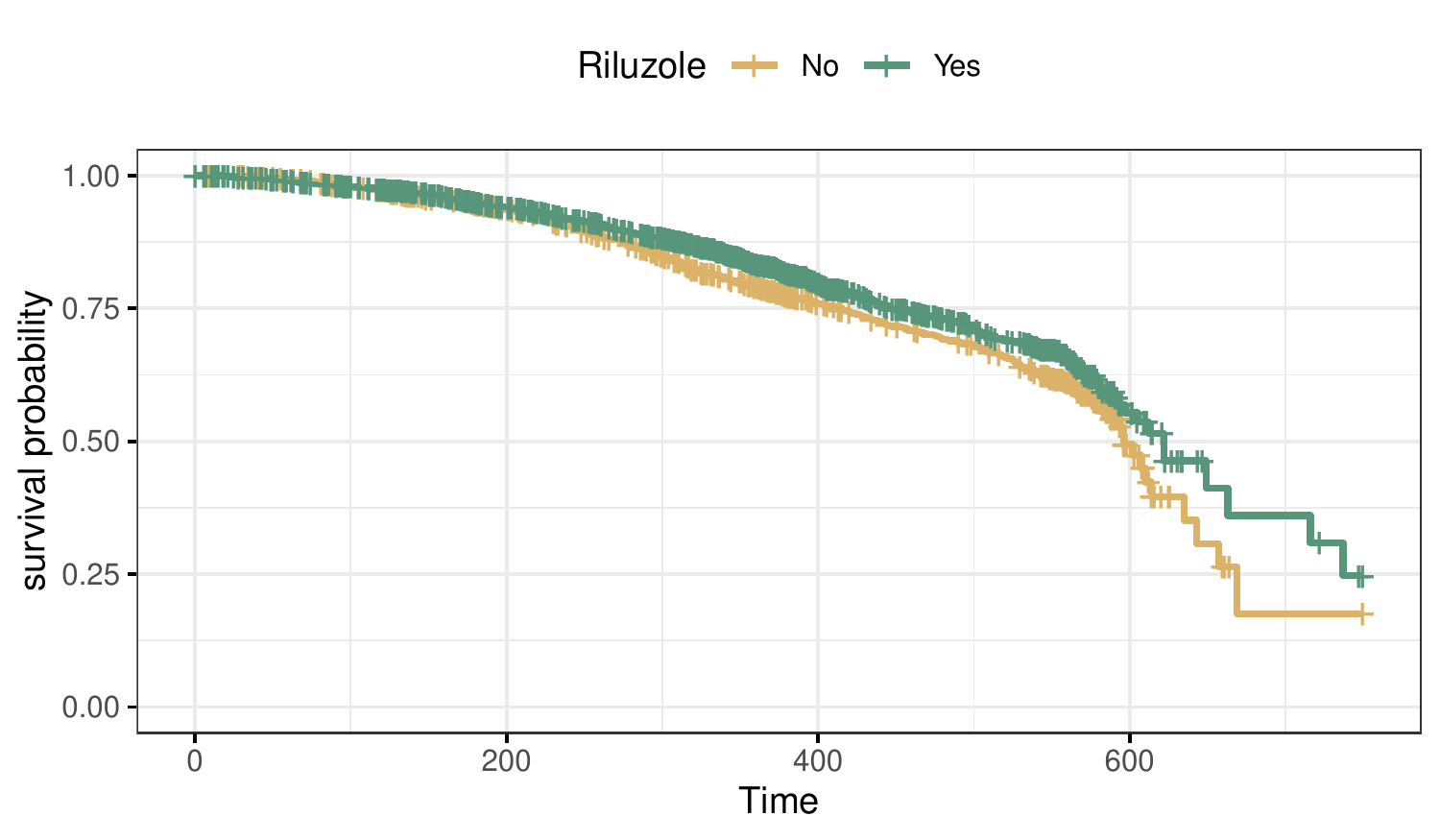}
	\caption{Kaplan-Meier curves of survival probability for both treatment arms. 
		\label{fig:baseprob}}
\end{figure}

\subsection{Survival Time}
The dataset for the survival time contains 3306 observations and 18 covariates.
Of the 3306 observations, 2199 received Riluzole. 
Because very few patients had event times that exceed those of the others by a factor of two, 
we artificially censored five observations with (censoring or event) times of more than 750 days.
The Kaplan-Meier estimates of survival probabilities for both treatment arms of the preprocessed dataset are shown in Figure~\ref{fig:baseprob}.  
Overall, the estimated survival curves are very close to each other, and the treated group has only a slight survival advantage compared to the untreated group. 
As a base model, we use a Cox proportional hazards model. 
We compared treatment effects from two approaches: the naive uncentered model-based forests (\textit{Naive}) and the model-based forest with Robinson's orthogonalization (\textit{Robinson}).

\subsubsection{Personalized models}
For the naive model-based forests, the underlying Cox proportional hazards
base model for the survival outcome $T$ was, on the hazard scale,
\begin{equation*}
\lambda(t) = \lambda_0(t) \exp(\mu + \tau w) 
\end{equation*}
Because $\lambda_0(t)$ contains an intercept term, $\mu$ is not identified
(and was constraint to $\mu \equiv 0$). 
The treatment effect $\tau$ is the log-hazard ratio of the treated versus untreated patients
and our aim is to replace a constant marginal effect $\tau$ with a
heterogenuous (and thus conditional) log-hazard ratio $\tau(\rx)$ and,
simultaneously, to estimate prognostic effects $\mu(\rx)$.

\begin{figure}[t!]
	\centering
	\includegraphics[width=.9\textwidth]{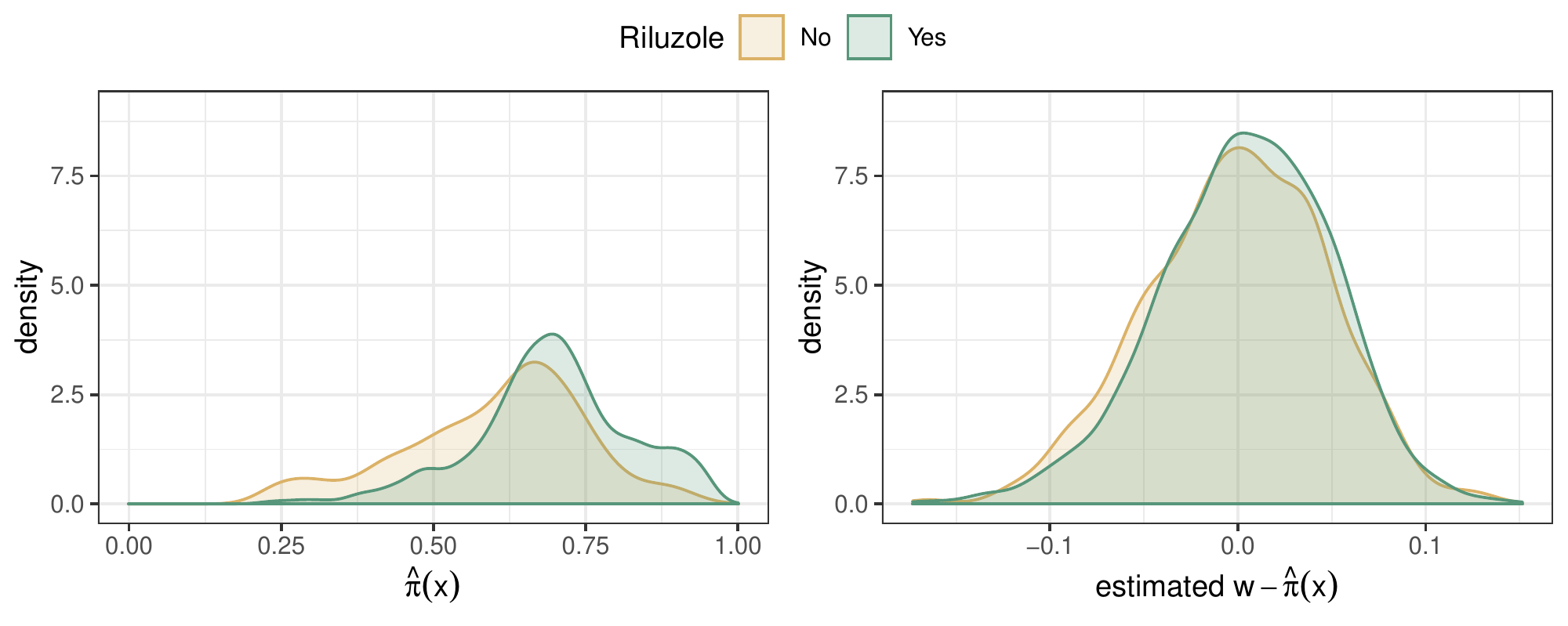}
	\caption{Distribution of estimated propensities $\hat{\pi}(\rx)$ (left) and estimated propensities of the centered treatment indicators (right, Robinson's strategy) as estimated by regression forests for the two
	treatment groups. 
		\label{fig:overlap}}
\end{figure}

For Robinson's strategy, we first centered the treatment indicator $W$ by estimating the propensity scores $\pi(\rx) = \Prob(W \mid \rX = \rx)$ using a regression forest. 
Figure~\ref{fig:overlap} compares the distributions of estimated propensity scores (left) and of the estimated centered treatment $W - \hat{\pi}(\rx)$ (Robinson's strategy, right), both obtained
from regression forests.  We can already see a decent overlap of propensity
scores in the two treatment arms without centering, but the overlap
increases if the strategy by Robinson was applied.

In addition to centering $W$, Robinson's strategy requires the estimation of $m(\rx)$ to use as an offset (see Section~\ref{subsec:ortho}). 
As in Section~\ref{sec:sim}, we used gradient boosting machines (with the negative log partial likelihood of the Cox proportional hazards model as a loss) to estimate the natural parameters $\eta_0(\rx)$ and $\eta_1(\rx)$ for the control and treatment group, respectively \citep{Friedman_2001}.
The offset $m(\rx)$ for each observation is equal to the sum of natural parameter estimates weighted by $\hat{\pi}(\rx)$  (see equation~\eqref{eq:robinsonm}).
The final base model for model-based forests using Robinson's orthogonalization is 
\begin{equation*}
\lambda_R(t) = \lambda_0(t) \exp(\mu + \tau (w - \hat{\pi}(\rx)) + \hat{m}(\rx)). 
\end{equation*}

\begin{figure}[t!]
	\centering
	\includegraphics[width=0.65\textwidth]{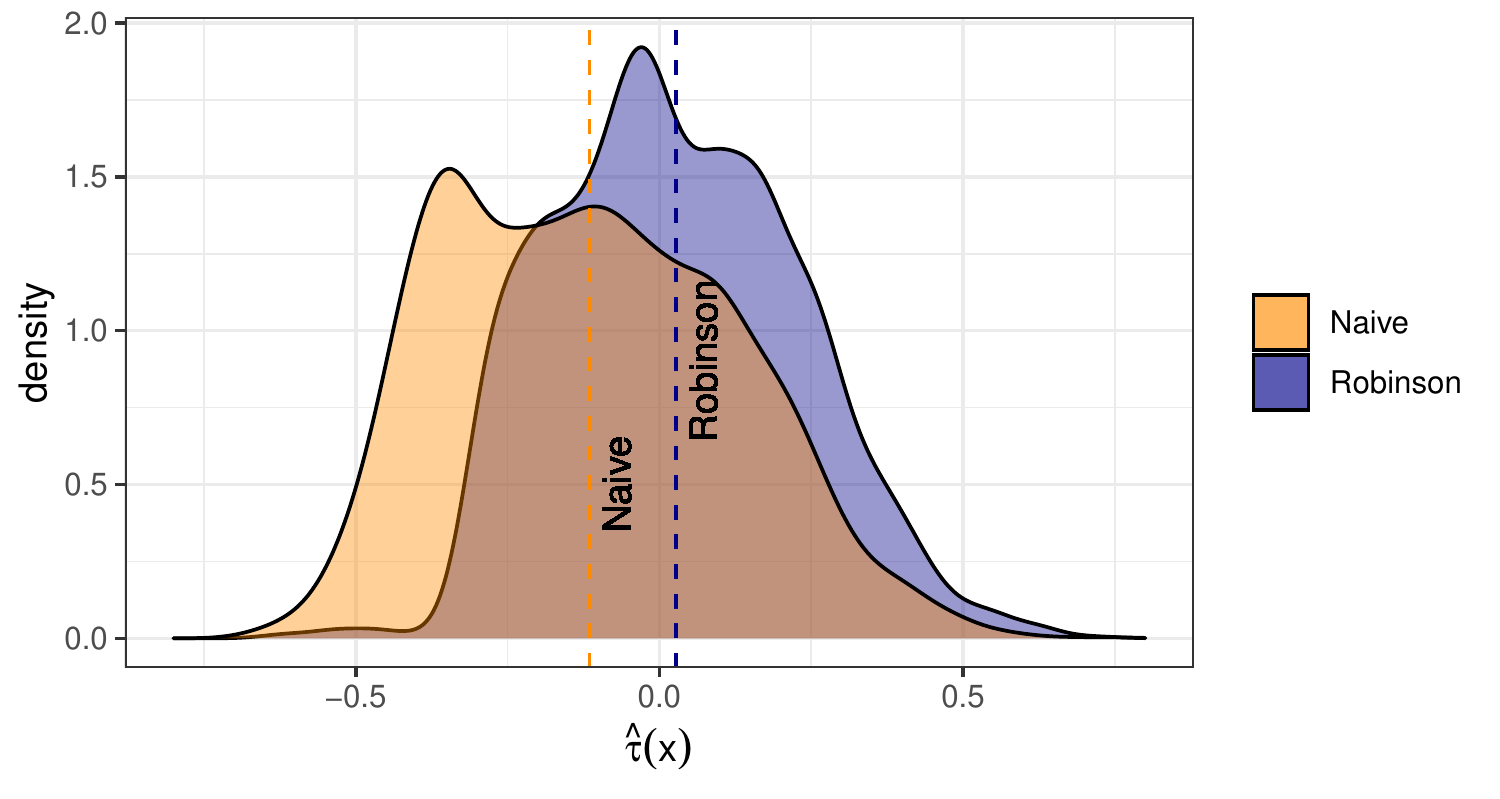}
	\caption{Kernel density estimates of the personalized treatment estimates for the naive model-based forest (\textit{Naive}) and for the model-based forest with Robinson orthogonalization (\textit{Robinson}).
		\label{fig:densalssurv}}
\end{figure}

\subsubsection{Model-based forests}
The corresponding base models serve as an input for the model-based forests to estimate personalized effects of Riluzole.
Figure~\ref{fig:densalssurv} compares the kernel density estimates of $\tau(\rx)$ for each forest version (\textit{Naive} and \textit{Robinson}).
The naive approach reveals that on average the treatment reduced the hazard compared to no treatment, whereas the model-based forest with centering according to Robinson obtained weaker effects of Riluzole with more mass centered around 0.

A meta-analysis of previous studies by \AYcite{Andrews et al.}{Andrews_Jackson_Heiman_2020}, also yielded a mixed picture: only eight of the 15 studies meeting their inclusion criteria showed a statistically significant increase of median survival time due to Riluzole.  

Over all strategies, for both approaches there were some patients for which Riluzole was estimated to increase the hazard.  
The dependency plots in Figures~S. 4 and S. 5 in the Supplementary Material provide indications of the characteristics of the group of harmed individuals. For example, both the naive and centering approach agree that for patients with atrophy or fasciculation, Riluzole intake would increase the hazard.
The estimated effects differed most between the uncentered forest (\textit{Naive}) and the orthogonalized forest (\textit{Robinson}) for the covariate sex (Figure~S. 4 (c)), the covariate of whether patients swallow, and for the covariate specifying whether cases in the same generation exist (Figure~S. 5 (f) and (i)). 

For the variables time onset treatment, age, height and weakness the dependency plots (Figure~S.~5~(a),~(d),~(e) and Figure~S.~6~(g)) of the \textit{Naive} forest agree with the ones of \AYcite{Seibold et al.}{Seibold_Zeileis_Hothorn_2017}: for middle-aged people with a longer time between disease onset and start of treatment, lower height, and no weakness, the treatment appears to be more beneficial.
By considering confounding effects due to orthogonalization (\textit{Robinson} forests), these effects diminished. 
For \AYcite{Korepanova et al.}{Korepanova_Seibold_Steffen_2019} the effect of Riluzole was also rather weak and showed low heterogeneity across covariates.

\subsection{Handwriting Ability Score}
\begin{figure}[t!]
	\centering
	\includegraphics[width=0.65\textwidth]{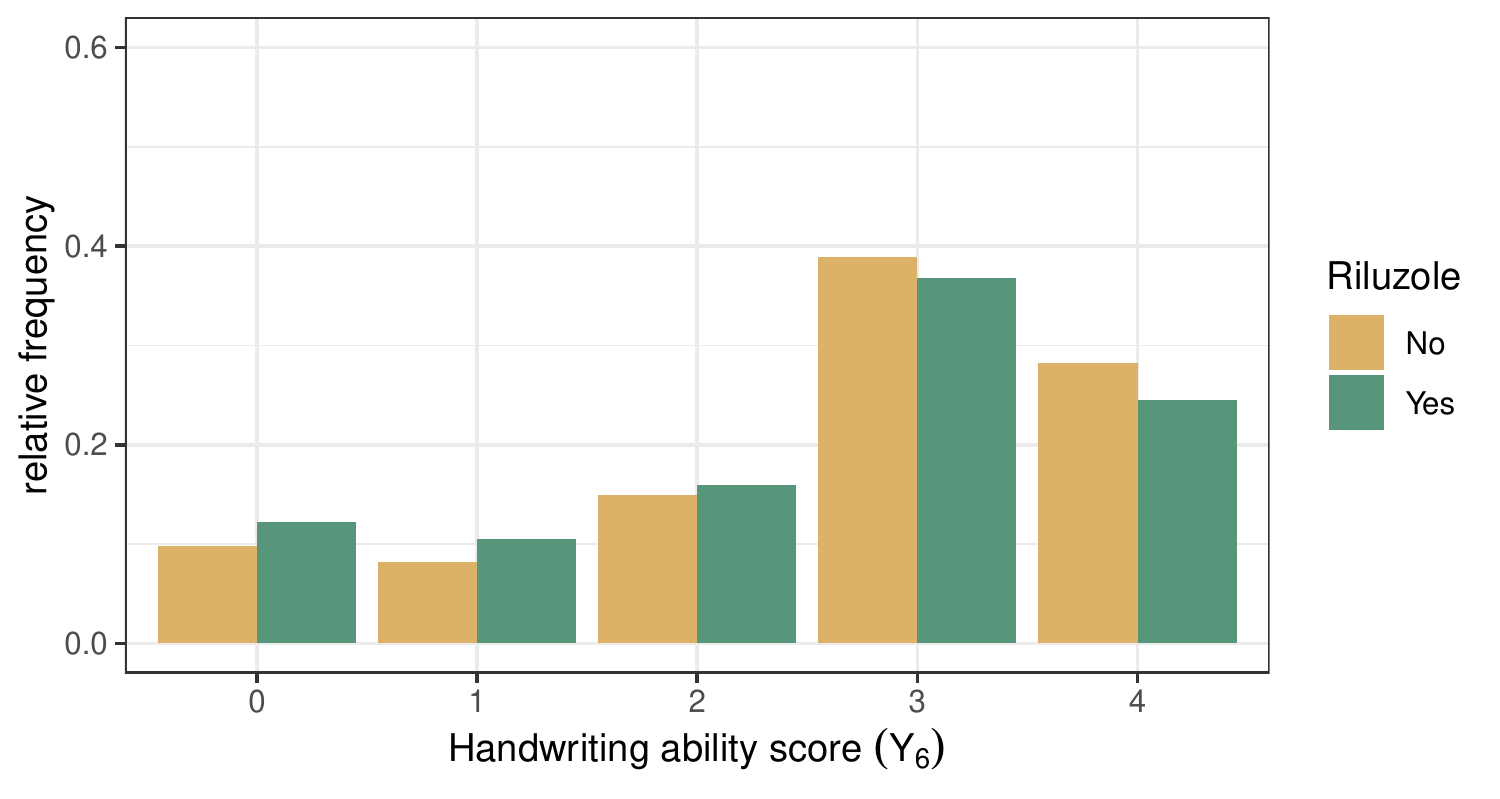}
	\caption{Relative frequency distribution plot of the handwriting ability score ($Y_6$) (left) and of changes of the handwriting ability score over six months ($Y_6 - Y_0$) (right) for both treatment arms. 
	Frequencies were calculated relative to the treatment indicator. 
		\label{fig:basebar}}
\end{figure}

The dataset for the handwriting ability score -- an ordinal outcome with five categories -- contains 2538 observations and 58 covariates. Besides the covariate age, all covariates had missing values (but less than 50 \% of the values were missing per variable enforced by the preprocessing step stated at the beginning of this section).
Of the 2538 observations, 1754 received Riluzole, and 784 did not.
Figure~\ref{fig:basebar} displays the frequency of the ability scores for both treatment groups. Most of the patients have an ability score of 3 or 4 (normal ability); only a few have ability scores less than 2. 
Note that the plot shows the conditional proportions given the treatment indicator. 
We chose a proportional odds logistic regression model as a base model for the model-based forests
-- once without further adaptions (\textit{Naive}), and once parameterized with centered $W$ and with an offset (\textit{Robinson}).

\begin{figure}[t!]
	\centering
	\includegraphics[width=0.9\textwidth]{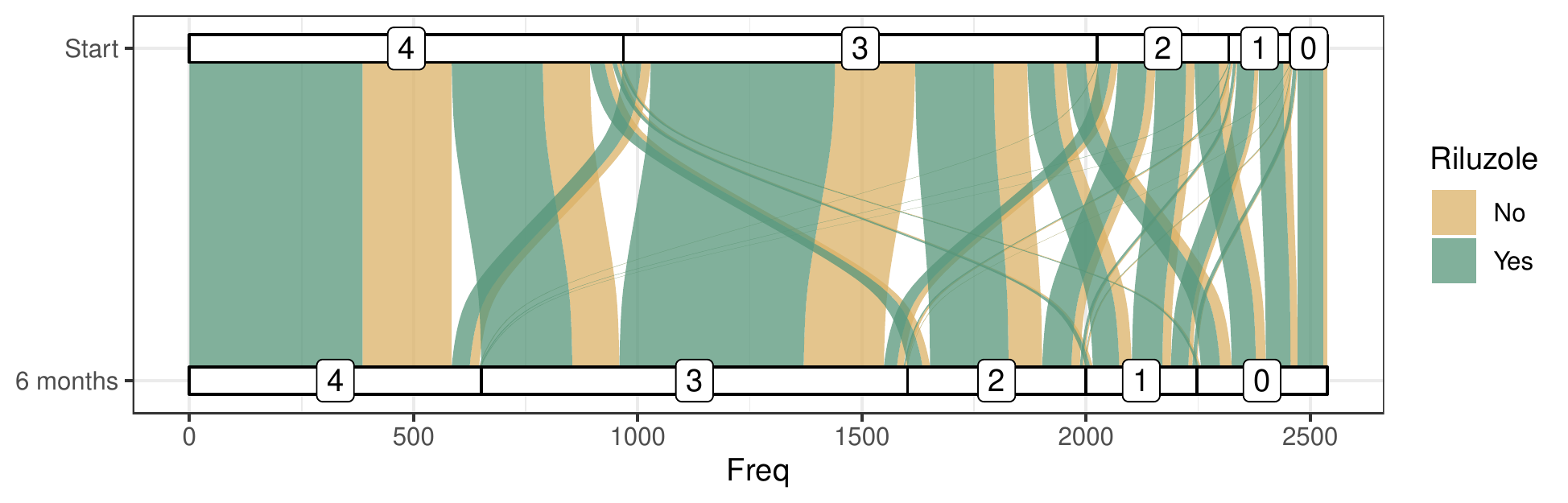}
	\caption{Alluvial plot of the progression of the handwriting ability score over six months for both treatment arms. 
		\label{fig:alluv}}
\end{figure}

In addition to the handwriting ability score after six months, the ability score values at treatment start are also available. In the following, we denote $Y_6$ as the handwriting score after six months and $Y_0$ as the handwriting score at the beginning of the treatment period. 
To account for the ability level at treatment start, $Y_0$ served as an additional splitting variable for both model-based forests (\textit{Naive} and \textit{Robinson}) and was included in $\mathbf{X}$.  
The alluvial plot in Figure~\ref{fig:alluv} breaks down the change in each ability class over six months.
Overall, for most patients, the handwriting ability remained constant over the six months or worsened slightly.
Rarely, patients experienced a progression to both extremes (0 to 4, or 4 to 0). 
These results hold regardless of whether patients received Riluzole or not.

\subsubsection{Personalized models}
The proportional odds logistic regression model for the naive model-based forests is defined as \citep{Agresti_2002,Venables_2002} 
\begin{equation*}
\logit(\Prob(Y_6 \le k|\rX = \rx, W = w, Y_0 = y_0)) = \vartheta_k(\rx, y_0) - \tau(\rx, y_0)w
\end{equation*}
with $k \in \{0, ..., 3\}$ as the ordinal ability score classes. The
parameters $\vartheta_k$ are increasing thresholds, depending on covariates
$\rx$ and the initial score $y_0$.
Due to the proportional odds assumption, the treatment effect $\tau(\rx, y_0)$ is the same for all scores $k$.
Negative $\tau(\rx, y_0)$ indicate a negative effect of Riluzole, as treated patients are expected to have a higher odds of low writing ability scores compared to untreated patients. 

As for the survival forest, we used regression forests to estimate propensity scores $\pi(\rx, y_0)$ and a gradient boosting machine (with adapted loss functions for the proportional odds model) to estimate the natural parameters $\eta_0(\rx, y_0)$ and $\eta_1(\rx, y_0)$. 
The personalized model for the model-based forest with Robinson orthogonalization was specified as 
\begin{equation*}
\logit(\Prob(Y_6 \le k|\rX = \rx, W = w, Y_0 = y_0)) = \vartheta_k(\rx, y_0) -
[\hat{m}(\rx, y_0) + \tau(\rx, y_0)\{w - \hat{\pi}(\rx, y_0)\}]
\end{equation*} 
with $\hat{m}(\rx, y_0)$ as defined in equation~\eqref{eq:robinsonm}.

\begin{figure}[t!]
	\centering
	\includegraphics[width=0.85\textwidth]{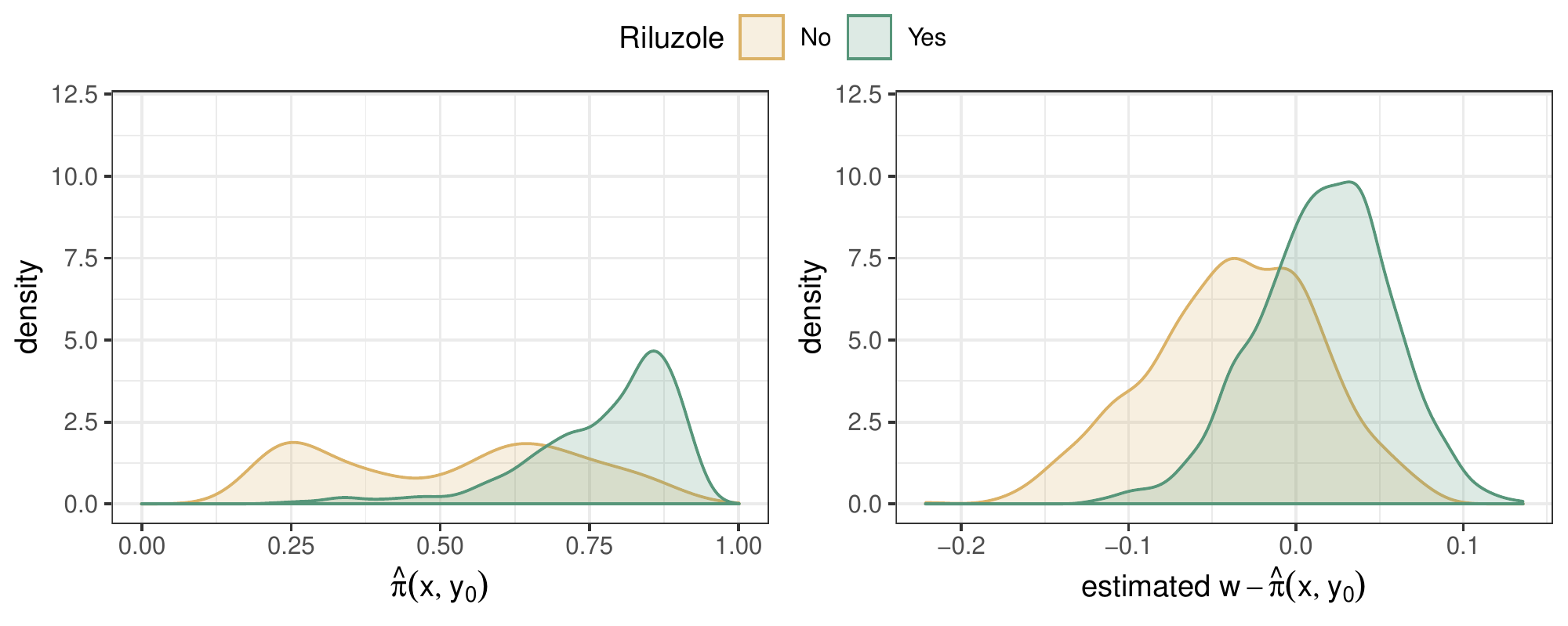}
	\caption{Estimates returned by the regression forest (rf) for orthogonalization of the treatment indicator: left for original $W$ as an outcome in the rf such that it estimates  propensity scores $\pi(\rx, y_0)$; right for the centered treatment indicator $W - \hat{\pi}(\rx, y_0)$ as an outcome in the rf.
		\label{fig:overlapalsfrs}}
\end{figure}

Figure~\ref{fig:overlapalsfrs} compares the estimated treatment indicators with $W$ as the outcome in the random forest without centering (left), with $(W - \hat{\pi}(\rx, y_0))$ as the outcome in the random forest (right). Before centering, there is a lack of overlap of the propensity scores; the distribution of $\hat{\pi}$ for the control group is bimodal, and the distribution for the treatment group is heavily left-skewed. After centering, the distributions of the estimated $W-\hat{\pi}(\rx, y_0)$ for the treatment groups move closer together and have a similar unimodal shape. 
However, there is still a lack of overlap of the groups, which indicates that important covariates to explain the remaining heterogeneity in the two treatment groups seem to be missing.

\subsubsection{Model-based forests}

\begin{figure}[t!]
	\centering
	\includegraphics[width=0.65\textwidth]{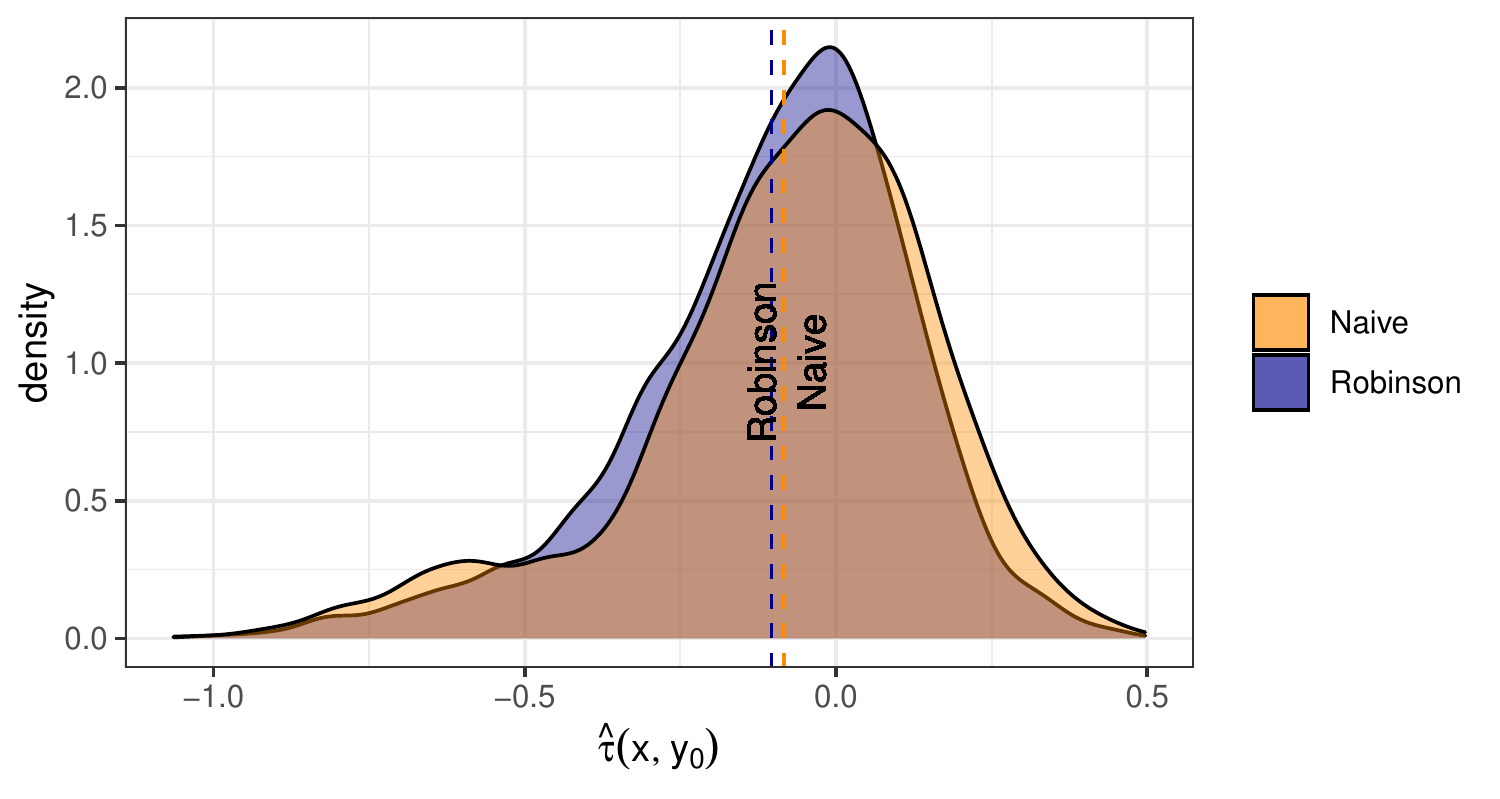}
	\caption{Kernel density estimates of the personalized treatment estimates for the naive model-based forest (\textit{Naive}) vs. the forest with Robinson orthogonalized (\textit{Robinson}).
		\label{fig:densalsfrs}}
\end{figure}

The proportional odds logistic regression models served as a base model for the (\textit{Naive} and \textit{Robinson}) model-based forests to derive personalized treatment effects. 
Figure~\ref{fig:densalsfrs} displays the kernel density estimates of $\tau(\rx, y_0)$ for each forest version (\textit{Naive} and \textit{Robinson}). 
Both random forests estimate on average a negative effect of Riluzole. 
Naive model-based forest estimated on average a log-odds of $\bar{\tau} = - 0.08$, which indicates that treated patients have a 0.08 points higher log-odds for low writing scores than untreated patients.
The distribution of $\hat{\tau}(\rx, y_0)$ for the model-based forest relying on the Robinson orthogonalization is slightly shifted to the left ($\bar{\tau} = -0.10$). 
For a larger subgroup of patients, the naive approach estimates a negative effect of Riluzole ($-1 \le \tau(\rx, y_0) \le -0.5$), meaning that patients receiving treatment with Riluzole have higher odds of low writing scores than untreated patients. 
According to the dependency plots (Figures~S.~6~to~S.~11 in the Supplementary Material), this subgroup could be identified as having the low initial ability scores (left side of Figure~S.~6~(a)).
For all other splitting variables, the distributions of estimated treatment effects are very similar. 

\section{Discussion and outlook}

HTE estimation is a challenging
problem, especially for observational studies and even more
when the outcome cannot be modeled by a linear model.
In this work, we investigated several versions of
model-based forests for the estimation of potentially complex HTEs $\tau(\rx)$ based on observational data with various outcome types based on the orthogonalization strategy by Robinson \citep{Robinson_1988}. 
These investigations suggest the
following workflow for model-based forests:
(1) estimate propensities $\pi(\rx)$ using some machine learning procedure
(binary random forests are a good default), (2) center the treatment
indicator $w - \hat{\pi}(\rx)$ for each observation, (3) setup an
appropriate model for the outcome conditioning on the centered treatment and -- if possible -- add an offset for centering $Y$, (4) use model-based forests to estimate
predictive and prognostic effects $\tau(\rx)$ and $\mu(\rx)$ simultaneously.
Notably, $\tau(\rx)$ is the CATE only in specific models,
especially a linear or log-linear model. 
We demonstrate these steps by estimating the individual effects of Riluzole for ALS patients using survival times and ordinal ability scores as outcomes. 
 
Our work still leaves open questions for example how model-based forests perform for survival data for which the censoring procedure is not randomized but depends on $\rX$, or how ($k$-fold) cross-fitting influences the performance, where only one part of the data is used to estimate nuisance parameters and the other part to estimate $\tau(\rx)$ \citep{Chernozhukov_Chetverikov_Demirer_2018}.
We leave investigations to these questions to future research. 

Last but not least, we want to emphasize that all approaches for estimating HTEs -- including those presented in this work -- rely on strong and typically untestable assumptions.
For example, for models beyond mean regression, $\hat{\tau}(\rx)$
cannot be expected to be robust against missing covariates or other
violations of model assumptions due to non-collapsibility.
Consequently, results from these approaches in practical applications should be evaluated with the utmost caution, reservation, and humility.

\section{Computational details}
\label{sec:comp}

For all computations, we used \textsf{R} version
4.1.1 \citep{Rcore},
with the following add-on packages:
\pkg{model4you} \citep{pkg:model4you}, \pkg{trtf} \citep{pkg:trtf},
\pkg{partykit} \citep{pkg:partykit}, \pkg{grf} \citep{pkg:grf}, \pkg{mboost}
\citep{pkg:mboost}, and \pkg{gbm} \citep{pkg:gbm}.

Model-based forests were always grown with \code{M} $= 500$ trees
(\code{model4you::pmforest} default) with a minimum node size of \code{node}
$= 14$, number of chosen variables per split \code{mtry} $=P$, and
subsampling. These settings were also used by \AYcite{Dandl et al.}{Dandl_Hothorn_Seibold_2022}.
Transformation forests implemented in the \pkg{trtf} package
fitted the Weibull transformation forests of Section~\ref{sec:sim} \citep{pkg:trtf, Hothorn_Zeileis_2021}.

Propensity scores $\pi(\rx)$ were estimated with \pkg{grf} (honest) regression forests
with 125 trees, a minimum node size of 5, and subsampling.
Natural parameters $\eta_0(\rx)$ and $\eta_1(\rx)$ and probability of
not being censored were estimated with gradient boosting machines implemented in the
\pkg{mboost} or \pkg{gbm} packages.
The used maximum tree depth was 2 (default of \code{mboost::blackboost}), and a loss function that differed depending on the outcome type was also employed \citep{pkg:mboost, pkg:gbm}.

Ratios and confidence intervals presented in Table~2 were calculated using generalized
linear mixed models of the \pkg{glmmTMB} package \citep{pkg:glmmTMB}.
Post-hoc inference relied on the \pkg{multcomp} package \citep{pkg:multcomp}.

All study settings are available in a dedicated \proglang{R} package called
\pkg{htesim} \citep{htesim}. It is published on Github: \url{https://github.com/dandls/htesim}.
 
\section*{Funding}
TH received funding from the Swiss National Science Foundation, Grant No.~200021\_184603, Horizon 2020 Research and Innovation
Programme of the European Union under grant agreement number 681094, and is
supported by the Swiss State Secretariat for Education, Research and
Innovation (SERI) under contract number 15.0137.

\newpage

\begin{appendix}
	
\renewcommand\thefigure{S.~\arabic{figure}}    
\renewcommand\thetable{S.~\arabic{table}}
\setcounter{table}{0}
\setcounter{figure}{0}

\section{Noncollapsibility}

As mentioned in Section~2.4, for members of the exponential family without an identity or linear link function the marginal and conditional treatment effects are not collapsible.
This means that the mean of the conditional treatment effects given a covariate are not equal to the marginal treatment effect estimate over the same covariate \citep{Greenland_Pearl_Robins_1999}.
This happens if the covariate conditioned on is associated with the outcome of interest. 
Caution is necessary on multiple stages of the estimation process of $\tau(\rx)$ as soon as we condition on other covariates, for example, because these covariates are assumed to be sufficient to control for confounding \citep{Daniel_Zhang_Farewell_2021}. 

In case of Robinson's orthogonalization, misspecification of $m(\rx)$
translates into biased estimators for $\tau(\rx)$, even under randomized
treatments. This also applies if one ignores the estimation of $\mu(\rx)$ 
at all and only concentrates on $\tau(\rx)$.
This is not the case for the linear model (identity link function) 
since misspecifications are absorbed in the additive error term and 
do not influence the estimation of $\tau(\rx)$ \citep{Gao_Hastie_2022}.

\subsection{Review Gao and Hastie (2022)}
\AYcite{Gao and Hastie}{Gao_Hastie_2022} extended the orthogonalization strategy of \AYcite{Robinson}{Robinson_1988} to improve robustness to both confounding and noncollapsibility.
The authors propose 
\begin{equation}
a(\rx) = \frac{\pi(\rx) \frac{\partial \gamma(\eta_1(\rx))}{\partial \eta}}{\pi(\rx) \frac{\partial \gamma(\eta_1(\rx))}{\partial \eta} + (1 - \pi(\rx)) \frac{\partial \gamma(\eta_0(\rx))}{\partial \eta}}
\label{eq:gaoa}
\end{equation}
and 
\begin{equation*}
\nu(\rx) = a(\rx) n_1(\rx) + (1-a(\rx)) n_0(\rx)
\label{eq:gaonu}
\end{equation*}
instead of  $\pi(\rx)$ (equation~(9)) and $m(\rx)$ (equation~(14)), respectively, where $\gamma(\eta)$ denotes the inverse of the canonical link function. Its derivative is equal to the variance function  of the exponential family. 
Therefore, $a(\rx)$ is larger if an observation is likely to be treated (which also holds for Robinson's orthogonalization) or if the response variance is higher under treatment compared to no treatment. As a consequence of the latter, the influence of spuriously influential natural parameter values is reduced for more robustness to misspecifications \citep{Gao_Hastie_2022}.

For Gaussian responses, $a(\rx) = \pi(\rx)$ and $\nu(\rx) = m(\rx)$ holds, while for other distributions the terms differ.
For example, for Bernoulli distributed $Y$, the closed form $a(\rx)$ is
\begin{equation}
a(\rx) = \frac{\pi(\rx)}{\pi(\rx) + (1-\pi(\rx)) \frac{p_0(\rx)(1-p_0(\rx))}{p_1(\rx)(1-p_1(\rx))}} 
\label{eq:binomgaoa}
\end{equation}
where $p_w(\rx) = \Prob(Y = 1|\rX = \rx, W = w)$. 

The noncollapsibility issue is not only present for distributions of the exponential family.
Also the Cox model suffers from noncollapsibility \citep{Greenland1996,AalenCookRoysland2015}.
This is in contrast to accelerated failure time models (such as the Weibull proportional hazards model), which can be rewritten as
location-scale models and therefore are indeed collapsible \citep{AalenCookRoysland2015}.  
For the Cox model, Gao and Hastie remark that with knowledge of the baseline hazard function and without censoring, the cumulative hazard function follows an exponential distribution. 
For the exponential distribution, $a(\rx)$ and $\nu(\rx)$ are equal to $\pi(\rx)$ and $m(\rx)$ \citep{Gao_Hastie_2022}.

In case of random censoring, the probability of not being censored under both treatment arms needs to be considered for the estimation of $a(\rx)$ and $\nu(\rx)$
\begin{equation}
a(\rx) = \frac{\pi(\rx) \Prob(C \ge Y |\rX = \rx,  W = 1)}{\pi(\rx) \Prob(C \ge Y | \rX = \rx,  W = 1) + (1 - \pi(\rx)) \Prob(C \ge Y | \rX = \rx, W = 0)} 
\label{eq:coxgaoa}
\end{equation}

\begin{equation}
\nu(\rx) = a(\rx) \eta_1(\rx) + (1- a(\rx)) \eta_0(\rx). 
\label{eq:coxgaonu}
\end{equation}
The nuisance parameter $a(\rx)$ is larger if an observation is likely to be treated or likely to be not censored.
Consequently, the influence of likely to be not censored observations for the estimation of $\tau(\rx)$ is increased. 
Above's $a(\rx)$ and $\nu(\rx)$ guarantee protection to misspecified nuisance parameter if the baseline hazard is known. 
If it is unknown and the partial likelihood is used -- this is not guaranteed.
Despite this lack of guarantee, Gao and Hastie, 2022, obtained promising results in their simulation study \citep{Gao_Hastie_2022}.

\subsection{Strategies against confounding and noncollapsibility}
An interesting question is if replacing $\hat{\pi}(\rx)$ and $\hat{m}(\rx)$ by 
$\hat{a}(\rx)$ and $\hat{\nu}(\rx)$, respectively, also helps to additionally tackle noncollapsibility when applying model-based forests. 
We can update the linear predictor for model-based forests in case of generalized linear models to
\begin{equation*}
g(\Ex(\rY \mid \rX = \rx, W = w)) =  \hat{\nu}(\rx) + \tilde{\mu}(\rx) + \tau(\rx)(w - \hat{a}(\rx)).
\label{eq:gaoglm}
\end{equation*}
Gao and Hastie additionally derived estimators for $a(\rx)$ and $\nu(\rx)$ for the Cox model which -- compared to the Weibull model -- is not collapsible.
For the Cox model, the natural parameter of equation~(8) could be updated to
\begin{equation*}
\eta_w(\rx) = \hat{\nu}(\rx) + \tau(\rx) (w - \hat{a}(\rx))
\label{eq:gaotf}
\end{equation*}     

with $a(\rx)$ and $\nu(\rx)$ as defined in equations~\eqref{eq:coxgaoa}~and~\eqref{eq:coxgaonu}. 

We call this version of model-based forests in the following \textit{Gao} approach. Before we apply model-based forests, 
we need to estimate $\pi(\rx)$, $\eta_0(\rx)$, $\eta_1(\rx)$ as well as $\frac{\partial \upsilon(\eta_1(\rx))}{\partial \eta}$ for exponential families and $\Prob(C \ge Y | \rX = \rx, W = w)$ for Cox models. As in Section~3.3, we state some research questions that are empirically inspected in the upcoming section.

\paragraph{RQ 4:}
	How do model-based forests centered according to Gao and Hastie (\textit{Gao}) perform compared to model-based forest with Robinson strategy (\textit{Robinson}) for the simulation settings of Section~4?

\noindent Similar to RQ 2, we could solely center $W$ by $a(\rx)$ without including an offset. We call this approach \textit{Gao}$_{\hat{W}}$ in the following.

\paragraph{RQ 5:}
	How do model-based forest with solely centered $W$ by $\hat{a}(\rx)$ (\textit{Gao}$_{\hat{W}}$) perform compared to model-based forests with solely centered $W$ by $\hat{\pi}(\rx)$ \textit{Robinson}$_{\hat{W}}$ for the simulation study settings of Section~4?

\begin{table}[ht]
	\renewcommand{\arraystretch}{1.2}
	\begin{center}
		\begin{threeparttable}
			\caption{Updated overview of proposed model-based forest versions (Table~1) for observational data.}
			\begin{tabular}{llll} \hline
				Method & Linear Predictor & Definitions  \\ \hline
				\textit{Naive} & $\mu(\rx) + \tau(\rx) \, \, w$ &  \\ \hline
				\textit{Robinson}$_{\hat{W}}$ & $\mu(\rx) + \tau(\rx) (w - \hat{\pi}(\rx))$ & $\pi(\rx) = \Prob(W = 1|\rX = \rx)$ \\
				\textit{Robinson} & $\tilde{\mu}(\rx) + \tau(\rx) (w - \hat{\pi}(\rx)) + \hat{m}(\rx)$ & $m(\rx) = \pi(\rx) \eta_1(\rx) - (1- \pi(\rx)) \eta_0(\rx)$  \\ \hline
				\textit{Gao}$_{\hat{W}}$ & $\mu(\rx) + \tau(\rx) (w - \hat{a}(\rx))$ & $a(\rx) = \frac{\pi(\rx) \frac{\partial \gamma(\eta_1(\rx))}{\partial \eta}}{\pi(\rx) \frac{\partial \gamma(\eta_1(\rx))}{\partial \eta} + (1 - \pi(\rx)) \frac{\partial \gamma(\eta_0(\rx))}{\partial \eta}}$  \\
				\textit{Gao} & $\tilde{\mu}(\rx) + \tau(\rx) (w - \hat{a}(\rx)) + \hat{\nu}(\rx)$ & $\nu(\rx) = a(\rx) n_1(\rx) + (1-a(\rx)) n_0(\rx)$\\
				\hline
			\end{tabular}
			\begin{tablenotes}
				\small
				\item Note: for the Cox model 
				$a(\rx) = \frac{\pi(\rx) \Prob(C \ge Y |\rX = \rx,  W = 1)}{\pi(\rx) \Prob(C \ge Y | \rX = \rx,  W = 1) + (1 - \pi(\rx)) \Prob(C \ge Y | \rX = \rx, W = 0)}$ is used.
			\end{tablenotes}
			\label{tab:strategies2}
		\end{threeparttable}
\end{center}
\end{table} \subsection{Data-generating process}

To investigate the research questions of Section~A.2., we compared the performance
of model-based forests with Gao's strategy proposed in Section~A.2 (\textit{Gao} and \textit{Gao}$_{\hat{W}}$) to model-based 
forests with Robinson's startegy (\textit{Robinson} and \textit{Robinson}$_{\hat{W}}$) for settings A, B, C, D described in Section~4.
Because we expect that the strategy of Gao is especially valuable for
settings with misspecified prognostic effect, e.g. because prognostic covariates are missing, 
we additionally created Setup A' from Setup A by removing covariate $\rX_3$ from the training data.
Therefore, the DGP of Setup A and Setup A' are identical, the only difference being 
that the training data did not contain $\rX_3$ although $\rX_3$ affects the prognostic effect.

Because the normal linear model and Weibull model are collapsible and Gao's strategy is equal to Robinson's strategy 
(Sections~2.4~and~A.1), we applied our proposed approaches based on \AYcite{Gao and Hastie}{Gao_Hastie_2022}
only to the binomial model and the Cox model. 
Transformation models such as the proportional odds model for multinomial data were not covered 
by the authors.

We used the same model-based forest parameter setup and evaluation scheme as in Section~4.

\begin{figure}

\includegraphics[width=\maxwidth]{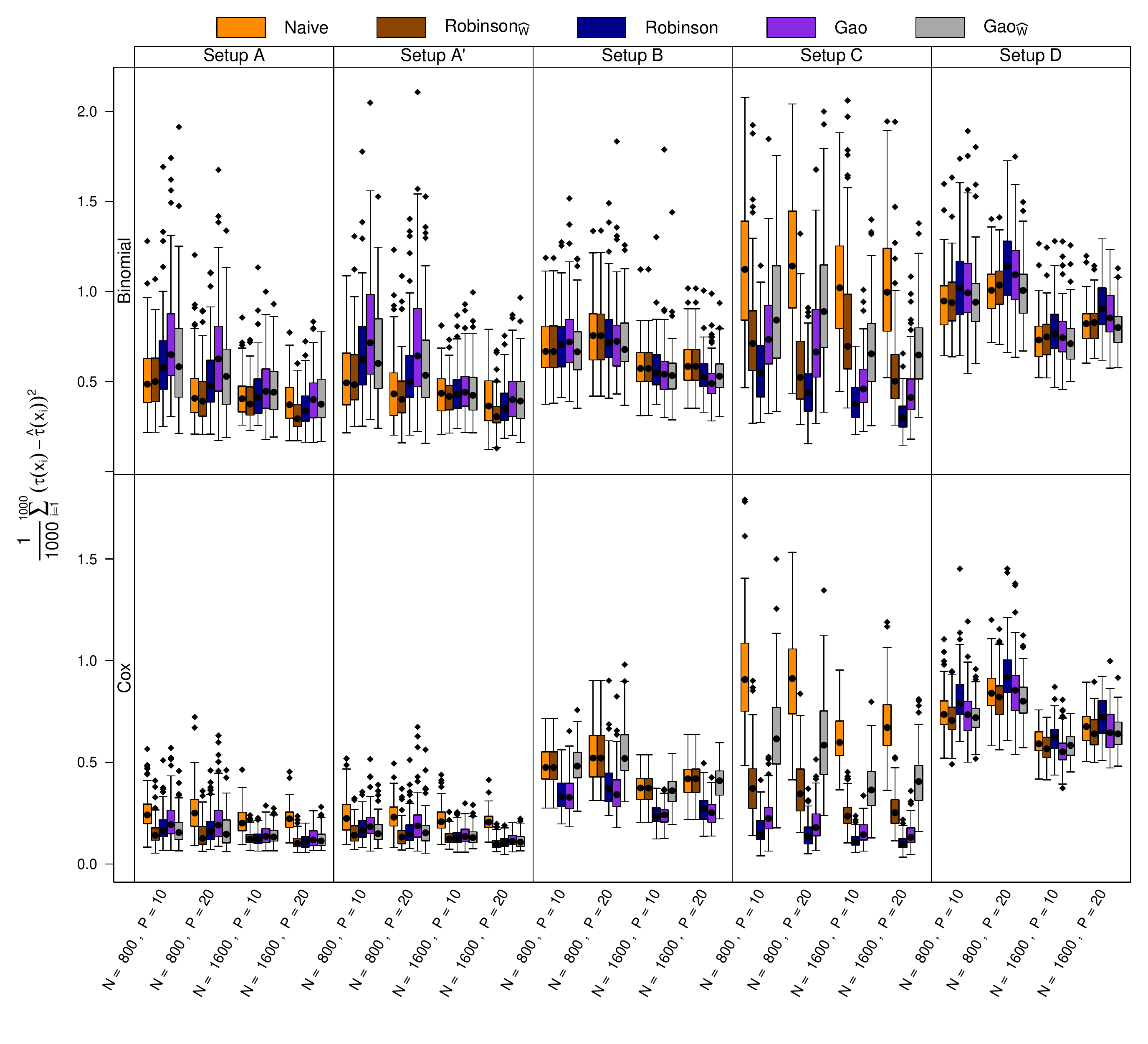} 

\caption{Model-based forest results for the empirical study (Section~4), Cox means a Cox model applied to the
Weibull data. For the Cox model, treatment effects $\tau(\rx)$
are estimated as conditional log hazard ratios.
Direct comparison of model-based forests without centering (Naive),
model-based forests with local centering according to \AYcite{Robinson}{Robinson_1988} or
\AYcite{Gao and Hastie}{Gao_Hastie_2022} of $Y$ and $W$ (originally proposed) (\textit{Robinson}, \textit{Gao}) or
only of $W$ (\textit{Robinson}$_{\widehat{W}}$, \textit{Gao}$_{\widehat{W}}$).
\label{fig:otherB-nie-collap}}
\end{figure}

\begin{table}
\caption{Results of \textbf{RQ 4} for the experimental setups in Section~4.  Comparison of mean
squared errors for $\hat{\tau}(\rx)$ in the different scenarios.  Estimates
and simultaneous $95$ \% confidence intervals were obtained from a normal
linear mixed model with log-link.  Cells printed in bold font correspond to
a superior reference of \textit{Robinson} forests, cells printed in italics indicate an inferior reference.
\label{tab:lmeradaptive-nie-noncoll1}}
\addtolength{\tabcolsep}{-2pt}
\tiny
\centering
\begin{tabular}{llllrr}
\hline
&& && \multicolumn{2}{c}{Mean squared error ratio for RQ 4: Gao vs. Robinson}\\
\cline{5-5}\cline{6-6}
DGP&N&P && \multicolumn{1}{c}{Binomial}&\multicolumn{1}{c}{Cox}\\
\hline

Setup A &800     &10       && \textbf{1.258 (1.152, 1.373)} & \textbf{1.203 (1.077, 1.344)}\\
        &        &20       && \textbf{1.307 (1.180, 1.449)} & \textbf{1.307 (1.170, 1.461)}\\
        &1600    &10       &&           1.067 (0.933, 1.220) &           1.121 (0.947, 1.326)\\
        &        &20       && \textbf{1.183 (1.009, 1.388)} &           1.155 (0.955, 1.398)\\
Setup A'&800     &10       && \textbf{1.201 (1.105, 1.304)} & \textbf{1.140 (1.011, 1.285)}\\
        &        &20       && \textbf{1.354 (1.233, 1.488)} & \textbf{1.272 (1.127, 1.435)}\\
        &1600    &10       &&           1.047 (0.915, 1.200) &           1.055 (0.895, 1.243)\\
        &        &20       && \textbf{1.184 (1.014, 1.382)} &           1.114 (0.911, 1.362)\\
Setup B &800     &10       &&           1.042 (0.958, 1.134) &           0.984 (0.920, 1.052)\\
        &        &20       &&           0.987 (0.909, 1.073) & \textit{0.906 (0.853, 0.963)}\\
        &1600    &10       &&           0.987 (0.885, 1.100) &           0.977 (0.889, 1.074)\\
        &        &20       &&           0.926 (0.824, 1.042) &           0.922 (0.845, 1.006)\\
Setup C &800     &10       && \textbf{1.388 (1.263, 1.524)} & \textbf{1.417 (1.261, 1.592)}\\
        &        &20       && \textbf{1.616 (1.448, 1.804)} & \textbf{1.401 (1.228, 1.598)}\\
        &1600    &10       && \textbf{1.276 (1.104, 1.476)} & \textbf{1.360 (1.146, 1.615)}\\
        &        &20       && \textbf{1.485 (1.255, 1.758)} & \textbf{1.400 (1.163, 1.686)}\\
Setup D &800     &10       &&           0.996 (0.939, 1.057) & \textit{0.916 (0.889, 0.943)}\\
        &        &20       &&           0.965 (0.913, 1.020) & \textit{0.925 (0.902, 0.949)}\\
        &1600    &10       &&           0.964 (0.890, 1.044) & \textit{0.910 (0.875, 0.946)}\\
        &        &20       &&           0.948 (0.884, 1.015) & \textit{0.907 (0.877, 0.938)}\\
\hline
\end{tabular}

\end{table}

\begin{table}
\caption{Results of \textbf{RQ 5} for the experimental setups in Section~4.  Comparison of mean
squared errors for $\hat{\tau}(\rx)$ in the different scenarios.  Estimates
and simultaneous $95$ \% confidence intervals were obtained from a normal
linear mixed model with log-link.  Cells printed in bold font correspond to
a superior reference of \textit{Robinson}$_{\hat{W}}$ forests, cells printed in italics indicate an inferior reference.
\label{tab:lmeradaptive-nie-noncoll2}}
\addtolength{\tabcolsep}{-2pt}
\tiny
\centering
\begin{tabular}{llllrr}
\hline
&& && \multicolumn{2}{c}{Mean squared error ratio for RQ 5: Gao$_{\hat{W}}$ vs. Robinson$_{\hat{W}}$}\\
\cline{5-5}\cline{6-6}
DGP&N&P && \multicolumn{1}{c}{Binomial}&\multicolumn{1}{c}{Cox}\\
\hline

Setup A &800     &10       && \textbf{1.299 (1.168, 1.445)} &           1.127 (0.986, 1.288)\\
        &        &20       && \textbf{1.425 (1.255, 1.618)} & \textbf{1.190 (1.038, 1.366)}\\
        &1600    &10       && \textbf{1.162 (1.009, 1.339)} &           1.110 (0.940, 1.310)\\
        &        &20       && \textbf{1.339 (1.128, 1.589)} &           1.144 (0.944, 1.386)\\
Setup A'&800     &10       && \textbf{1.261 (1.139, 1.397)} &           1.096 (0.952, 1.263)\\
        &        &20       && \textbf{1.427 (1.264, 1.610)} & \textbf{1.195 (1.033, 1.382)}\\
        &1600    &10       &&           1.096 (0.950, 1.264) &           1.060 (0.896, 1.255)\\
        &        &20       && \textbf{1.305 (1.101, 1.548)} &           1.114 (0.906, 1.370)\\
Setup B &800     &10       &&           0.988 (0.904, 1.079) &           1.005 (0.959, 1.053)\\
        &        &20       &&           0.959 (0.883, 1.042) &           1.037 (0.995, 1.081)\\
        &1600    &10       &&           0.947 (0.849, 1.056) &           0.968 (0.910, 1.031)\\
        &        &20       &&           0.905 (0.811, 1.009) &           0.982 (0.929, 1.038)\\
Setup C &800     &10       && \textbf{1.228 (1.141, 1.323)} & \textbf{1.636 (1.561, 1.715)}\\
        &        &20       && \textbf{1.658 (1.524, 1.804)} & \textbf{1.585 (1.510, 1.664)}\\
        &1600    &10       && \textit{0.716 (0.660, 0.776)} & \textbf{1.552 (1.437, 1.677)}\\
        &        &20       && \textbf{1.272 (1.149, 1.408)} & \textbf{1.588 (1.481, 1.702)}\\
Setup D &800     &10       &&           1.011 (0.948, 1.079) &           1.004 (0.973, 1.037)\\
        &        &20       &&           0.981 (0.923, 1.042) &           0.987 (0.960, 1.016)\\
        &1600    &10       &&           0.969 (0.891, 1.054) &           1.027 (0.987, 1.069)\\
        &        &20       &&           0.970 (0.899, 1.048) &           1.003 (0.968, 1.039)\\
\hline
\end{tabular}

\end{table}

\subsection{Results}

For Setup A, solely centering $W$ by $\hat{a}(\rx)$ (\textit{Gao}$_{\hat{W}}$) achieved better results than additionally adding the offset $\hat{\nu}(\rx)$ (\textit{Gao}).
Model-based forests with Robinson's strategy (\textit{Robinson}, \textit{Robinson}$_{\hat{W}}$) overall performed better than
model-based forests with Gao's strategy (\textit{Gao}, \textit{Gao}$_{\hat{W}}$).
Surpressing $X_3$ in the training dataset (Setup A'), did not deteriorate the performance
of all methods such that the ranking of methods was retained.

For Setup B, model-based forests centered by
\textit{Gao} and \textit{Robinson} model-based forests performed akin for binary outcomes. 
Also \textit{Robinson}$_{\hat{W}}$ and \textit{Gao}$_{\hat{W}}$ model-based forests achieved similar performance.

In Setup C, Gao's strategy for the Cox and logistic regression model overall fare worse than Robinson's
strategy.
In Setup D, \textit{Gao}$_{\hat{W}}$ forests performed as good as \textit{Robinson}$_{\hat{W}}$ forests for the Cox and logistic regression models.
Notably, for the Cox model, \textit{Gao} forests outperformed \textit{Robinson} forests.

Overall, the orthogonalization strategy of Gao for the exponential family
-- that aims at addressing the noncollapsibility issue -- did not perform as well as
expected.
Our expectation was that the strategy would reduce the effect of overfitting the marginal effect $\hat{m}(\rx)$ on the treatment effect estimate.
Overall, however, the estimation of additional nuisance parameters tended to worsen the performance results on average
-- at least for the binomial model.
For the Cox model, Gao's strategy, which additionally takes the
probability for not getting censored into account, did not worsen performance.
Further experiments are necessary in which the censoring probability is not
constant but depends on covariates $\rx$.
 
\clearpage

\newpage

\section{Empirical evaluation based on Wager and Athey (2018)}
\label{sec:simathey}

We evaluated the performance of our proposed model-based forest versions also 
with the study setting of \AYcite{Wager and Athey}{Wager_Athey_2018}, which were later reused by \AYcite{Athey et al.}{Athey_Tibshirani_Wager_2019}.
Given uniformly distributed covariates $\rX \sim U([0, 1]^P)$ of dimensionality $P \in \{10, 20\}$ and
a binomially distributed treatment indicator $W \mid \rX = \rx \sim \BD(1, \pi(\rx))$, the propensity function $\pi(\cdot)$ either did or did not depend on $\rx$
\begin{eqnarray*}
\pi(\rx) = \left\{
  \begin{array}{l} \pi \equiv 0.5 \\
                   \pi(x_1) = \nicefrac{1}{4}(1 + \beta_{2, 4}(x_1)) \\
                   \pi(x_3) = \nicefrac{1}{4}(1 + \beta_{2, 4}(x_3)) \\
                   \pi(x_4) = \nicefrac{1}{4}(1 + \beta_{2, 4}(x_4))
  \end{array} \right.
\end{eqnarray*}
where $\beta_{2, 4}$ is the $\beta$-density with shape $2$ and scale $4$.
The probability $\pi \equiv 0.5$ indicates no confounding and thus a randomized
trial.
The treatment effect function $\tau(\cdot)$ was either $0$ (no treatment effect) or
depended on a smooth interaction function of $x_1$ and $x_2$
\begin{eqnarray*}
\tau(\rx) = \left\{ \begin{array}{l} \tau \equiv 0 \\
  \tau(x_1, x_2) = \prod_{p = 1, 2} \left(1 + \left(1 + \exp\left(-20\left(x_p -
  \nicefrac{1}{3}\right)\right)\right)^{-1}\right).
\end{array} \right.
\end{eqnarray*}
The prognostic effect function $\mu(\cdot)$ was either $0$ (no prognostic effect) or
linear in $x_1$ or $x_3$
\begin{eqnarray*}
\mu(\rx) = \left\{
  \begin{array}{l} \mu \equiv 0 \\
                   \mu(x_1) = 2 x_1 - 1 \\
                   \mu(x_3) = 2 x_3 - 1.
  \end{array} \right.
\end{eqnarray*}
We studied four different simulation models
\begin{subnumcases}{(\rY \mid \rX = \rx, W = w) \sim}
 \ND(\mu(\rx) + \tau(\rx)w, 1)  \label{m1a} \\
 \BD(1, \text{expit}(\mu(\rx) + \tau(\rx)w))  \label{m2a}  \\
 \MD \text{with} \log(O(\ry_k \mid \rx, w)) = \eparm_k - \mu(\rx) - \tau(\rx)w \label{m3a} \\
 \WD \text{with} \log(H(\ry \mid \rx, w)) = 2 \log(y) - \mu(\rx) - \tau(\rx)w
  \label{m4a}
\end{subnumcases}
Model (\ref{m1a}) is a normal linear regression model, model (\ref{m2a}) a binary
logistic regression model,
model (\ref{m3a}) is a 4-nomial model with log-odds function
$\eparm_k - \mu(\rx) - \tau(\rx)w$ with threshold parameters
$\eparm_k = \text{logit}(k / 4)$ for $k = 1, 2, 3$, and
model (\ref{m4a}) is a Weibull model with
log-cumulative hazard function $2 \log(y) - \mu(\rx) - \tau(\rx)w$.
We added $50\,\%$ random right-censoring to the Weibull-generated data and also
applied a Cox proportional hazards model in addition to the Weibull model.

For the additive predictor $\mu(\rx) + \tau(\rx)w$ we considered the 16
scenarios as specified in Table~\ref{DGP}.
Compared to Part~A of this table, in Part~B half of the (negative) predictive effect is
added to the prognostic effect.
We term the implied scenario where at least one variable exists which is both prognostic (impact in $\mu(\rx)$) and predictive (impact in $\tau(\rx)$)
as overlay.
$W(x_1)$, $W(x_3)$ and $W(x_4)$ depict that $W$ was drawn from a Bernoulli
distribution with $\pi(x_1)$, $\pi(x_3)$ or $\pi(x_4)$, respectively.

\begin{table}[ht]
\caption{\label{DGP}Experimental setup~\ref{sec:simathey}. Confounding
is present for non-constant propensities $\pi(\rx)$, an instrumental
variable impacts $\pi(\rx)$ exclusively, heterogeneity of the treatment
effect $\tau(\rx)$ is present when $\tau$ is non-constant, and overlay
refers to variables being prognostic (impact in $\mu(\rx)$) and predictive
(impact in $\tau(\rx)$) at the same time.
}
\centering
\begin{tabular}{llllll} \hline
& Additive Predictor & Confounding & Instrument & Heterogeneity & Overlay \\ \hline
& $\mu(x_3) + 0 \cdot W(x_3)$ & yes & no & no & no \\
\parbox[t]{2mm}{\multirow{3}{*}{\rotatebox[origin=c]{90}{Part A}}}
& $\tau(x_1, x_2) W$ & no & no & yes & no \\
& $\mu(x_1) + \tau(x_1, x_2) W(x_1)$ & yes & no & yes & yes \\
& $\mu(x_1) + \tau(x_1, x_2) W$ & no & no & yes & yes \\
& $\mu(x_3) + \tau(x_1, x_2) W$ & no & no & yes & no \\
& $\mu(x_3) + \tau(x_1, x_2) W(x_3)$ & yes & no & yes & no \\
& $\tau(x_1, x_2) W(x_3)$ & no & yes & yes & no \\
& $\mu(x_3) + \tau(x_1, x_2) W(x_4)$ & no & yes & yes & no \\ \hline
\parbox[t]{2mm}{\multirow{3}{*}{\rotatebox[origin=c]{90}{Part B}}}
& $\mu(x_3) + 0 \cdot (W(x_3) - 0.5)$ & yes & no & no & no \\
& $\tau(x_1, x_2) (W - 0.5)$ & no & no & yes & yes \\
& $\mu(x_1) + \tau(x_1, x_2) (W(x_1) - 0.5)$ & yes & no & yes & yes \\
& $\mu(x_1) + \tau(x_1, x_2) (W - 0.5)$ & no & no & yes & yes \\
& $\mu(x_3) + \tau(x_1, x_2) (W - 0.5)$ & no & no & yes & yes \\
& $\mu(x_3) + \tau(x_1, x_2) (W(x_3) - 0.5)$ & yes & no & yes & yes \\
& $\tau(x_1, x_2) (W(x_3) - 0.5)$ & no & yes & yes & yes \\
& $\mu(x_3) + \tau(x_1, x_2) (W(x_4) - 0.5)$ & no & yes & yes & yes \\ \hline
\end{tabular}
\end{table}
In Part~A of Table~\ref{DGP}, the prognostic term and the predictive term are
separate and there is only overlay of prognostic and predictive effects when
both terms depend on $x_1$, \ie $x_1$ is both prognostic and predictive in
this scenario.  The treatment assignment probability may depend on $x_1$,
$x_3$, or $x_4$. In the third scenario, $x_1$ is a predictive
confounder (with impact on $\mu$, $\tau$, and $\pi$) and in the last two
scenarios, $x_3$ and $x_4$ can be understood as instruments with direct impact on
treatment assignment but without direct impact on the response.
In Part~B of this table, half of the predictive effect is
added to the prognostic effect, so there is always overlay of both types of
effects.

Again, we used random forests to estimate
$\pi(\rx)$ and gradient boosting machines to estimate $\eta_0(\rx)$ and $\eta_1(\rx)$
as described in Section~4.
We also applied the same performance assessment (mean squared error evaluated on 1000 test samples).
The results are presented in Figures~\ref{fig:otherA} and \ref{fig:otherB}.
The results for the statistical analysis of RQ 1 to RQ 3 based on a normal linear mixed model
are presented in Table~\ref{tab:lmeradaptive} to \ref{tab:lmeradaptive3}.

\subsubsection{Results}

\begin{figure}

\includegraphics[width=\maxwidth]{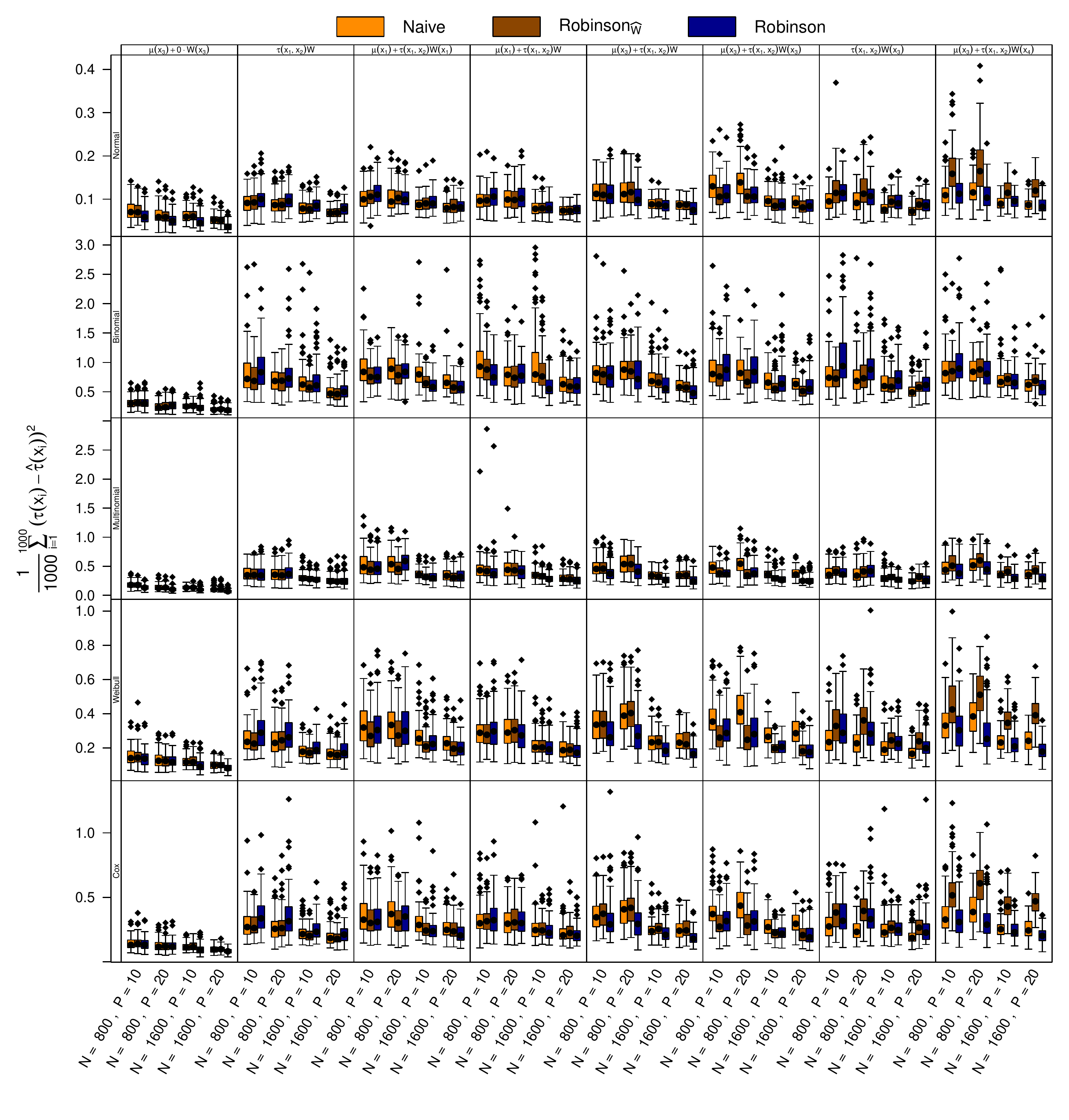} 

\caption{Model-based forest results for Part  A (Table~\ref{DGP}), Cox means a Cox model applied to the
Weibull data. For the Weibull and Cox model, treatment effects $\tau(\rx)$
are estimated as conditional log hazard ratios.
Direct comparison of model-based forests without centering (Naive),
model-based forests with local centering according to \AYcite{Robinson}{Robinson_1988} of $Y$ and $W$ (\textit{Robinson}) or
only of $W$ (\textit{Robinson}$_{\widehat{W}}$).
\label{fig:otherA}}
\end{figure}

\begin{figure}

\includegraphics[width=\maxwidth]{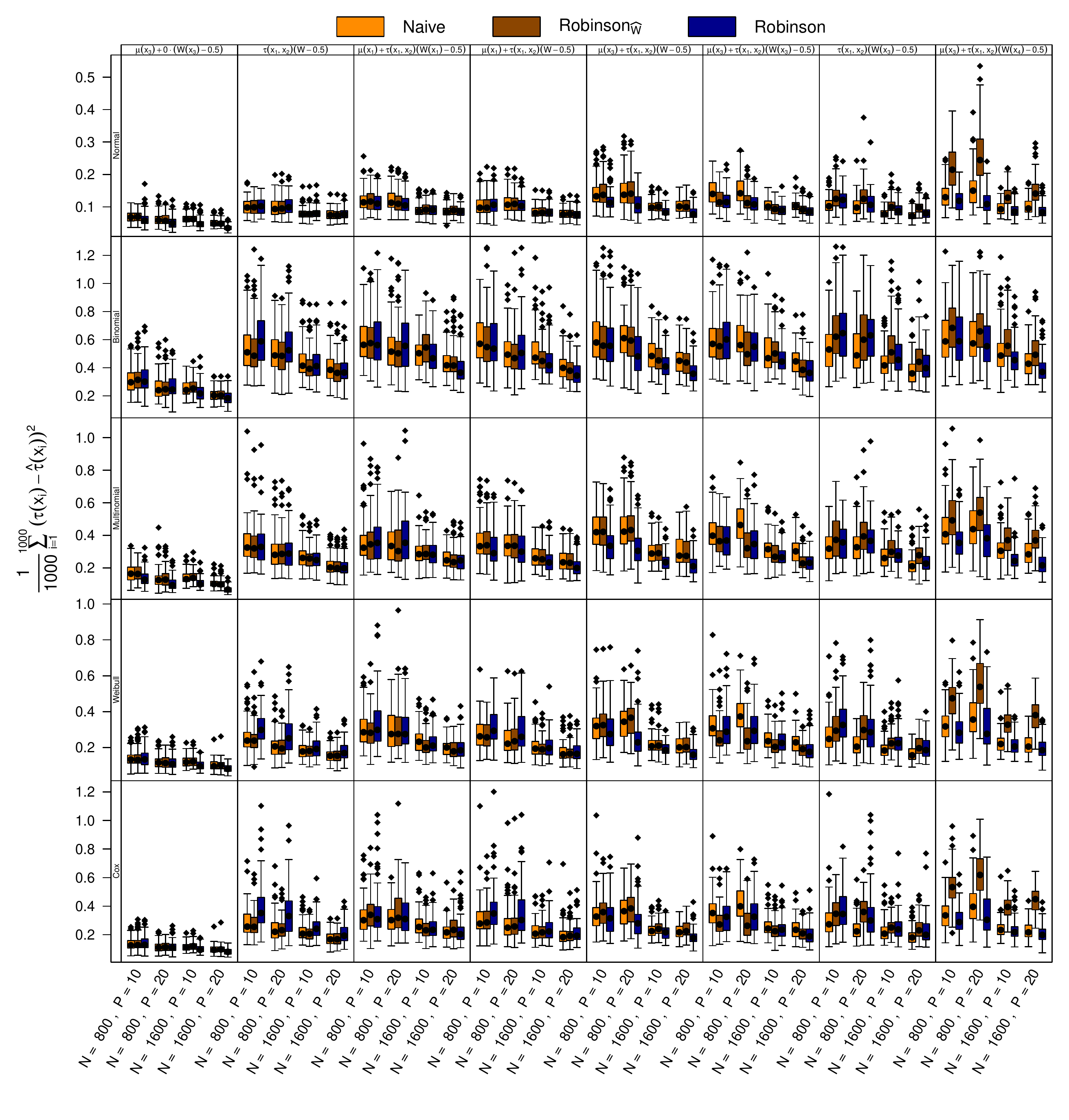} 

\caption{Model-based forest results for Part B (Table~\ref{DGP}), Cox means a Cox model applied to the
Weibull data. For the Weibull and Cox model, treatment effects $\tau(\rx)$
are estimated as conditional log hazard ratios.
Direct comparison of model-based forests without centering (Naive),
model-based forests with local centering according to \AYcite{Robinson}{Robinson_1988}
of $Y$ and $W$ (\textit{Robinson}) or
only of $W$ (\textit{Robinson}$_{\widehat{W}}$).
\label{fig:otherB}}
\end{figure}

\begin{sidewaystable}
\caption{Results of \textbf{RQ 1} for the experimental setups in Section~\ref{sec:simathey}.  Comparison of mean
squared errors for $\hat{\tau}(\rx)$ in the different scenarios.  Estimates
and simultaneous $95$ \% confidence intervals were obtained from a normal
linear mixed model with log-link.  Cells printed in bold font correspond to
a superior reference of the naive model-based forests, cells printed in italics indicate an inferior reference of naive model-based forests.
\label{tab:lmeradaptive}}
\addtolength{\tabcolsep}{-2pt}
\tiny
\begin{tabular}{llllrrrrr}
\hline
&& && \multicolumn{5}{c}{Mean squared error ratio for RQ 1: Robinson vs. Naive}\\
\cline{5-5}\cline{6-6}\cline{7-7}\cline{8-8}\cline{9-9}
\rule{0pt}{6ex} Part \hspace{.4cm} DGP&N&P && \multicolumn{1}{c}{Normal}&\multicolumn{1}{c}{Binomial}&\multicolumn{1}{c}{Multinomial}&\multicolumn{1}{c}{Weibull}&\multicolumn{1}{c}{Cox}\\
\hline

A \hspace{.8cm} $\mu(x_3) + 0 \cdot W(x_3)$                 &800                                                            &10                                                              && \textit{0.837 (0.762, 0.919)} &           1.025 (0.713, 1.474) & \textit{0.766 (0.660, 0.888)} &           0.927 (0.769, 1.118) &           0.975 (0.759, 1.254)\\
                                                               &                                                               &20                                                              && \textit{0.808 (0.723, 0.903)} &           1.123 (0.726, 1.735) & \textit{0.761 (0.624, 0.928)} &           0.931 (0.755, 1.149) &           0.975 (0.738, 1.287)\\
                                                               &1600                                                           &10                                                              && \textit{0.795 (0.707, 0.894)} &           0.912 (0.570, 1.462) & \textit{0.742 (0.601, 0.915)} &           0.821 (0.641, 1.053) &           0.848 (0.609, 1.180)\\
                                                               &                                                               &20                                                              && \textit{0.697 (0.601, 0.809)} &           0.942 (0.526, 1.686) & \textit{0.696 (0.533, 0.907)} &           0.829 (0.613, 1.123) &           0.852 (0.569, 1.276)\\
\hspace{1.1cm} $\tau(x_1, x_2) W$                            &800                                                            &10                                                              && \textbf{1.106 (1.036, 1.180)} & \textbf{3.295 (2.924, 3.714)} &           0.976 (0.918, 1.037) & \textbf{1.222 (1.111, 1.344)} & \textbf{1.220 (1.098, 1.356)}\\
                                                               &                                                               &20                                                              && \textbf{1.094 (1.023, 1.171)} & \textbf{1.210 (1.044, 1.402)} &           1.047 (0.986, 1.112) & \textbf{1.127 (1.019, 1.248)} & \textbf{1.379 (1.234, 1.541)}\\
                                                               &1600                                                           &10                                                              && \textbf{1.085 (1.005, 1.171)} &           1.012 (0.878, 1.167) & \textit{0.897 (0.827, 0.972)} &           1.080 (0.938, 1.245) &           1.093 (0.946, 1.262)\\
                                                               &                                                               &20                                                              && \textbf{1.128 (1.034, 1.231)} &           1.021 (0.826, 1.261) &           1.004 (0.918, 1.098) &           1.108 (0.954, 1.287) &           1.140 (0.969, 1.342)\\
\hspace{1.1cm} $\mu(x_1) + \tau(x_1, x_2) W(x_1)$           &800                                                            &10                                                              && \textbf{1.118 (1.054, 1.185)} & \textit{0.805 (0.706, 0.917)} & \textit{0.924 (0.885, 0.966)} &           0.970 (0.895, 1.052) &           0.916 (0.833, 1.008)\\
                                                               &                                                               &20                                                              &&           0.988 (0.929, 1.051) &           0.943 (0.831, 1.070) &           1.032 (0.990, 1.074) & \textit{0.923 (0.853, 0.998)} &           0.912 (0.831, 1.002)\\
                                                               &1600                                                           &10                                                              && \textbf{1.078 (1.005, 1.156)} & \textit{0.188 (0.157, 0.225)} & \textit{0.872 (0.814, 0.935)} & \textit{0.875 (0.787, 0.973)} & \textit{0.805 (0.712, 0.910)}\\
                                                               &                                                               &20                                                              &&           1.049 (0.970, 1.133) & \textit{0.633 (0.532, 0.753)} &           0.981 (0.919, 1.048) & \textit{0.845 (0.747, 0.956)} & \textit{0.857 (0.742, 0.990)}\\
\hspace{1.1cm} $\mu(x_1) + \tau(x_1, x_2) W$                &800                                                            &10                                                              && \textbf{1.076 (1.011, 1.146)} & \textit{0.465 (0.408, 0.531)} &           0.964 (0.917, 1.013) &           1.013 (0.927, 1.106) &           1.059 (0.956, 1.173)\\
                                                               &                                                               &20                                                              &&           1.063 (1.000, 1.130) &           0.991 (0.862, 1.139) & \textit{0.877 (0.831, 0.926)} & \textit{0.911 (0.832, 0.997)} &           0.990 (0.889, 1.104)\\
                                                               &1600                                                           &10                                                              &&           1.025 (0.947, 1.109) & \textit{0.307 (0.259, 0.363)} & \textit{0.795 (0.738, 0.857)} &           0.946 (0.834, 1.072) &           0.901 (0.788, 1.031)\\
                                                               &                                                               &20                                                              &&           1.027 (0.943, 1.119) &           0.918 (0.769, 1.096) & \textit{0.880 (0.809, 0.958)} &           0.933 (0.812, 1.072) &           0.863 (0.740, 1.007)\\
\hspace{1.1cm} $\mu(x_3) + \tau(x_1, x_2) W$                &800                                                            &10                                                              &&           0.984 (0.931, 1.041) & \textbf{1.166 (1.032, 1.318)} & \textit{0.802 (0.760, 0.846)} & \textit{0.821 (0.753, 0.894)} &           0.951 (0.860, 1.052)\\
                                                               &                                                               &20                                                              && \textit{0.875 (0.827, 0.926)} &           0.897 (0.789, 1.020) & \textit{0.800 (0.763, 0.839)} & \textit{0.690 (0.635, 0.751)} & \textit{0.724 (0.654, 0.801)}\\
                                                               &1600                                                           &10                                                              &&           0.951 (0.883, 1.023) &           0.946 (0.807, 1.110) & \textit{0.776 (0.716, 0.840)} & \textit{0.820 (0.723, 0.930)} &           0.880 (0.759, 1.019)\\
                                                               &                                                               &20                                                              && \textit{0.902 (0.834, 0.976)} &           0.848 (0.693, 1.038) & \textit{0.741 (0.685, 0.802)} & \textit{0.688 (0.598, 0.790)} & \textit{0.735 (0.622, 0.868)}\\
\hspace{1.1cm} $\mu(x_3) + \tau(x_1, x_2) W(x_3)$           &800                                                            &10                                                              && \textit{0.865 (0.821, 0.911)} & \textbf{1.161 (1.031, 1.308)} & \textit{0.778 (0.736, 0.823)} & \textit{0.820 (0.757, 0.888)} & \textit{0.844 (0.765, 0.931)}\\
                                                               &                                                               &20                                                              && \textit{0.764 (0.726, 0.804)} &           1.102 (0.974, 1.248) & \textit{0.747 (0.711, 0.786)} & \textit{0.717 (0.663, 0.774)} & \textit{0.731 (0.666, 0.803)}\\
                                                               &1600                                                           &10                                                              &&           0.944 (0.882, 1.011) &           0.867 (0.744, 1.011) & \textit{0.744 (0.690, 0.803)} & \textit{0.748 (0.666, 0.840)} & \textit{0.804 (0.698, 0.926)}\\
                                                               &                                                               &20                                                              && \textit{0.918 (0.855, 0.986)} &           0.982 (0.821, 1.174) & \textit{0.697 (0.645, 0.754)} & \textit{0.632 (0.559, 0.715)} & \textit{0.701 (0.608, 0.809)}\\
\hspace{1.1cm} $\tau(x_1, x_2) W(x_3)$                       &800                                                            &10                                                              && \textbf{1.213 (1.142, 1.289)} & \textbf{1.537 (1.372, 1.722)} &           1.013 (0.954, 1.076) & \textbf{1.265 (1.154, 1.386)} & \textbf{1.221 (1.096, 1.360)}\\
                                                               &                                                               &20                                                              && \textbf{1.148 (1.080, 1.220)} & \textbf{1.349 (1.189, 1.530)} & \textbf{1.179 (1.112, 1.250)} & \textbf{1.335 (1.207, 1.477)} & \textbf{1.468 (1.301, 1.657)}\\
                                                               &1600                                                           &10                                                              && \textbf{1.227 (1.138, 1.323)} & \textbf{1.178 (1.011, 1.372)} & \textit{0.916 (0.844, 0.993)} & \textbf{1.150 (1.015, 1.304)} &           0.965 (0.843, 1.105)\\
                                                               &                                                               &20                                                              && \textbf{1.200 (1.106, 1.301)} & \textbf{1.278 (1.052, 1.553)} & \textbf{1.152 (1.051, 1.262)} & \textbf{1.265 (1.094, 1.462)} & \textbf{1.366 (1.167, 1.599)}\\
\hspace{1.1cm} $\mu(x_3) + \tau(x_1, x_2) W(x_4)$           &800                                                            &10                                                              &&           1.017 (0.961, 1.076) & \textbf{1.191 (1.063, 1.334)} & \textit{0.877 (0.833, 0.924)} & \textit{0.913 (0.843, 0.990)} &           0.942 (0.854, 1.040)\\
                                                               &                                                               &20                                                              && \textit{0.907 (0.857, 0.960)} &           0.999 (0.880, 1.135) & \textit{0.901 (0.860, 0.943)} & \textit{0.780 (0.720, 0.845)} & \textit{0.810 (0.736, 0.891)}\\
                                                               &1600                                                           &10                                                              &&           1.048 (0.978, 1.123) & \textit{0.841 (0.722, 0.980)} & \textit{0.847 (0.788, 0.909)} &           0.946 (0.841, 1.063) &           0.932 (0.811, 1.071)\\
                                                               &                                                               &20                                                              &&           0.967 (0.898, 1.042) &           0.943 (0.785, 1.133) & \textit{0.847 (0.790, 0.910)} & \textit{0.752 (0.661, 0.857)} & \textit{0.782 (0.669, 0.913)}\\
\hline B \hspace{.8cm} $\mu(x_3) + 0 \cdot (W(x_3) - 0.5)$ &800                                                            &10                                                              &&           0.907 (0.820, 1.002) &           1.033 (0.925, 1.154) & \textit{0.820 (0.723, 0.930)} &           1.028 (0.867, 1.219) &           1.069 (0.850, 1.346)\\
                                                               &                                                               &20                                                              && \textit{0.818 (0.727, 0.921)} &           1.047 (0.918, 1.194) & \textit{0.739 (0.627, 0.872)} &           0.980 (0.797, 1.205) &           1.025 (0.773, 1.359)\\
                                                               &1600                                                           &10                                                              && \textit{0.783 (0.693, 0.885)} &           0.896 (0.772, 1.041) & \textit{0.746 (0.634, 0.879)} &           0.827 (0.662, 1.032) &           0.869 (0.642, 1.177)\\
                                                               &                                                               &20                                                              && \textit{0.705 (0.597, 0.831)} &           0.908 (0.755, 1.092) & \textit{0.643 (0.503, 0.821)} &           0.801 (0.610, 1.050) &           0.833 (0.577, 1.204)\\
\hspace{1.1cm} $\tau(x_1, x_2) (W - 0.5)$                    &800                                                            &10                                                              &&           1.013 (0.948, 1.082) & \textbf{1.112 (1.050, 1.177)} &           0.973 (0.922, 1.026) & \textbf{1.222 (1.123, 1.330)} & \textbf{1.414 (1.286, 1.555)}\\
                                                               &                                                               &20                                                              &&           1.037 (0.971, 1.107) & \textbf{1.118 (1.047, 1.194)} &           0.987 (0.928, 1.050) & \textbf{1.313 (1.189, 1.449)} & \textbf{1.440 (1.291, 1.606)}\\
                                                               &1600                                                           &10                                                              &&           1.039 (0.956, 1.129) &           0.964 (0.890, 1.043) &           0.964 (0.896, 1.036) &           1.102 (0.974, 1.245) & \textbf{1.147 (1.004, 1.311)}\\
                                                               &                                                               &20                                                              &&           1.033 (0.947, 1.127) &           0.971 (0.888, 1.062) &           0.983 (0.900, 1.074) &           1.140 (0.987, 1.317) & \textbf{1.240 (1.047, 1.468)}\\
\hspace{1.1cm} $\mu(x_1) + \tau(x_1, x_2) (W(x_1) - 0.5)$   &800                                                            &10                                                              &&           0.954 (0.901, 1.011) & \textbf{1.074 (1.016, 1.135)} & \textbf{1.082 (1.028, 1.139)} & \textbf{1.130 (1.048, 1.218)} &           1.094 (0.999, 1.198)\\
                                                               &                                                               &20                                                              && \textit{0.931 (0.879, 0.986)} & \textbf{1.131 (1.063, 1.203)} & \textbf{1.140 (1.083, 1.201)} &           1.032 (0.953, 1.118) &           1.069 (0.970, 1.179)\\
                                                               &1600                                                           &10                                                              &&           1.017 (0.944, 1.096) &           0.994 (0.926, 1.068) &           0.987 (0.924, 1.054) &           0.947 (0.854, 1.051) &           0.944 (0.835, 1.067)\\
                                                               &                                                               &20                                                              &&           0.974 (0.900, 1.053) &           0.935 (0.858, 1.018) &           0.977 (0.904, 1.057) &           0.984 (0.870, 1.113) &           1.029 (0.892, 1.187)\\
\hspace{1.1cm} $\mu(x_1) + \tau(x_1, x_2) (W - 0.5)$        &800                                                            &10                                                              &&           1.024 (0.962, 1.089) &           0.969 (0.916, 1.024) & \textit{0.874 (0.827, 0.923)} & \textbf{1.152 (1.061, 1.250)} & \textbf{1.179 (1.078, 1.289)}\\
                                                               &                                                               &20                                                              && \textit{0.940 (0.884, 1.000)} & \textbf{1.071 (1.004, 1.143)} & \textit{0.942 (0.889, 0.997)} & \textbf{1.192 (1.091, 1.303)} & \textbf{1.257 (1.138, 1.388)}\\
                                                               &1600                                                           &10                                                              &&           1.002 (0.925, 1.085) & \textit{0.871 (0.809, 0.937)} & \textit{0.902 (0.838, 0.971)} &           1.028 (0.917, 1.153) &           1.101 (0.963, 1.259)\\
                                                               &                                                               &20                                                              &&           0.976 (0.895, 1.064) & \textit{0.871 (0.794, 0.956)} & \textit{0.867 (0.796, 0.944)} &           1.001 (0.872, 1.149) &           1.039 (0.892, 1.211)\\
\hspace{1.1cm} $\mu(x_3) + \tau(x_1, x_2) (W - 0.5)$        &800                                                            &10                                                              && \textit{0.821 (0.780, 0.865)} &           0.957 (0.905, 1.012) & \textit{0.792 (0.752, 0.833)} & \textit{0.872 (0.805, 0.944)} & \textit{0.903 (0.820, 0.994)}\\
                                                               &                                                               &20                                                              && \textit{0.705 (0.669, 0.744)} & \textit{0.845 (0.793, 0.900)} & \textit{0.728 (0.691, 0.766)} & \textit{0.730 (0.671, 0.793)} & \textit{0.775 (0.702, 0.856)}\\
                                                               &1600                                                           &10                                                              && \textit{0.850 (0.790, 0.915)} & \textit{0.840 (0.777, 0.909)} & \textit{0.787 (0.732, 0.847)} & \textit{0.863 (0.766, 0.974)} &           0.935 (0.810, 1.081)\\
                                                               &                                                               &20                                                              && \textit{0.791 (0.734, 0.851)} & \textit{0.805 (0.737, 0.878)} & \textit{0.721 (0.668, 0.779)} & \textit{0.780 (0.682, 0.892)} & \textit{0.807 (0.687, 0.948)}\\
\hspace{1.1cm} $\mu(x_3) + \tau(x_1, x_2) (W(x_3) - 0.5)$   &800                                                            &10                                                              && \textit{0.810 (0.770, 0.853)} & \textbf{1.084 (1.022, 1.149)} & \textit{0.927 (0.882, 0.973)} &           0.979 (0.908, 1.056) &           0.927 (0.844, 1.017)\\
                                                               &                                                               &20                                                              && \textit{0.720 (0.684, 0.759)} & \textit{0.911 (0.857, 0.968)} & \textit{0.757 (0.721, 0.794)} & \textit{0.787 (0.733, 0.844)} & \textit{0.783 (0.719, 0.853)}\\
                                                               &1600                                                           &10                                                              && \textit{0.884 (0.825, 0.948)} & \textit{0.909 (0.844, 0.978)} & \textit{0.870 (0.814, 0.930)} &           0.946 (0.855, 1.045) &           0.945 (0.831, 1.075)\\
                                                               &                                                               &20                                                              && \textit{0.829 (0.772, 0.890)} & \textit{0.847 (0.776, 0.925)} & \textit{0.755 (0.701, 0.813)} & \textit{0.830 (0.738, 0.934)} & \textit{0.832 (0.718, 0.965)}\\
\hspace{1.1cm} $\tau(x_1, x_2) (W(x_3) - 0.5)$               &800                                                            &10                                                              && \textbf{1.152 (1.085, 1.222)} & \textbf{1.277 (1.206, 1.353)} & \textbf{1.082 (1.023, 1.144)} & \textbf{1.300 (1.200, 1.409)} & \textbf{1.212 (1.101, 1.334)}\\
                                                               &                                                               &20                                                              && \textbf{1.107 (1.040, 1.179)} & \textbf{1.228 (1.154, 1.306)} & \textbf{1.127 (1.069, 1.190)} & \textbf{1.470 (1.335, 1.618)} & \textbf{1.498 (1.335, 1.680)}\\
                                                               &1600                                                           &10                                                              && \textbf{1.132 (1.045, 1.226)} & \textbf{1.136 (1.054, 1.225)} &           1.040 (0.970, 1.116) & \textbf{1.228 (1.095, 1.377)} &           1.123 (0.980, 1.286)\\
                                                               &                                                               &20                                                              && \textbf{1.126 (1.032, 1.227)} & \textbf{1.122 (1.026, 1.226)} & \textbf{1.096 (1.006, 1.194)} & \textbf{1.233 (1.077, 1.410)} & \textbf{1.262 (1.078, 1.477)}\\
\hspace{1.1cm} $\mu(x_3) + \tau(x_1, x_2)(W(x_4) - 0.5)$    &800                                                            &10                                                              && \textit{0.864 (0.819, 0.911)} &           1.021 (0.966, 1.080) & \textit{0.823 (0.784, 0.864)} & \textit{0.907 (0.838, 0.982)} & \textit{0.891 (0.806, 0.986)}\\
                                                               &                                                               &20                                                              && \textit{0.718 (0.682, 0.755)} &           0.976 (0.921, 1.035) & \textit{0.843 (0.806, 0.883)} & \textit{0.863 (0.805, 0.925)} & \textit{0.850 (0.780, 0.925)}\\
                                                               &1600                                                           &10                                                              &&           0.935 (0.869, 1.006) & \textit{0.903 (0.841, 0.971)} & \textit{0.790 (0.737, 0.846)} &           0.935 (0.837, 1.045) &           0.954 (0.833, 1.093)\\
                                                               &                                                               &20                                                              && \textit{0.861 (0.801, 0.925)} & \textit{0.897 (0.822, 0.978)} & \textit{0.762 (0.706, 0.821)} &           0.899 (0.799, 1.011) &           0.911 (0.786, 1.055)\\
\hline
\end{tabular}

\end{sidewaystable}

\begin{sidewaystable}
\caption{Results of \textbf{RQ 2} for the experimental setups in Section~\ref{sec:simathey}.  Comparison of mean
squared errors for $\hat{\tau}(\rx)$ in the different scenarios.  Estimates
and simultaneous $95$ \% confidence intervals were obtained from a normal
linear mixed model with log-link.  Cells printed in bold font correspond to
a superior reference of the naive model-based forests, cells printed in italics indicate an inferior reference of naive model-based forests.
\label{tab:lmeradaptive2}}
\addtolength{\tabcolsep}{-2pt}
\tiny
\begin{tabular}{llllrrrrr}
\hline
&& && \multicolumn{5}{c}{Mean squared error ratio for RQ 2: Robinson$_{\hat{W}}$ vs. Naive}\\
\cline{5-5}\cline{6-6}\cline{7-7}\cline{8-8}\cline{9-9}
DGP&N&P && \multicolumn{1}{c}{Normal}&\multicolumn{1}{c}{Binomial}&\multicolumn{1}{c}{Multinomial}&\multicolumn{1}{c}{Weibull}&\multicolumn{1}{c}{Cox}\\
\hline

$\mu(x_3) + 0 \cdot W(x_3)$               &800                                         &10                                           && \textbf{1.182 (1.076, 1.298)} &           1.012 (0.709, 1.446) & \textbf{1.284 (1.106, 1.491)} &           1.109 (0.922, 1.335) &           1.063 (0.831, 1.361)\\
                                            &                                            &20                                           && \textbf{1.186 (1.059, 1.328)} &           0.917 (0.598, 1.408) & \textbf{1.304 (1.069, 1.591)} &           1.018 (0.821, 1.263) &           0.987 (0.744, 1.310)\\
                                            &1600                                        &10                                           && \textbf{1.277 (1.137, 1.436)} &           1.131 (0.711, 1.800) & \textbf{1.375 (1.116, 1.693)} &           1.238 (0.967, 1.584) &           1.224 (0.883, 1.695)\\
                                            &                                            &20                                           && \textbf{1.392 (1.197, 1.618)} &           1.084 (0.608, 1.930) & \textbf{1.430 (1.096, 1.866)} &           1.203 (0.889, 1.629) &           1.200 (0.804, 1.790)\\
$\tau(x_1, x_2) W$                         &800                                         &10                                           && \textit{0.913 (0.856, 0.974)} & \textit{0.537 (0.499, 0.577)} &           0.998 (0.939, 1.061) & \textit{0.785 (0.712, 0.866)} & \textit{0.787 (0.706, 0.876)}\\
                                            &                                            &20                                           && \textit{0.913 (0.853, 0.976)} & \textit{0.808 (0.695, 0.938)} &           0.954 (0.898, 1.013) &           0.906 (0.820, 1.002) & \textit{0.766 (0.688, 0.854)}\\
                                            &1600                                        &10                                           && \textit{0.920 (0.853, 0.993)} & \textit{0.822 (0.703, 0.961)} &           1.060 (0.976, 1.151) &           0.879 (0.760, 1.017) & \textit{0.816 (0.700, 0.952)}\\
                                            &                                            &20                                           && \textit{0.888 (0.814, 0.969)} &           0.938 (0.756, 1.164) &           0.989 (0.904, 1.081) &           0.877 (0.753, 1.022) & \textit{0.826 (0.697, 0.978)}\\
$\mu(x_1) + \tau(x_1, x_2) W(x_1)$        &800                                         &10                                           &&           0.948 (0.896, 1.004) &           1.000 (0.867, 1.153) &           0.959 (0.916, 1.005) & \textit{0.898 (0.824, 0.979)} &           0.986 (0.892, 1.089)\\
                                            &                                            &20                                           &&           1.038 (0.977, 1.103) &           0.941 (0.823, 1.076) & \textit{0.817 (0.781, 0.854)} & \textit{0.918 (0.843, 0.999)} &           0.941 (0.851, 1.040)\\
                                            &1600                                        &10                                           &&           0.951 (0.887, 1.018) &           1.006 (0.785, 1.289) &           1.026 (0.954, 1.103) &           0.911 (0.810, 1.025) &           0.972 (0.848, 1.114)\\
                                            &                                            &20                                           &&           1.001 (0.928, 1.080) &           0.959 (0.782, 1.175) & \textit{0.924 (0.862, 0.991)} &           1.020 (0.894, 1.164) &           1.148 (0.993, 1.327)\\
$\mu(x_1) + \tau(x_1, x_2) W$             &800                                         &10                                           && \textit{0.938 (0.881, 0.998)} & \textbf{1.826 (1.593, 2.093)} & \textbf{1.107 (1.055, 1.161)} &           0.964 (0.881, 1.054) &           0.932 (0.841, 1.033)\\
                                            &                                            &20                                           &&           0.945 (0.889, 1.004) &           0.984 (0.854, 1.133) & \textbf{1.096 (1.037, 1.158)} & \textbf{1.108 (1.013, 1.213)} &           1.052 (0.946, 1.170)\\
                                            &1600                                        &10                                           &&           0.978 (0.904, 1.058) & \textbf{2.057 (1.722, 2.456)} & \textbf{1.196 (1.108, 1.291)} &           1.036 (0.912, 1.177) &           1.051 (0.916, 1.205)\\
                                            &                                            &20                                           &&           0.975 (0.895, 1.063) &           1.018 (0.848, 1.221) & \textbf{1.134 (1.042, 1.234)} &           1.058 (0.919, 1.217) &           1.115 (0.955, 1.303)\\
$\mu(x_3) + \tau(x_1, x_2) W$             &800                                         &10                                           &&           1.018 (0.963, 1.076) & \textit{0.823 (0.727, 0.933)} & \textbf{1.257 (1.191, 1.326)} & \textbf{1.231 (1.130, 1.340)} & \textbf{1.107 (1.003, 1.221)}\\
                                            &                                            &20                                           && \textbf{1.147 (1.084, 1.213)} &           1.064 (0.933, 1.214) & \textbf{1.267 (1.208, 1.329)} & \textbf{1.473 (1.356, 1.601)} & \textbf{1.435 (1.298, 1.586)}\\
                                            &1600                                        &10                                           &&           1.049 (0.975, 1.130) &           1.014 (0.863, 1.192) & \textbf{1.276 (1.179, 1.382)} & \textbf{1.217 (1.072, 1.380)} & \textbf{1.172 (1.013, 1.355)}\\
                                            &                                            &20                                           && \textbf{1.108 (1.024, 1.199)} &           1.138 (0.927, 1.397) & \textbf{1.358 (1.256, 1.469)} & \textbf{1.479 (1.288, 1.698)} & \textbf{1.434 (1.217, 1.690)}\\
$\mu(x_3) + \tau(x_1, x_2) W(x_3)$        &800                                         &10                                           &&           0.971 (0.918, 1.027) & \textit{0.794 (0.702, 0.898)} &           1.061 (0.998, 1.127) & \textit{0.894 (0.815, 0.980)} &           0.938 (0.841, 1.045)\\
                                            &                                            &20                                           &&           0.980 (0.925, 1.038) & \textit{0.777 (0.679, 0.889)} &           0.946 (0.893, 1.003) & \textit{0.877 (0.798, 0.964)} &           0.904 (0.809, 1.011)\\
                                            &1600                                        &10                                           &&           0.989 (0.922, 1.061) & \textit{0.819 (0.686, 0.977)} & \textbf{1.127 (1.039, 1.222)} &           0.961 (0.841, 1.099) &           0.996 (0.852, 1.164)\\
                                            &                                            &20                                           &&           0.972 (0.902, 1.049) &           0.843 (0.692, 1.026) &           1.002 (0.916, 1.096) &           1.021 (0.884, 1.181) &           1.006 (0.854, 1.186)\\
$\tau(x_1, x_2) W(x_3)$                    &800                                         &10                                           &&           1.043 (0.988, 1.100) & \textit{0.570 (0.502, 0.646)} & \textbf{1.074 (1.014, 1.138)} &           1.027 (0.948, 1.112) &           1.055 (0.961, 1.159)\\
                                            &                                            &20                                           && \textbf{1.103 (1.045, 1.165)} & \textit{0.721 (0.634, 0.820)} &           0.982 (0.930, 1.036) & \textbf{1.155 (1.065, 1.251)} & \textbf{1.116 (1.018, 1.223)}\\
                                            &1600                                        &10                                           &&           1.014 (0.948, 1.084) & \textit{0.774 (0.660, 0.910)} & \textbf{1.140 (1.053, 1.235)} &           1.072 (0.958, 1.200) &           1.056 (0.925, 1.206)\\
                                            &                                            &20                                           &&           1.021 (0.949, 1.098) &           0.883 (0.736, 1.058) & \textbf{1.151 (1.063, 1.246)} & \textbf{1.158 (1.028, 1.305)} &           1.028 (0.901, 1.173)\\
$\mu(x_3) + \tau(x_1, x_2) W(x_4)$        &800                                         &10                                           && \textbf{1.490 (1.420, 1.563)} & \textit{0.814 (0.725, 0.914)} & \textbf{1.295 (1.233, 1.360)} & \textbf{1.476 (1.374, 1.586)} & \textbf{1.597 (1.466, 1.738)}\\
                                            &                                            &20                                           && \textbf{1.706 (1.625, 1.792)} &           1.019 (0.898, 1.156) & \textbf{1.272 (1.219, 1.329)} & \textbf{1.701 (1.581, 1.831)} & \textbf{1.810 (1.661, 1.972)}\\
                                            &1600                                        &10                                           && \textbf{1.268 (1.193, 1.348)} &           1.026 (0.873, 1.205) & \textbf{1.381 (1.292, 1.477)} & \textbf{1.604 (1.451, 1.774)} & \textbf{1.782 (1.585, 2.003)}\\
                                            &                                            &20                                           && \textbf{1.467 (1.375, 1.565)} &           1.154 (0.967, 1.377) & \textbf{1.427 (1.336, 1.524)} & \textbf{2.160 (1.927, 2.422)} & \textbf{2.323 (2.033, 2.653)}\\
$\mu(x_3) + 0 \cdot (W(x_3) - 0.5)$       &800                                         &10                                           &&           1.100 (0.995, 1.215) &           1.000 (0.897, 1.115) & \textbf{1.211 (1.068, 1.374)} &           1.007 (0.852, 1.190) &           0.985 (0.787, 1.233)\\
                                            &                                            &20                                           && \textbf{1.194 (1.059, 1.345)} &           0.956 (0.839, 1.091) & \textbf{1.352 (1.147, 1.594)} &           1.017 (0.827, 1.250) &           0.987 (0.746, 1.306)\\
                                            &1600                                        &10                                           && \textbf{1.309 (1.160, 1.478)} & \textbf{1.161 (1.003, 1.345)} & \textbf{1.372 (1.167, 1.613)} &           1.232 (0.989, 1.535) &           1.203 (0.893, 1.619)\\
                                            &                                            &20                                           && \textbf{1.395 (1.181, 1.647)} &           1.119 (0.932, 1.344) & \textbf{1.526 (1.193, 1.951)} &           1.245 (0.949, 1.634) &           1.227 (0.852, 1.768)\\
$\tau(x_1, x_2) (W - 0.5)$                 &800                                         &10                                           &&           0.994 (0.931, 1.061) & \textit{0.843 (0.794, 0.895)} &           0.999 (0.946, 1.055) & \textit{0.791 (0.725, 0.862)} & \textit{0.682 (0.618, 0.752)}\\
                                            &                                            &20                                           &&           0.975 (0.913, 1.040) & \textit{0.871 (0.814, 0.931)} &           1.017 (0.956, 1.082) & \textit{0.746 (0.675, 0.825)} & \textit{0.668 (0.597, 0.747)}\\
                                            &1600                                        &10                                           &&           0.965 (0.888, 1.048) &           0.969 (0.893, 1.052) &           0.994 (0.924, 1.071) &           0.894 (0.790, 1.012) & \textit{0.826 (0.720, 0.949)}\\
                                            &                                            &20                                           &&           0.974 (0.893, 1.062) &           0.966 (0.881, 1.060) &           0.997 (0.912, 1.090) &           0.868 (0.751, 1.004) & \textit{0.813 (0.688, 0.962)}\\
$\mu(x_1) + \tau(x_1, x_2) (W(x_1) - 0.5)$&800                                         &10                                           &&           1.054 (0.996, 1.116) &           0.998 (0.946, 1.053) &           0.953 (0.906, 1.002) & \textit{0.853 (0.789, 0.921)} &           0.947 (0.866, 1.036)\\
                                            &                                            &20                                           &&           1.035 (0.976, 1.097) & \textit{0.881 (0.828, 0.937)} & \textit{0.878 (0.834, 0.924)} &           1.019 (0.943, 1.102) &           1.091 (0.996, 1.196)\\
                                            &1600                                        &10                                           &&           1.022 (0.950, 1.100) & \textbf{1.104 (1.031, 1.182)} &           1.029 (0.964, 1.098) &           0.933 (0.835, 1.042) &           0.937 (0.822, 1.067)\\
                                            &                                            &20                                           &&           1.061 (0.983, 1.146) & \textbf{1.113 (1.024, 1.210)} &           1.020 (0.944, 1.103) &           0.960 (0.846, 1.090) &           1.063 (0.927, 1.220)\\
$\mu(x_1) + \tau(x_1, x_2) (W - 0.5)$     &800                                         &10                                           &&           0.986 (0.926, 1.048) &           0.952 (0.899, 1.009) & \textbf{1.128 (1.068, 1.192)} & \textit{0.849 (0.782, 0.923)} & \textit{0.843 (0.770, 0.922)}\\
                                            &                                            &20                                           && \textbf{1.066 (1.003, 1.134)} & \textit{0.875 (0.818, 0.935)} &           1.057 (0.999, 1.120) & \textit{0.856 (0.784, 0.935)} & \textit{0.811 (0.735, 0.895)}\\
                                            &1600                                        &10                                           &&           1.004 (0.928, 1.087) & \textbf{1.084 (1.005, 1.169)} & \textbf{1.078 (1.001, 1.162)} &           0.934 (0.831, 1.051) &           0.904 (0.791, 1.035)\\
                                            &                                            &20                                           &&           1.038 (0.952, 1.131) &           1.091 (0.993, 1.200) & \textbf{1.144 (1.051, 1.246)} &           0.993 (0.865, 1.141) &           0.964 (0.827, 1.123)\\
$\mu(x_3) + \tau(x_1, x_2) (W - 0.5)$     &800                                         &10                                           && \textbf{1.237 (1.175, 1.302)} &           1.003 (0.948, 1.062) & \textbf{1.260 (1.197, 1.326)} & \textbf{1.175 (1.086, 1.271)} & \textbf{1.151 (1.047, 1.266)}\\
                                            &                                            &20                                           && \textbf{1.421 (1.348, 1.498)} & \textbf{1.167 (1.095, 1.243)} & \textbf{1.391 (1.321, 1.464)} & \textbf{1.414 (1.302, 1.535)} & \textbf{1.342 (1.217, 1.480)}\\
                                            &1600                                        &10                                           && \textbf{1.192 (1.108, 1.283)} & \textbf{1.124 (1.036, 1.218)} & \textbf{1.250 (1.161, 1.345)} & \textbf{1.168 (1.036, 1.317)} &           1.132 (0.983, 1.303)\\
                                            &                                            &20                                           && \textbf{1.261 (1.171, 1.358)} & \textbf{1.216 (1.113, 1.328)} & \textbf{1.393 (1.291, 1.504)} & \textbf{1.320 (1.156, 1.508)} & \textbf{1.319 (1.127, 1.543)}\\
$\mu(x_3) + \tau(x_1, x_2) (W(x_3) - 0.5)$&800                                         &10                                           &&           1.028 (0.973, 1.087) &           0.967 (0.913, 1.024) &           1.003 (0.954, 1.055) & \textit{0.821 (0.755, 0.893)} & \textit{0.857 (0.772, 0.951)}\\
                                            &                                            &20                                           && \textbf{1.067 (1.007, 1.131)} &           0.982 (0.921, 1.047) &           0.999 (0.947, 1.055) & \textit{0.845 (0.776, 0.921)} & \textit{0.859 (0.775, 0.952)}\\
                                            &1600                                        &10                                           &&           1.027 (0.955, 1.105) & \textbf{1.129 (1.050, 1.214)} &           1.044 (0.974, 1.118) &           0.902 (0.809, 1.006) &           0.973 (0.850, 1.113)\\
                                            &                                            &20                                           &&           1.074 (0.997, 1.157) &           1.074 (0.981, 1.177) &           1.024 (0.943, 1.111) &           0.994 (0.874, 1.129) &           1.038 (0.886, 1.215)\\
$\tau(x_1, x_2) (W(x_3) - 0.5)$            &800                                         &10                                           && \textbf{1.078 (1.022, 1.137)} &           1.007 (0.957, 1.058) & \textbf{1.115 (1.060, 1.174)} & \textit{0.914 (0.850, 0.982)} &           0.925 (0.846, 1.011)\\
                                            &                                            &20                                           && \textbf{1.204 (1.140, 1.271)} &           1.001 (0.947, 1.058) & \textbf{1.088 (1.037, 1.142)} &           0.987 (0.913, 1.066) &           1.006 (0.919, 1.101)\\
                                            &1600                                        &10                                           && \textbf{1.152 (1.075, 1.236)} & \textbf{1.111 (1.039, 1.188)} & \textbf{1.140 (1.068, 1.216)} &           0.983 (0.886, 1.089) &           1.024 (0.903, 1.161)\\
                                            &                                            &20                                           && \textbf{1.192 (1.106, 1.286)} & \textbf{1.116 (1.031, 1.208)} & \textbf{1.192 (1.105, 1.285)} &           1.044 (0.928, 1.174) &           1.063 (0.929, 1.216)\\
$\mu(x_3) + \tau(x_1, x_2)(W(x_4) - 0.5)$ &800                                         &10                                           && \textbf{1.908 (1.823, 1.997)} & \textbf{1.124 (1.066, 1.184)} & \textbf{1.438 (1.374, 1.505)} & \textbf{1.613 (1.505, 1.728)} & \textbf{1.774 (1.627, 1.934)}\\
                                            &                                            &20                                           && \textbf{2.266 (2.167, 2.370)} & \textbf{1.147 (1.085, 1.212)} & \textbf{1.390 (1.332, 1.452)} & \textbf{1.724 (1.622, 1.831)} & \textbf{1.811 (1.681, 1.951)}\\
                                            &1600                                        &10                                           && \textbf{1.484 (1.391, 1.583)} & \textbf{1.291 (1.206, 1.381)} & \textbf{1.482 (1.389, 1.581)} & \textbf{1.587 (1.442, 1.746)} & \textbf{1.700 (1.516, 1.906)}\\
                                            &                                            &20                                           && \textbf{1.728 (1.622, 1.840)} & \textbf{1.325 (1.222, 1.437)} & \textbf{1.686 (1.573, 1.808)} & \textbf{1.935 (1.754, 2.136)} & \textbf{2.129 (1.888, 2.402)}\\
\hline
\end{tabular}

\end{sidewaystable}

\begin{sidewaystable}
\caption{Results of \textbf{RQ 3} for the experimental setups in Section~\ref{sec:simathey}.  Comparison of mean
squared errors for $\hat{\tau}(\rx)$ in the different scenarios.  Estimates
and simultaneous $95$ \% confidence intervals were obtained from a normal
linear mixed model with log-link.  Cells printed in bold font correspond to
a superior reference of \textit{Robinson}$_{\hat{W}}$, cells printed in italics indicate an inferior reference of \textit{Robinson}$_{\hat{W}}$.
\label{tab:lmeradaptive3}}
\addtolength{\tabcolsep}{-2pt}
\tiny
\begin{tabular}{llllrrrrr}
\hline
&& && \multicolumn{5}{c}{Mean squared error ratio for RQ 3: Robinson vs. Robinson$_{\hat{W}}$}\\
\cline{5-5}\cline{6-6}\cline{7-7}\cline{8-8}\cline{9-9}
\rule{0pt}{6ex} Part \hspace{.4cm} DGP&N&P && \multicolumn{1}{c}{Normal}&\multicolumn{1}{c}{Binomial}&\multicolumn{1}{c}{Multinomial}&\multicolumn{1}{c}{Weibull}&\multicolumn{1}{c}{Cox}\\
\hline

A \hspace{.8cm} $\mu(x_3) + 0 \cdot W(x_3)$                 &800                                                            &10                                                              && \textit{0.846 (0.770, 0.929)} &           0.988 (0.692, 1.410) & \textit{0.779 (0.671, 0.905)} &           0.902 (0.749, 1.085) &           0.941 (0.735, 1.204)\\
                                                               &                                                               &20                                                              && \textit{0.843 (0.753, 0.944)} &           1.090 (0.710, 1.673) & \textit{0.767 (0.628, 0.935)} &           0.982 (0.792, 1.218) &           1.013 (0.763, 1.345)\\
                                                               &1600                                                           &10                                                              && \textit{0.783 (0.697, 0.880)} &           0.884 (0.555, 1.406) & \textit{0.727 (0.591, 0.896)} &           0.808 (0.631, 1.034) &           0.817 (0.590, 1.132)\\
                                                               &                                                               &20                                                              && \textit{0.719 (0.618, 0.835)} &           0.923 (0.518, 1.644) & \textit{0.699 (0.536, 0.913)} &           0.831 (0.614, 1.125) &           0.834 (0.559, 1.244)\\
\hspace{1.1cm} $\tau(x_1, x_2) W$                            &800                                                            &10                                                              && \textbf{1.095 (1.026, 1.168)} & \textbf{1.863 (1.732, 2.004)} &           1.002 (0.942, 1.065) & \textbf{1.274 (1.155, 1.404)} & \textbf{1.271 (1.141, 1.416)}\\
                                                               &                                                               &20                                                              && \textbf{1.096 (1.024, 1.172)} & \textbf{1.238 (1.066, 1.438)} &           1.049 (0.987, 1.114) &           1.103 (0.998, 1.220) & \textbf{1.305 (1.172, 1.453)}\\
                                                               &1600                                                           &10                                                              && \textbf{1.086 (1.007, 1.173)} & \textbf{1.217 (1.041, 1.423)} &           0.943 (0.869, 1.024) &           1.138 (0.983, 1.317) & \textbf{1.225 (1.051, 1.428)}\\
                                                               &                                                               &20                                                              && \textbf{1.126 (1.032, 1.228)} &           1.066 (0.859, 1.323) &           1.012 (0.925, 1.106) &           1.140 (0.979, 1.328) & \textbf{1.211 (1.023, 1.434)}\\
\hspace{1.1cm} $\mu(x_1) + \tau(x_1, x_2) W(x_1)$           &800                                                            &10                                                              &&           1.055 (0.996, 1.116) &           1.000 (0.867, 1.153) &           1.042 (0.995, 1.092) & \textbf{1.114 (1.022, 1.214)} &           1.014 (0.918, 1.121)\\
                                                               &                                                               &20                                                              &&           0.963 (0.906, 1.023) &           1.062 (0.929, 1.214) & \textbf{1.225 (1.171, 1.281)} & \textbf{1.090 (1.001, 1.187)} &           1.063 (0.961, 1.175)\\
                                                               &1600                                                           &10                                                              &&           1.052 (0.982, 1.127) &           0.994 (0.776, 1.273) &           0.975 (0.906, 1.049) &           1.098 (0.975, 1.235) &           1.029 (0.898, 1.180)\\
                                                               &                                                               &20                                                              &&           0.999 (0.926, 1.077) &           1.043 (0.851, 1.279) & \textbf{1.082 (1.009, 1.159)} &           0.981 (0.859, 1.119) &           0.871 (0.754, 1.007)\\
\hspace{1.1cm} $\mu(x_1) + \tau(x_1, x_2) W$                &800                                                            &10                                                              && \textbf{1.066 (1.002, 1.135)} & \textit{0.548 (0.478, 0.628)} & \textit{0.903 (0.861, 0.948)} &           1.038 (0.949, 1.135) &           1.073 (0.968, 1.189)\\
                                                               &                                                               &20                                                              &&           1.059 (0.996, 1.125) &           1.017 (0.883, 1.171) & \textit{0.912 (0.863, 0.964)} & \textit{0.902 (0.824, 0.988)} &           0.950 (0.854, 1.057)\\
                                                               &1600                                                           &10                                                              &&           1.022 (0.945, 1.106) & \textit{0.486 (0.407, 0.581)} & \textit{0.836 (0.775, 0.902)} &           0.965 (0.850, 1.096) &           0.952 (0.830, 1.092)\\
                                                               &                                                               &20                                                              &&           1.025 (0.941, 1.117) &           0.983 (0.819, 1.179) & \textit{0.882 (0.810, 0.960)} &           0.945 (0.822, 1.088) &           0.897 (0.767, 1.048)\\
\hspace{1.1cm} $\mu(x_3) + \tau(x_1, x_2) W$                &800                                                            &10                                                              &&           0.982 (0.930, 1.038) & \textbf{1.215 (1.072, 1.376)} & \textit{0.796 (0.754, 0.839)} & \textit{0.813 (0.746, 0.885)} & \textit{0.904 (0.819, 0.997)}\\
                                                               &                                                               &20                                                              && \textit{0.872 (0.824, 0.922)} &           0.940 (0.824, 1.072) & \textit{0.789 (0.752, 0.828)} & \textit{0.679 (0.624, 0.738)} & \textit{0.697 (0.631, 0.770)}\\
                                                               &1600                                                           &10                                                              &&           0.953 (0.885, 1.026) &           0.986 (0.839, 1.159) & \textit{0.784 (0.723, 0.849)} & \textit{0.822 (0.725, 0.933)} & \textit{0.853 (0.738, 0.987)}\\
                                                               &                                                               &20                                                              && \textit{0.903 (0.834, 0.976)} &           0.879 (0.716, 1.079) & \textit{0.736 (0.681, 0.796)} & \textit{0.676 (0.589, 0.777)} & \textit{0.697 (0.592, 0.821)}\\
\hspace{1.1cm} $\mu(x_3) + \tau(x_1, x_2) W(x_3)$           &800                                                            &10                                                              &&           1.030 (0.974, 1.090) & \textbf{1.260 (1.114, 1.425)} &           0.943 (0.888, 1.002) & \textbf{1.118 (1.020, 1.226)} &           1.066 (0.957, 1.189)\\
                                                               &                                                               &20                                                              &&           1.021 (0.963, 1.081) & \textbf{1.287 (1.125, 1.472)} &           1.057 (0.997, 1.120) & \textbf{1.140 (1.037, 1.253)} &           1.106 (0.989, 1.236)\\
                                                               &1600                                                           &10                                                              &&           1.011 (0.942, 1.085) & \textbf{1.222 (1.024, 1.458)} & \textit{0.888 (0.818, 0.963)} &           1.040 (0.910, 1.189) &           1.004 (0.859, 1.174)\\
                                                               &                                                               &20                                                              &&           1.028 (0.954, 1.109) &           1.187 (0.975, 1.445) &           0.998 (0.912, 1.092) &           0.979 (0.847, 1.132) &           0.994 (0.843, 1.171)\\
\hspace{1.1cm} $\tau(x_1, x_2) W(x_3)$                       &800                                                            &10                                                              &&           0.959 (0.909, 1.012) & \textbf{1.756 (1.548, 1.991)} & \textit{0.931 (0.879, 0.986)} &           0.974 (0.899, 1.055) &           0.948 (0.863, 1.041)\\
                                                               &                                                               &20                                                              && \textit{0.906 (0.858, 0.957)} & \textbf{1.386 (1.219, 1.576)} &           1.019 (0.965, 1.075) & \textit{0.866 (0.799, 0.939)} & \textit{0.896 (0.818, 0.983)}\\
                                                               &1600                                                           &10                                                              &&           0.986 (0.923, 1.055) & \textbf{1.291 (1.099, 1.516)} & \textit{0.877 (0.810, 0.950)} &           0.933 (0.834, 1.044) &           0.947 (0.829, 1.081)\\
                                                               &                                                               &20                                                              &&           0.979 (0.911, 1.053) &           1.133 (0.945, 1.359) & \textit{0.869 (0.803, 0.941)} & \textit{0.864 (0.766, 0.973)} &           0.973 (0.852, 1.110)\\
\hspace{1.1cm} $\mu(x_3) + \tau(x_1, x_2) W(x_4)$           &800                                                            &10                                                              && \textit{0.671 (0.640, 0.704)} & \textbf{1.229 (1.095, 1.380)} & \textit{0.772 (0.735, 0.811)} & \textit{0.677 (0.630, 0.728)} & \textit{0.626 (0.575, 0.682)}\\
                                                               &                                                               &20                                                              && \textit{0.586 (0.558, 0.615)} &           0.981 (0.865, 1.113) & \textit{0.786 (0.753, 0.821)} & \textit{0.588 (0.546, 0.633)} & \textit{0.552 (0.507, 0.602)}\\
                                                               &1600                                                           &10                                                              && \textit{0.789 (0.742, 0.838)} &           0.975 (0.830, 1.146) & \textit{0.724 (0.677, 0.774)} & \textit{0.623 (0.564, 0.689)} & \textit{0.561 (0.499, 0.631)}\\
                                                               &                                                               &20                                                              && \textit{0.682 (0.639, 0.727)} &           0.867 (0.726, 1.034) & \textit{0.701 (0.656, 0.748)} & \textit{0.463 (0.413, 0.519)} & \textit{0.431 (0.377, 0.492)}\\
\hline B \hspace{.8cm} $\mu(x_3) + 0 \cdot (W(x_3) - 0.5)$ &800                                                            &10                                                              &&           0.909 (0.823, 1.005) &           1.000 (0.897, 1.115) & \textit{0.826 (0.728, 0.937)} &           0.993 (0.840, 1.173) &           1.015 (0.811, 1.270)\\
                                                               &                                                               &20                                                              && \textit{0.838 (0.743, 0.944)} &           1.046 (0.917, 1.192) & \textit{0.740 (0.627, 0.872)} &           0.984 (0.800, 1.209) &           1.013 (0.766, 1.341)\\
                                                               &1600                                                           &10                                                              && \textit{0.764 (0.677, 0.862)} & \textit{0.861 (0.744, 0.997)} & \textit{0.729 (0.620, 0.857)} &           0.812 (0.652, 1.012) &           0.832 (0.618, 1.120)\\
                                                               &                                                               &20                                                              && \textit{0.717 (0.607, 0.846)} &           0.894 (0.744, 1.073) & \textit{0.655 (0.513, 0.838)} &           0.803 (0.612, 1.054) &           0.815 (0.566, 1.174)\\
\hspace{1.1cm} $\tau(x_1, x_2) (W - 0.5)$                    &800                                                            &10                                                              &&           1.006 (0.942, 1.075) & \textbf{1.186 (1.118, 1.259)} &           1.001 (0.948, 1.057) & \textbf{1.265 (1.160, 1.379)} & \textbf{1.467 (1.331, 1.617)}\\
                                                               &                                                               &20                                                              &&           1.026 (0.961, 1.096) & \textbf{1.149 (1.074, 1.228)} &           0.983 (0.925, 1.046) & \textbf{1.340 (1.213, 1.482)} & \textbf{1.497 (1.338, 1.675)}\\
                                                               &1600                                                           &10                                                              &&           1.037 (0.954, 1.126) &           1.032 (0.951, 1.120) &           1.006 (0.934, 1.082) &           1.119 (0.989, 1.266) & \textbf{1.210 (1.054, 1.389)}\\
                                                               &                                                               &20                                                              &&           1.027 (0.942, 1.120) &           1.035 (0.944, 1.136) &           1.003 (0.917, 1.096) &           1.151 (0.996, 1.331) & \textbf{1.229 (1.039, 1.454)}\\
\hspace{1.1cm} $\mu(x_1) + \tau(x_1, x_2) (W(x_1) - 0.5)$   &800                                                            &10                                                              &&           0.948 (0.896, 1.004) &           1.002 (0.950, 1.057) &           1.050 (0.998, 1.104) & \textbf{1.173 (1.086, 1.267)} &           1.056 (0.965, 1.154)\\
                                                               &                                                               &20                                                              &&           0.967 (0.912, 1.025) & \textbf{1.135 (1.067, 1.207)} & \textbf{1.139 (1.082, 1.199)} &           0.981 (0.907, 1.061) &           0.916 (0.836, 1.004)\\
                                                               &1600                                                           &10                                                              &&           0.978 (0.909, 1.053) & \textit{0.906 (0.846, 0.970)} &           0.972 (0.911, 1.038) &           1.072 (0.960, 1.197) &           1.067 (0.937, 1.216)\\
                                                               &                                                               &20                                                              &&           0.942 (0.872, 1.018) & \textit{0.898 (0.826, 0.976)} &           0.980 (0.907, 1.060) &           1.042 (0.917, 1.183) &           0.940 (0.819, 1.079)\\
\hspace{1.1cm} $\mu(x_1) + \tau(x_1, x_2) (W - 0.5)$        &800                                                            &10                                                              &&           1.015 (0.954, 1.079) &           1.050 (0.991, 1.113) & \textit{0.887 (0.839, 0.937)} & \textbf{1.178 (1.084, 1.280)} & \textbf{1.187 (1.084, 1.299)}\\
                                                               &                                                               &20                                                              && \textit{0.938 (0.882, 0.997)} & \textbf{1.143 (1.069, 1.222)} &           0.946 (0.893, 1.001) & \textbf{1.168 (1.070, 1.275)} & \textbf{1.233 (1.118, 1.360)}\\
                                                               &1600                                                           &10                                                              &&           0.996 (0.920, 1.078) & \textit{0.923 (0.855, 0.995)} & \textit{0.927 (0.861, 0.999)} &           1.070 (0.952, 1.203) &           1.106 (0.967, 1.265)\\
                                                               &                                                               &20                                                              &&           0.964 (0.884, 1.050) &           0.916 (0.833, 1.007) & \textit{0.874 (0.802, 0.951)} &           1.007 (0.876, 1.156) &           1.037 (0.890, 1.209)\\
\hspace{1.1cm} $\mu(x_3) + \tau(x_1, x_2) (W - 0.5)$        &800                                                            &10                                                              && \textit{0.809 (0.768, 0.851)} &           0.997 (0.941, 1.055) & \textit{0.794 (0.754, 0.835)} & \textit{0.851 (0.787, 0.921)} & \textit{0.869 (0.790, 0.955)}\\
                                                               &                                                               &20                                                              && \textit{0.704 (0.668, 0.742)} & \textit{0.857 (0.804, 0.913)} & \textit{0.719 (0.683, 0.757)} & \textit{0.707 (0.651, 0.768)} & \textit{0.745 (0.676, 0.822)}\\
                                                               &1600                                                           &10                                                              && \textit{0.839 (0.780, 0.903)} & \textit{0.890 (0.821, 0.965)} & \textit{0.800 (0.743, 0.861)} & \textit{0.856 (0.759, 0.965)} &           0.883 (0.767, 1.017)\\
                                                               &                                                               &20                                                              && \textit{0.793 (0.737, 0.854)} & \textit{0.823 (0.753, 0.899)} & \textit{0.718 (0.665, 0.775)} & \textit{0.758 (0.663, 0.865)} & \textit{0.758 (0.648, 0.887)}\\
\hspace{1.1cm} $\mu(x_3) + \tau(x_1, x_2) (W(x_3) - 0.5)$   &800                                                            &10                                                              &&           0.973 (0.920, 1.028) &           1.034 (0.977, 1.095) &           0.997 (0.948, 1.048) & \textbf{1.218 (1.120, 1.325)} & \textbf{1.167 (1.051, 1.296)}\\
                                                               &                                                               &20                                                              && \textit{0.937 (0.884, 0.993)} &           1.018 (0.955, 1.085) &           1.001 (0.948, 1.056) & \textbf{1.183 (1.086, 1.288)} & \textbf{1.164 (1.050, 1.291)}\\
                                                               &1600                                                           &10                                                              &&           0.974 (0.905, 1.047) & \textit{0.885 (0.823, 0.952)} &           0.958 (0.894, 1.027) &           1.109 (0.994, 1.236) &           1.028 (0.899, 1.176)\\
                                                               &                                                               &20                                                              &&           0.931 (0.864, 1.003) &           0.931 (0.850, 1.020) &           0.977 (0.900, 1.060) &           1.006 (0.885, 1.144) &           0.964 (0.823, 1.128)\\
\hspace{1.1cm} $\tau(x_1, x_2) (W(x_3) - 0.5)$               &800                                                            &10                                                              && \textit{0.928 (0.880, 0.979)} &           0.993 (0.945, 1.045) & \textit{0.897 (0.852, 0.944)} & \textbf{1.095 (1.018, 1.177)} &           1.081 (0.989, 1.182)\\
                                                               &                                                               &20                                                              && \textit{0.831 (0.786, 0.877)} &           0.999 (0.946, 1.056) & \textit{0.919 (0.876, 0.964)} &           1.014 (0.938, 1.095) &           0.994 (0.908, 1.088)\\
                                                               &1600                                                           &10                                                              && \textit{0.868 (0.809, 0.931)} & \textit{0.900 (0.842, 0.963)} & \textit{0.877 (0.822, 0.936)} &           1.018 (0.918, 1.128) &           0.977 (0.862, 1.108)\\
                                                               &                                                               &20                                                              && \textit{0.839 (0.778, 0.904)} & \textit{0.896 (0.828, 0.970)} & \textit{0.839 (0.778, 0.905)} &           0.958 (0.852, 1.077) &           0.941 (0.822, 1.077)\\
\hspace{1.1cm} $\mu(x_3) + \tau(x_1, x_2)(W(x_4) - 0.5)$    &800                                                            &10                                                              && \textit{0.524 (0.501, 0.548)} & \textit{0.890 (0.845, 0.938)} & \textit{0.695 (0.664, 0.728)} & \textit{0.620 (0.579, 0.664)} & \textit{0.564 (0.517, 0.615)}\\
                                                               &                                                               &20                                                              && \textit{0.441 (0.422, 0.461)} & \textit{0.872 (0.825, 0.921)} & \textit{0.719 (0.689, 0.751)} & \textit{0.580 (0.546, 0.616)} & \textit{0.552 (0.513, 0.595)}\\
                                                               &1600                                                           &10                                                              && \textit{0.674 (0.632, 0.719)} & \textit{0.775 (0.724, 0.829)} & \textit{0.675 (0.632, 0.720)} & \textit{0.630 (0.573, 0.694)} & \textit{0.588 (0.525, 0.659)}\\
                                                               &                                                               &20                                                              && \textit{0.579 (0.543, 0.617)} & \textit{0.754 (0.696, 0.818)} & \textit{0.593 (0.553, 0.636)} & \textit{0.517 (0.468, 0.570)} & \textit{0.470 (0.416, 0.530)}\\
\hline
\end{tabular}

\end{sidewaystable}

For the normal distribution (first row of Figures~\ref{fig:otherA}~and~\ref{fig:otherB}),
model-based forests with centered $W$
(\textit{Robinson}$_{\hat{W}}$) performed better than naive model-based forests without centering 
in case of confounding (columns 1 and 6).
If predictive covariates were also prognostic (column 3), the effect of local centering
on performance diminished. In case of variables that only influence
the treatment assignment but not the outcome (column 7 and 8), solely centering $W$
led to biased results.
Especially in this scenario, additional adding $\hat{m}(\rx)$ as an offset (\textit{Robinson})
is recommended. However, also in all other scenarios \textit{Robinson} model-based forests
perform at least as well as \textit{Robinson}$_{\hat{W}}$ forests -- except for the setup without a
prognostic effect ($\mu(\rx) \equiv 0$, column 2, see also Table~ref{tab:lmeradaptive3}).

We obtained similar results for the other distributions as shown in Figures~\ref{fig:otherA}~and~\ref{fig:otherB}.
Overlay of prognostic and predictive effects (Part B compared to Part A)
did slightly worsen the performance of all methods in smaller samples
(except in the absence of a predictive effect, see first column of both figures).

We also inspected if the performance of model-based forests degrades for the Weibull
data when the forests do not take the true underlying model as their base model.
We compared the performance of model-based forests when using a Cox
model compared to a Weibull model (Last row of
Figures~\ref{fig:otherA}~\&~\ref{fig:otherB}).
Although knowledge of the true functional form does not enter the Cox modeling process,
it did not lead to a major decrease in performance.

\clearpage

\newpage

\section{Dependence plots} 
\label{ap:dependence}
Dependence plots depict the treatment effect $\tau$ on the prepartum variables - scatter plots for continuous covariates and boxplots for categorical covariates.
For categorical covariates, diamonds display the mean effect per group, and for continuous covariates, we provide the smooth conditional mean effect function calculated by a generalized additive model (GAM) with a single smooth term - the covariate under consideration.
This evaluation scheme closely follows \AYcite{Dandl et al.}{Dandl_Hothorn_Seibold_2022}.

\begin{figure}[h!]
	\subfigure[time onset until treatment]{ 
		\includegraphics[width=0.48\textwidth, page=1]{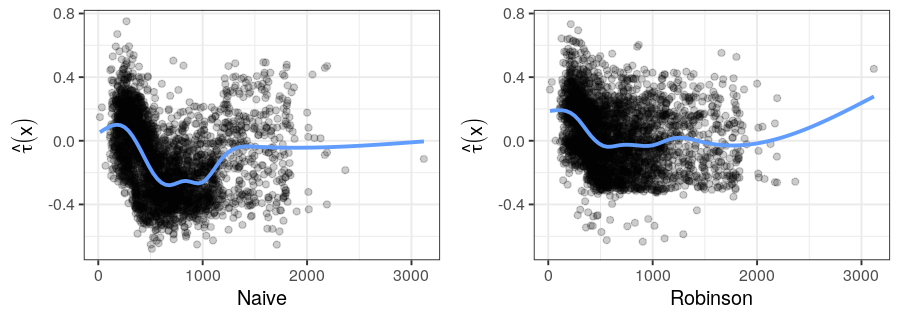}}
	\subfigure[race]{
		\includegraphics[width=0.48\textwidth, page=1]{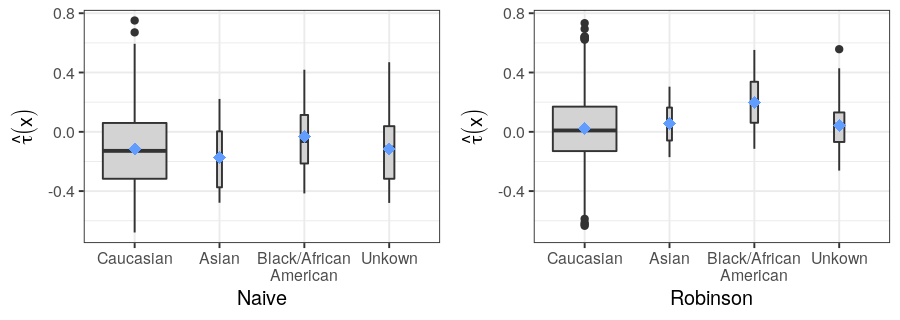}}
	\subfigure[sex]{
		\includegraphics[width=0.48\textwidth, page=1]{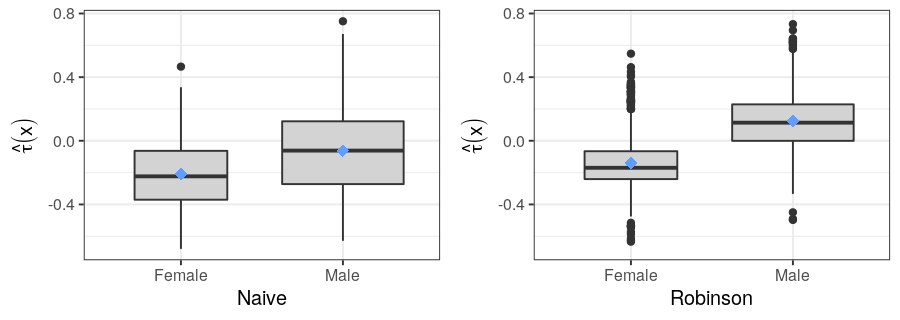}}
	\subfigure[age]{ 
		\includegraphics[width=0.48\textwidth, page=1]{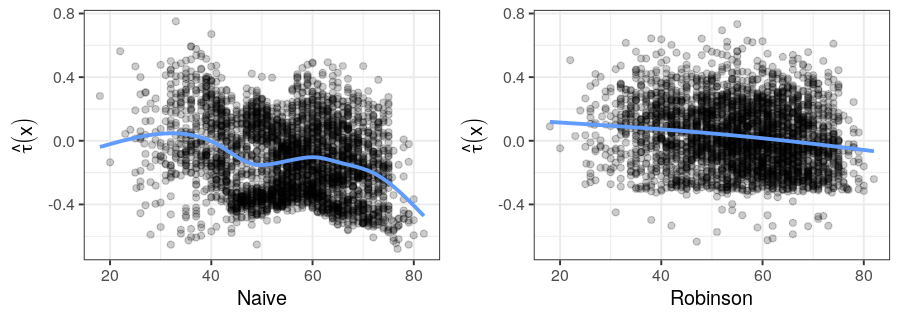}}
	\subfigure[height]{
		\includegraphics[width=0.48\textwidth, page=1]{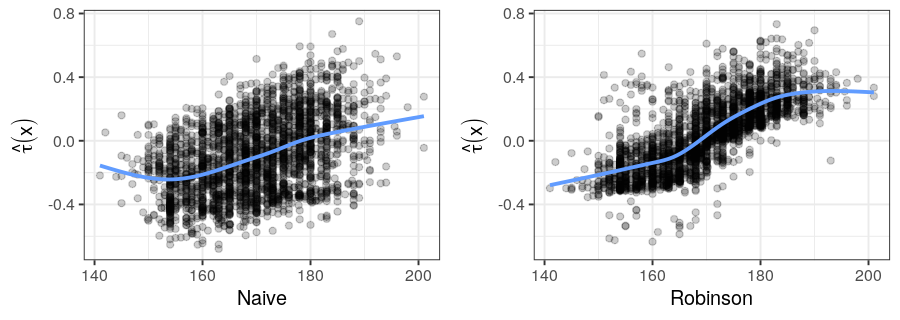}}
	\subfigure[atrophy]{
		\includegraphics[width=0.48\textwidth, page=1]{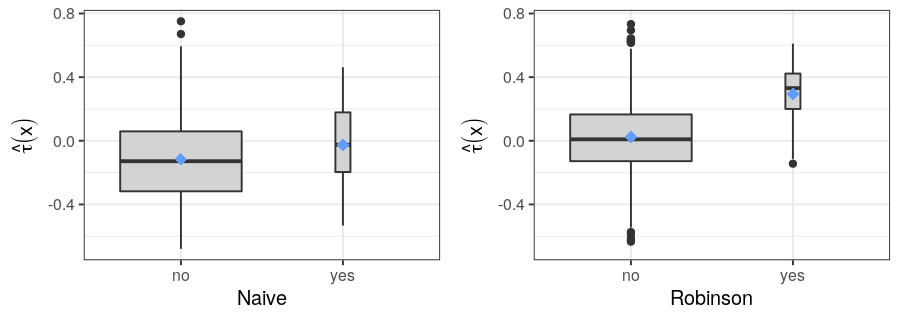}}
	\subfigure[cramps]{
		\includegraphics[width=0.48\textwidth, page=1]{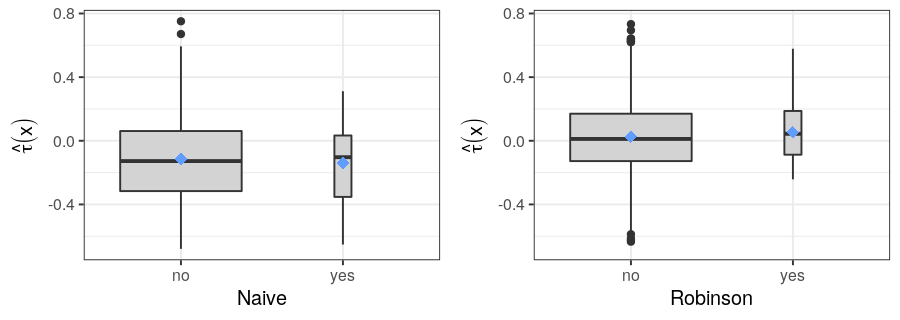}}
	\subfigure[fasciculations]{
		\includegraphics[width=0.48\textwidth, page=1]{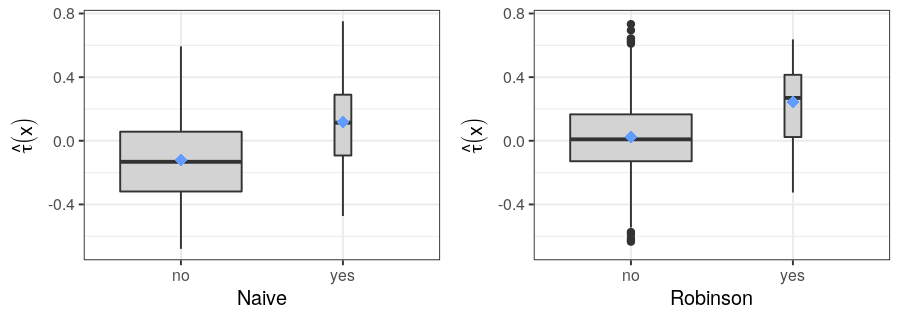}}
	\caption{Survival time: dependency plot of individual average treatment effects calculated by model-based forest without orthogonalization (left), with Robinson orthogonalization (right). Blue lines and diamond points depict (smooth conditional) mean effects.}
	\label{fig:dependplotsurv1}
\end{figure}

\begin{figure}[h!]
	\subfigure[gait changes]{
		\includegraphics[width=0.48\textwidth, page=1]{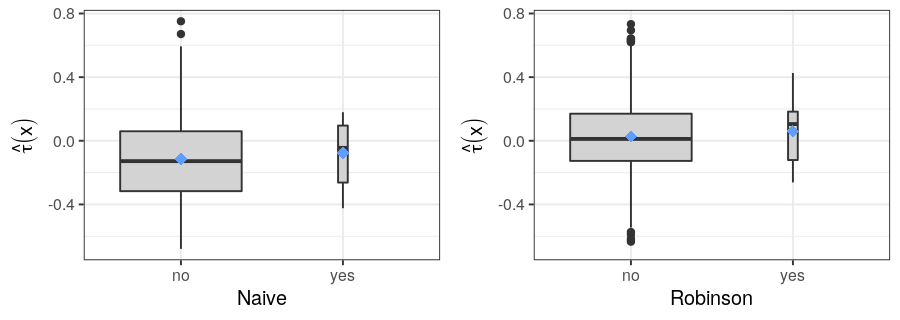}}
	\subfigure[other]{
		\includegraphics[width=0.48\textwidth, page=1]{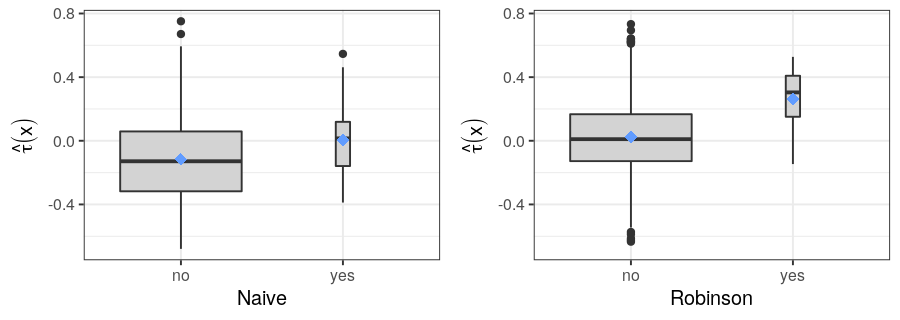}}
	\subfigure[sensory changes]{
		\includegraphics[width=0.48\textwidth, page=1]{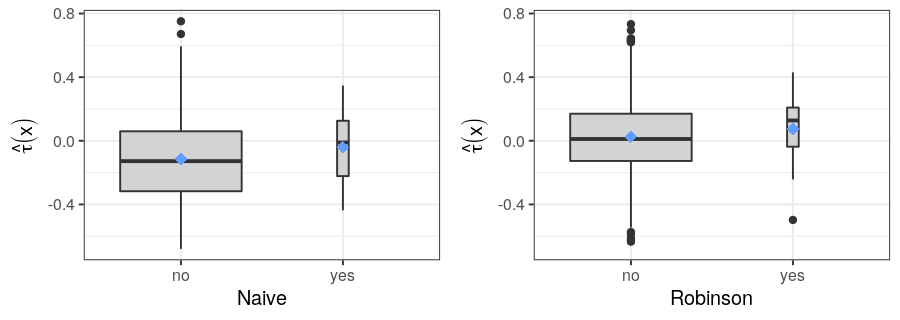}}
	\subfigure[speech]{
		\includegraphics[width=0.48\textwidth, page=1]{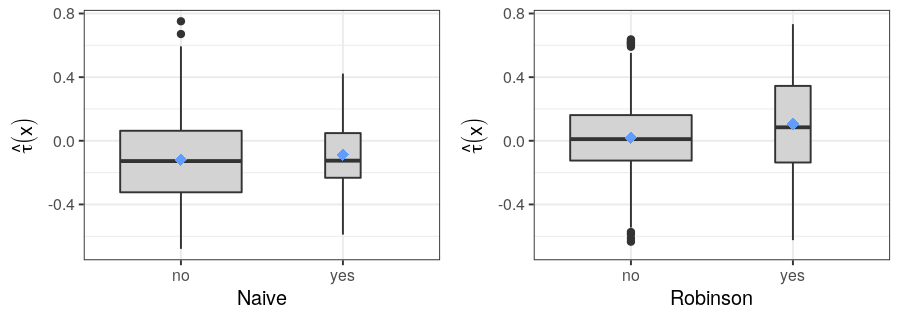}}
	\subfigure[stiffness]{
		\includegraphics[width=0.48\textwidth, page=1]{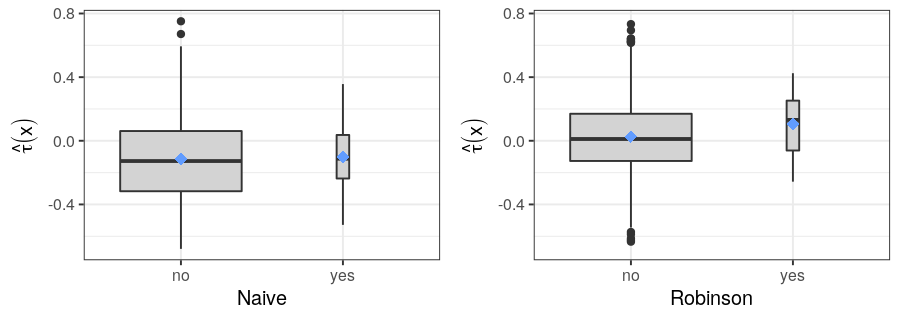}}
	\subfigure[swallowing]{
		\includegraphics[width=0.48\textwidth, page=1]{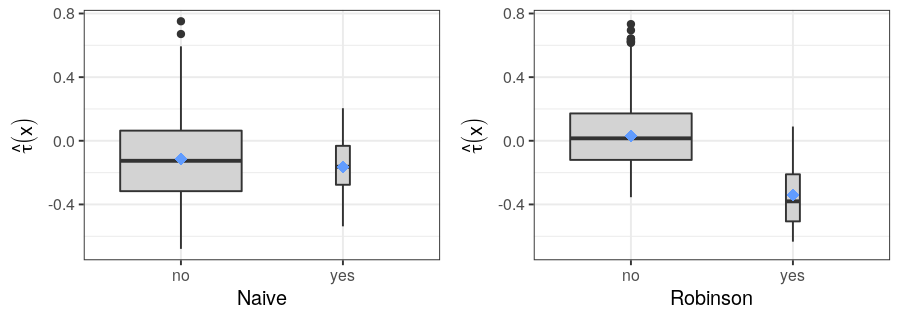}}
	\subfigure[weakness]{
		\includegraphics[width=0.48\textwidth, page=1]{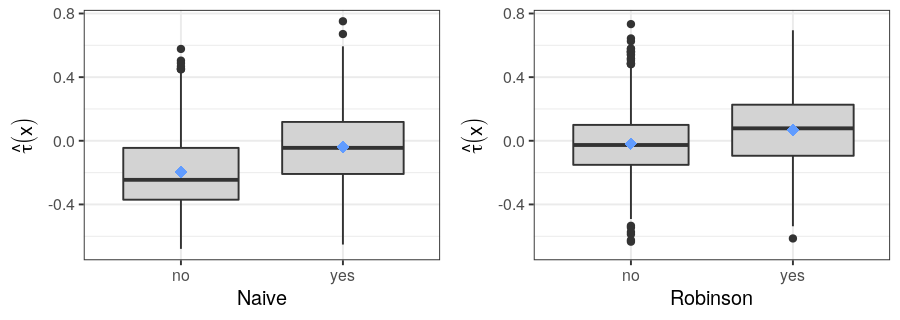}}
	\subfigure[family history (older)]{
		\includegraphics[width=0.48\textwidth, page=1]{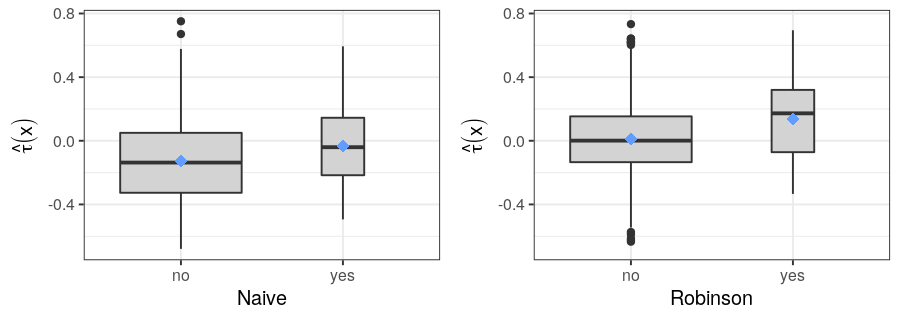}}
	\subfigure[family history (same)]{
		\includegraphics[width=0.48\textwidth, page=1]{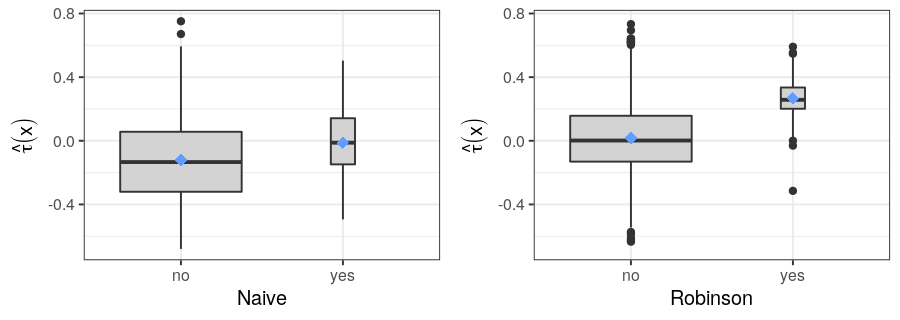}}
	\subfigure[family history (younger)]{
		\includegraphics[width=0.48\textwidth, page=1]{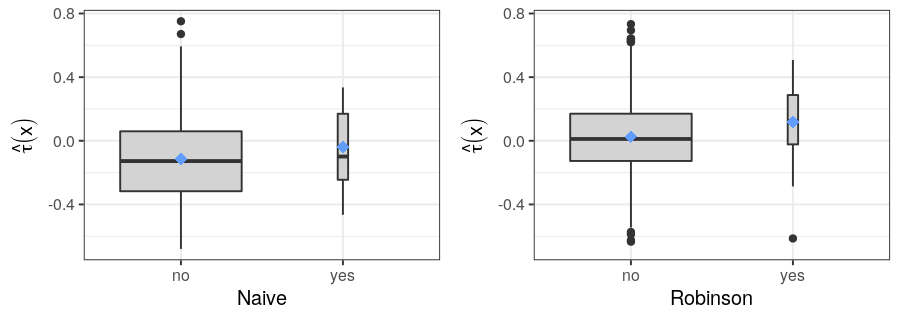}}
	\caption{Survival time: dependency plot of individual average treatment effects calculated by model-based forest without orthogonalization (left), with Robinson orthogonalization (right). Blue lines and diamond points depict (smooth conditional) mean effects.}
	\label{fig:dependplotsurv2}
\end{figure}

\begin{figure}[h!]
	\subfigure[$y_0$]{
		\includegraphics[width=0.48\textwidth, page=1]{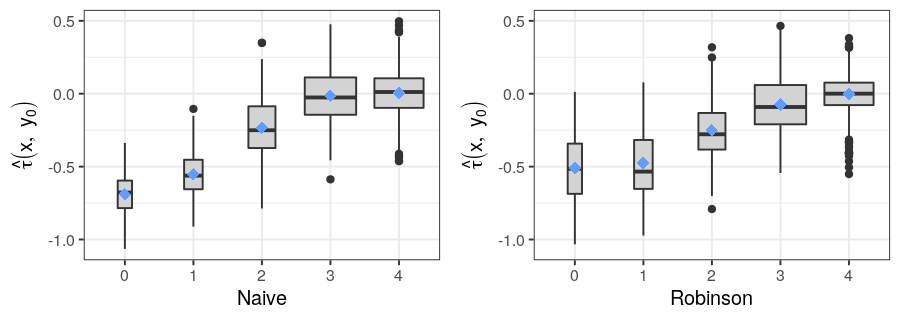}}
	\subfigure[time onset until treatment]{ 
		\includegraphics[width=0.48\textwidth, page=1]{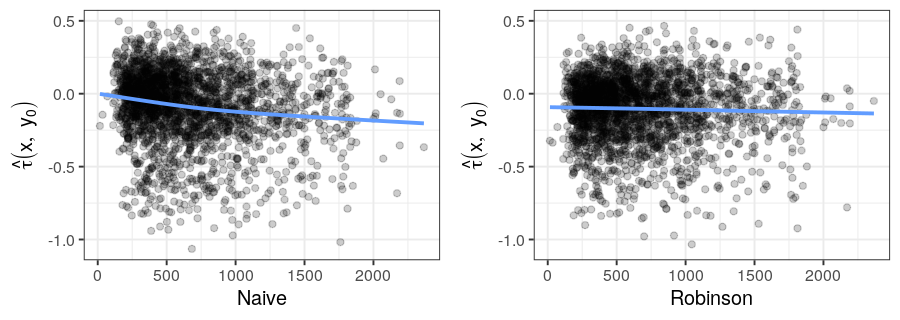}}
	\subfigure[race]{
		\includegraphics[width=0.48\textwidth, page=1]{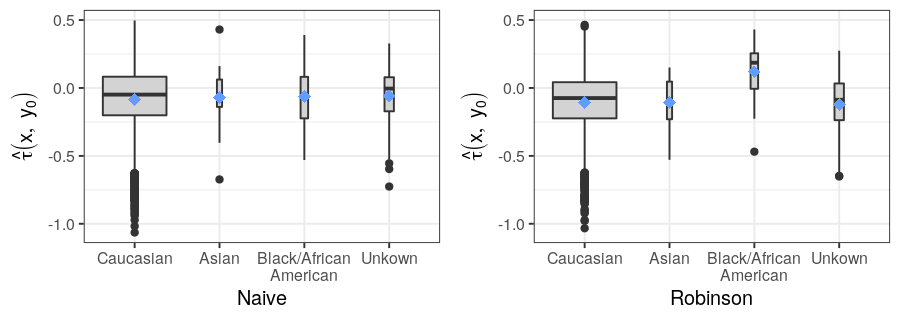}}
	\subfigure[sex]{
		\includegraphics[width=0.48\textwidth, page=1]{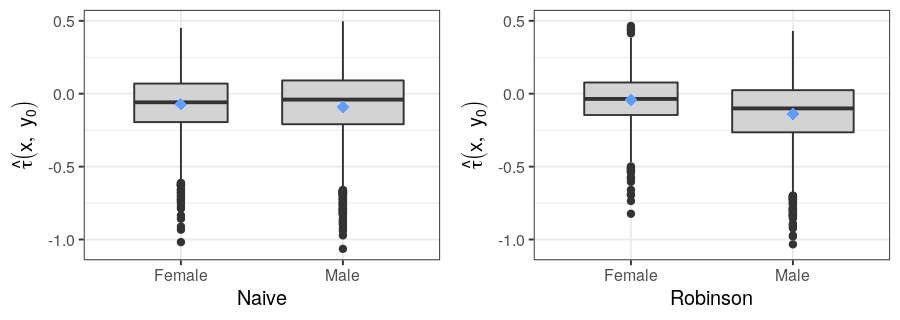}}
	\subfigure[age]{ 
		\includegraphics[width=0.48\textwidth, page=1]{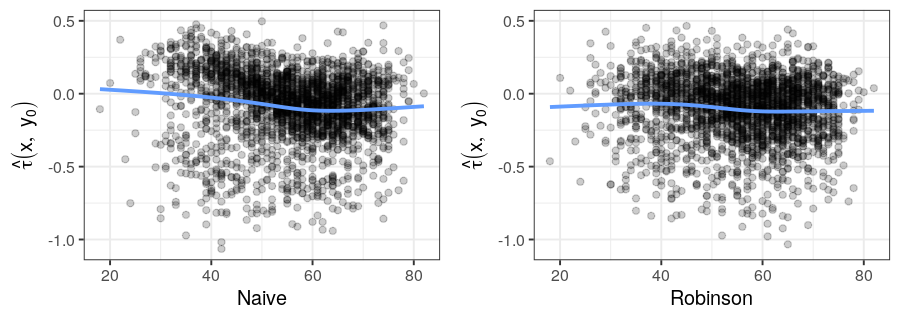}}
	\subfigure[height]{
		\includegraphics[width=0.48\textwidth, page=1]{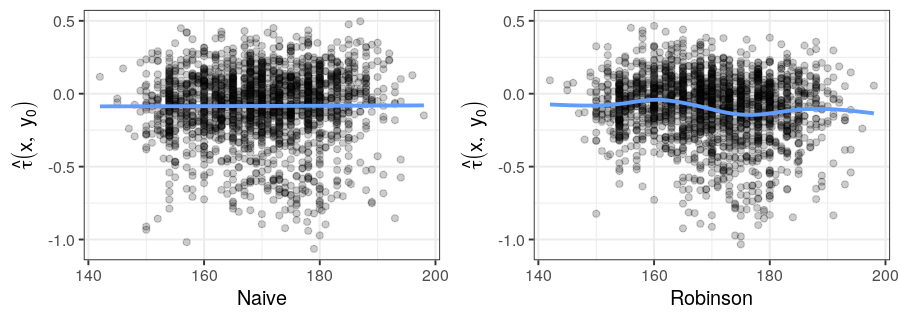}}
	\subfigure[atrophy]{
		\includegraphics[width=0.48\textwidth, page=1]{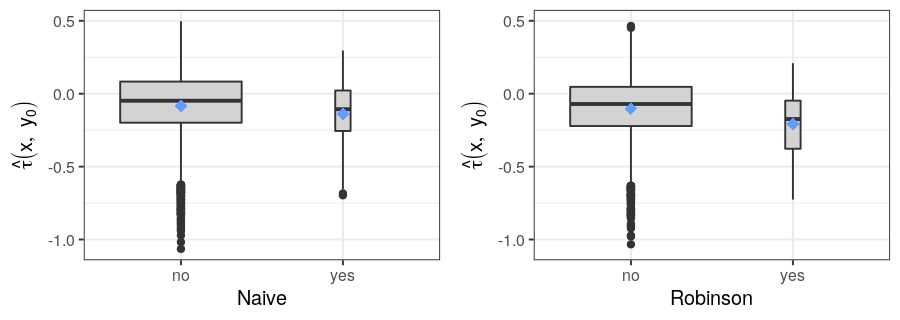}}
	\subfigure[cramps]{
		\includegraphics[width=0.48\textwidth, page=1]{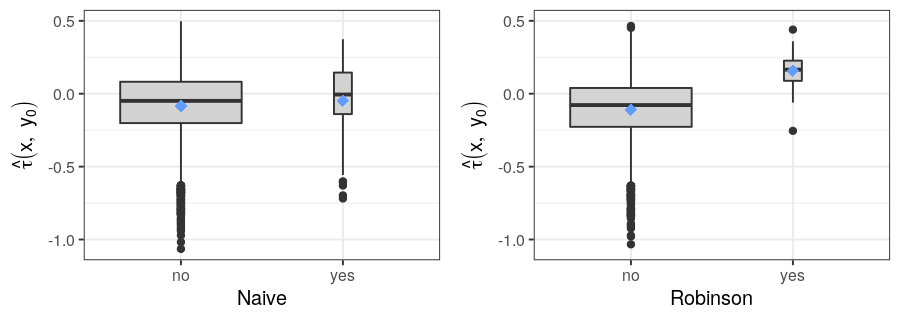}}
	\subfigure[fasciculations]{
		\includegraphics[width=0.48\textwidth, page=1]{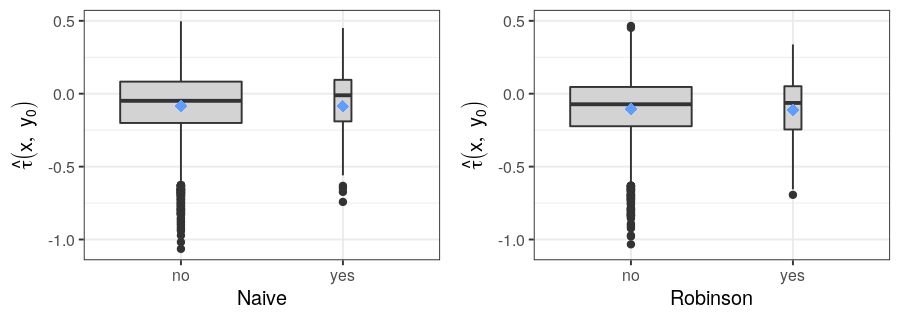}}
	\subfigure[gait changes]{
		\includegraphics[width=0.48\textwidth, page=1]{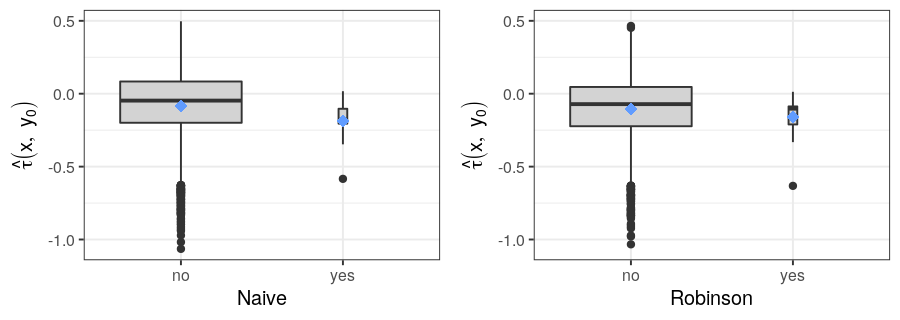}}
	\caption{Handwriting ability score: dependency plot of individual average treatment effects calculated by model-based forest without (left) and with Robinson centering (right). Blue lines and diamond points depict (smooth conditional) mean effects.}
	\label{fig:dependplotalsfrs1}
\end{figure}

\begin{figure}[h!]
	\subfigure[other]{
		\includegraphics[width=0.48\textwidth, page=1]{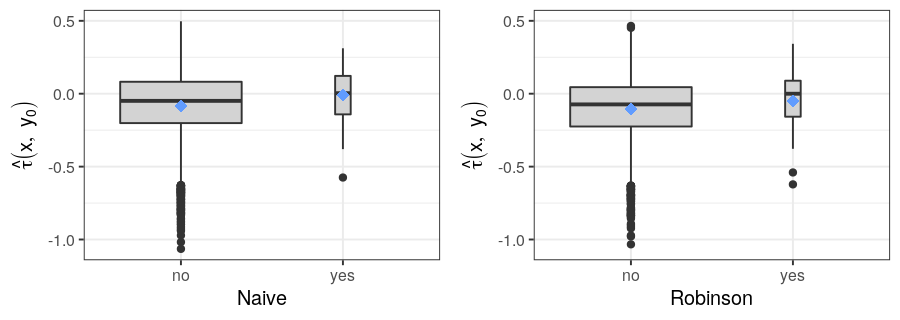}}
	\subfigure[sensory changes]{
		\includegraphics[width=0.48\textwidth, page=1]{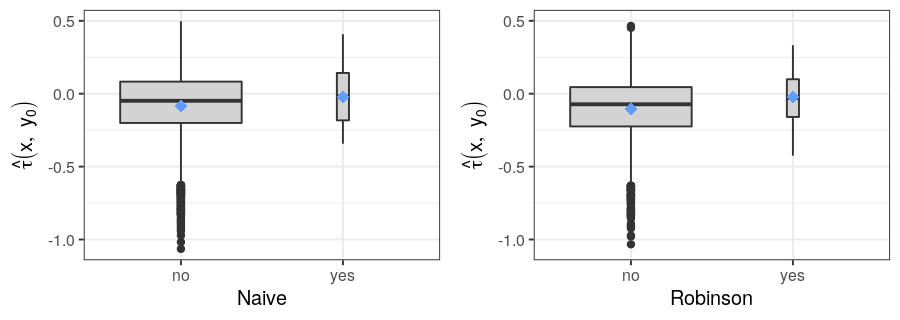}}
	\subfigure[speech]{
		\includegraphics[width=0.48\textwidth, page=1]{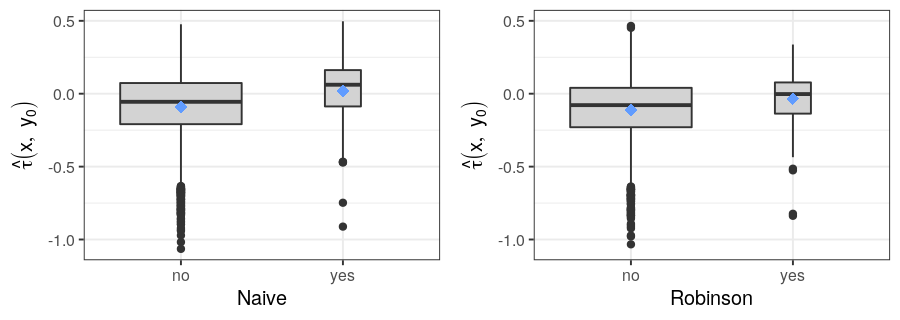}}
	\subfigure[stiffness]{
		\includegraphics[width=0.48\textwidth, page=1]{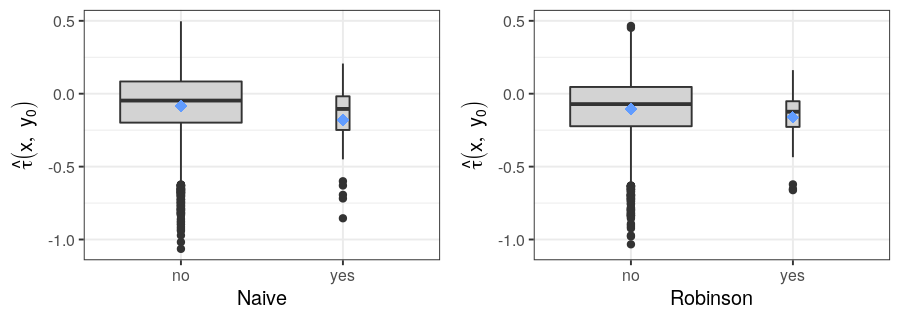}}
	\subfigure[swallowing]{
		\includegraphics[width=0.48\textwidth, page=1]{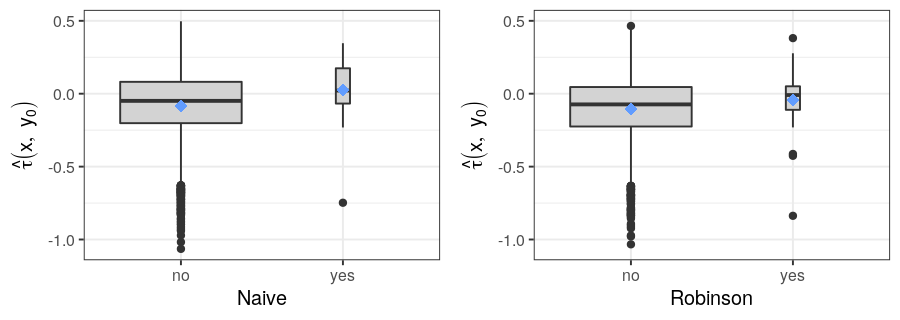}}
	\subfigure[weakness]{
		\includegraphics[width=0.48\textwidth, page=1]{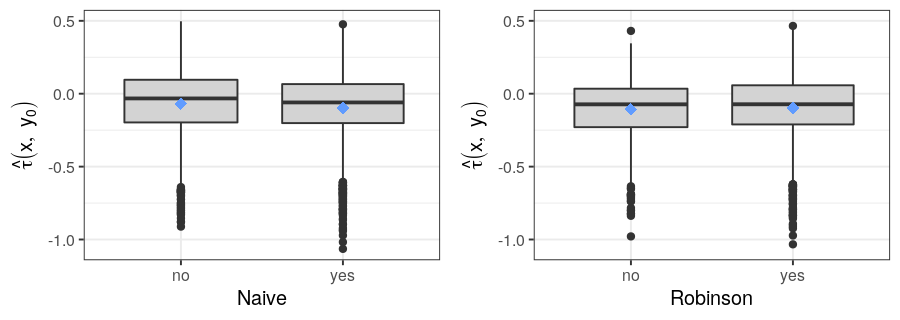}}
	\subfigure[family history (older)]{
		\includegraphics[width=0.48\textwidth, page=1]{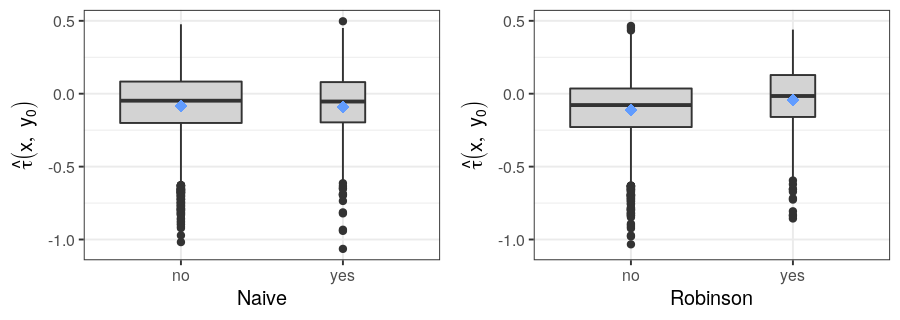}}
	\subfigure[family history (same)]{
		\includegraphics[width=0.48\textwidth, page=1]{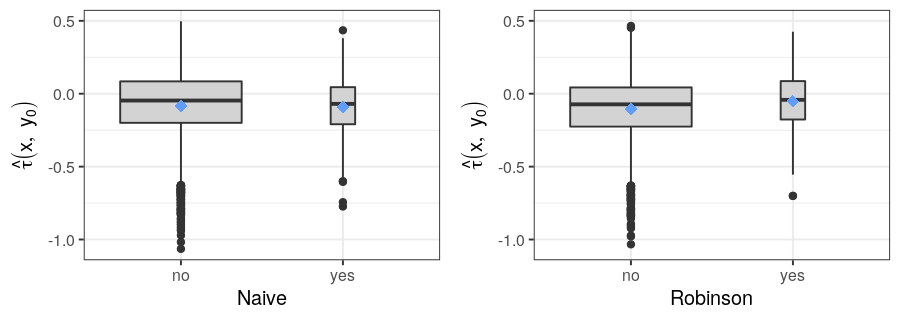}}
	\subfigure[family history (younger)]{
		\includegraphics[width=0.48\textwidth, page=1]{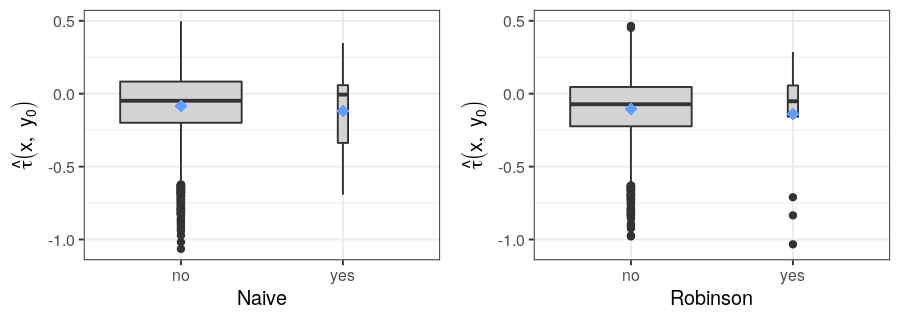}}
	\subfigure[blood pressure (diastolic)]{
		\includegraphics[width=0.48\textwidth, page=1]{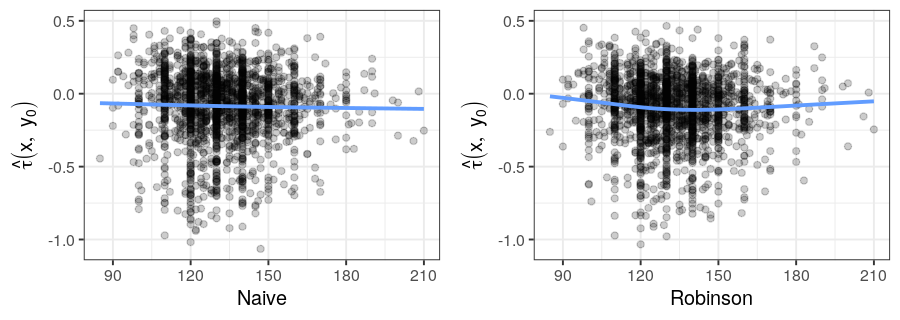}}
	\caption{Handwriting ability score: dependency plot of individual average treatment effects calculated by model-based forest without (left) and with Robinson centering (right). Blue lines and diamond points depict (smooth conditional) mean effects.}
	\label{fig:dependplotalsfrs2}
\end{figure}

\begin{figure}[h!]
	\subfigure[blood pressure (systolic)]{
		\includegraphics[width=0.48\textwidth, page=1]{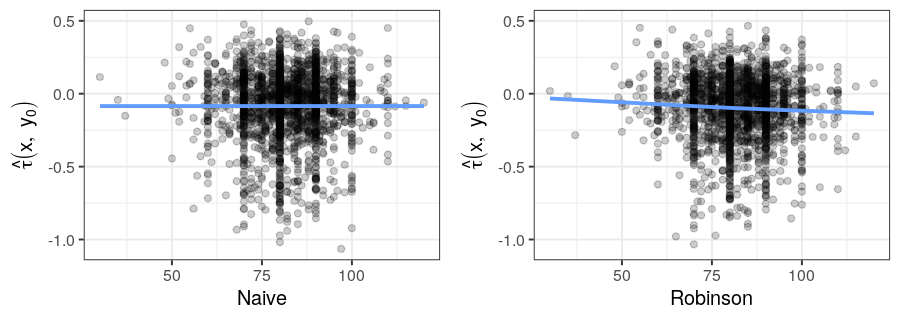}}
	\subfigure[weight]{ 
		\includegraphics[width=0.48\textwidth, page=1]{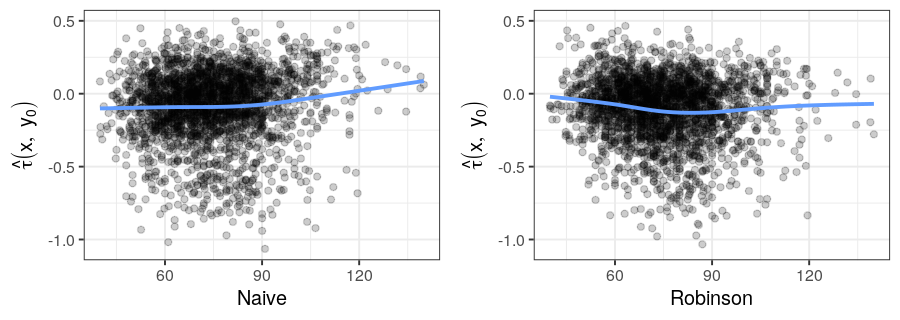}}
	\subfigure[forced vital capacity]{
		\includegraphics[width=0.48\textwidth, page=1]{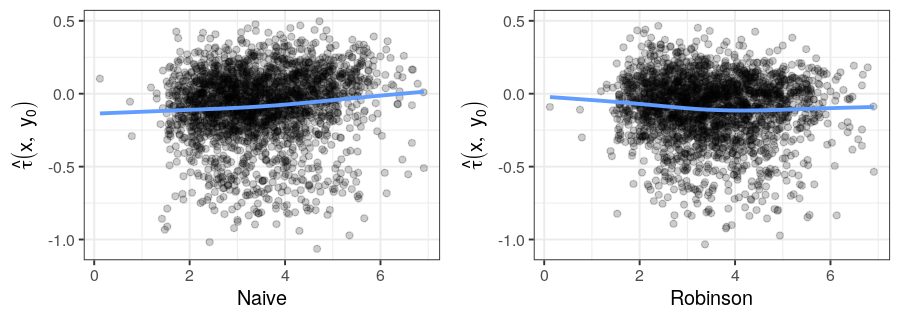}}
	\subfigure[monocytes]{
		\includegraphics[width=0.48\textwidth, page=1]{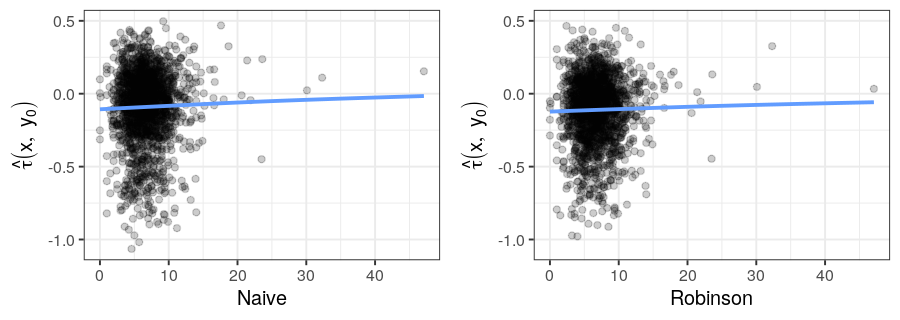}}
	\subfigure[chloride]{
		\includegraphics[width=0.48\textwidth, page=1]{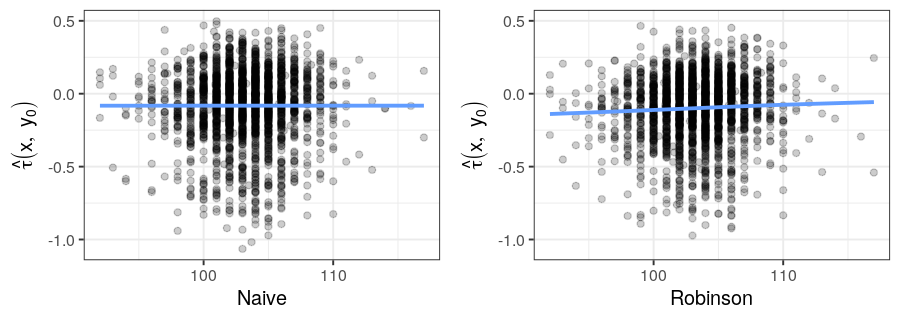}}
	\subfigure[astsgot]{
		\includegraphics[width=0.48\textwidth, page=1]{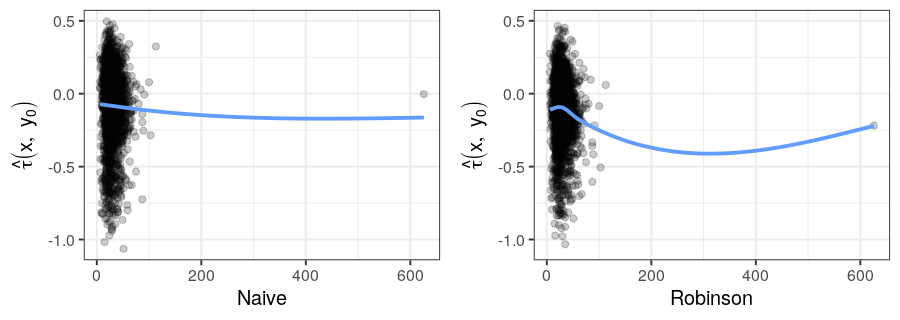}}
	\subfigure[ck]{
		\includegraphics[width=0.48\textwidth, page=1]{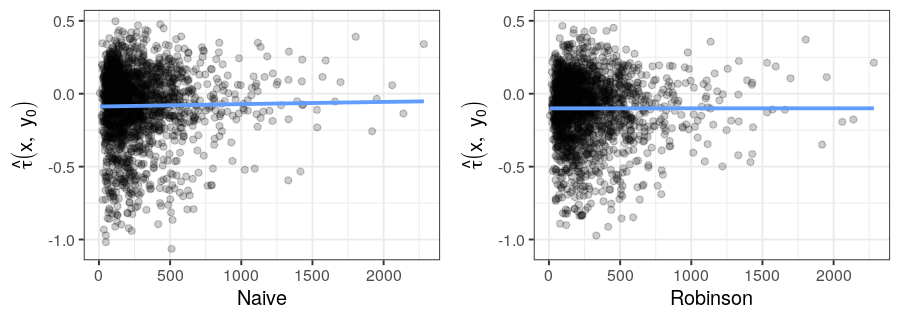}}
	\subfigure[white blood cells]{
		\includegraphics[width=0.48\textwidth, page=1]{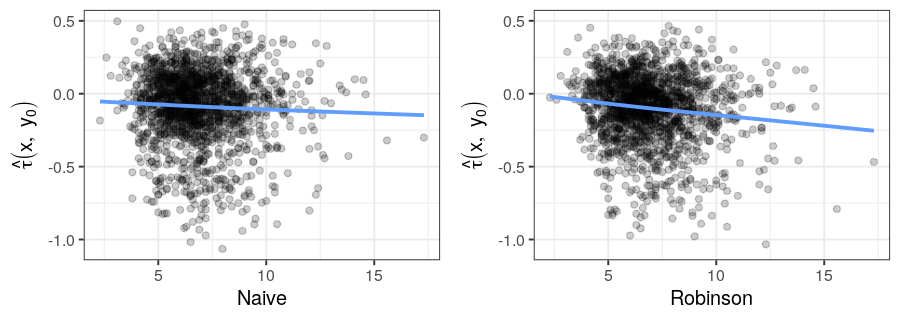}}
	\subfigure[glucose]{
		\includegraphics[width=0.48\textwidth, page=1]{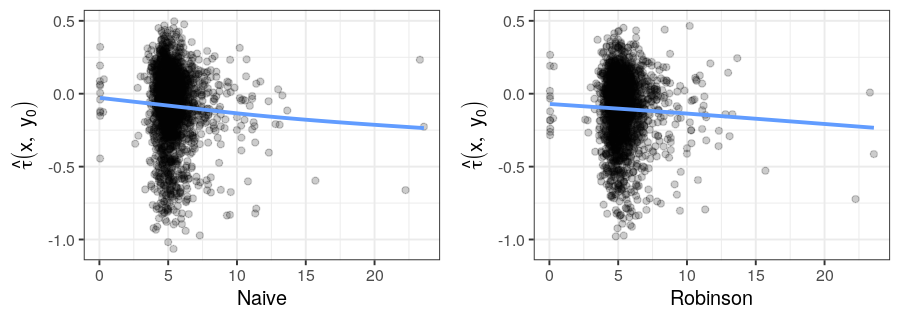}}
	\subfigure[alkaline phosphatase]{ 
		\includegraphics[width=0.48\textwidth, page=1]{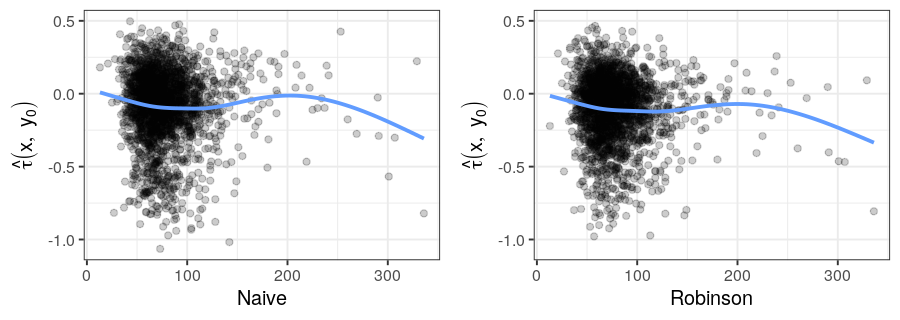}}
	\caption{Handwriting ability score: dependency plot of individual average treatment effects calculated by model-based forest without (left) and with Robinson centering (right). Blue lines and diamond points depict (smooth conditional) mean effects.}
	\label{fig:dependplotalsfrs3}
\end{figure}

\begin{figure}[h!]
	\subfigure[basophils]{
		\includegraphics[width=0.48\textwidth, page=1]{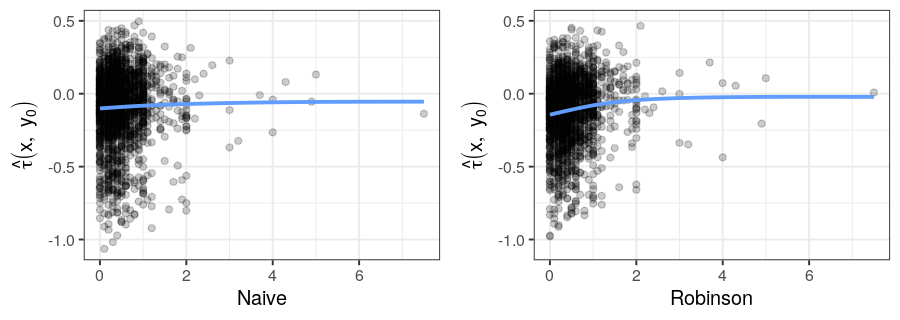}}
	\subfigure[calcium]{
		\includegraphics[width=0.48\textwidth, page=1]{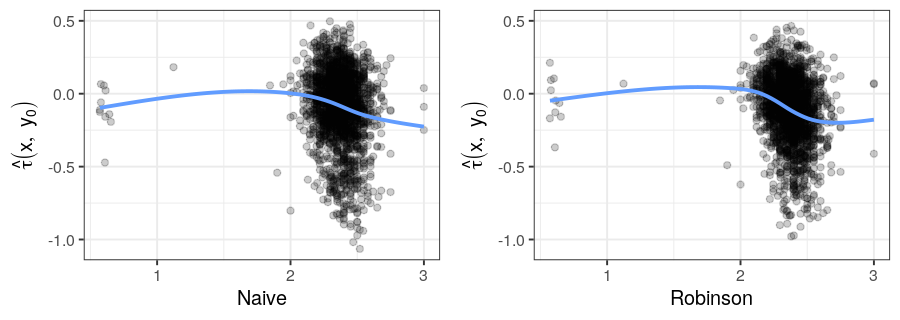}}
	\subfigure[hemoglobin]{
		\includegraphics[width=0.48\textwidth, page=1]{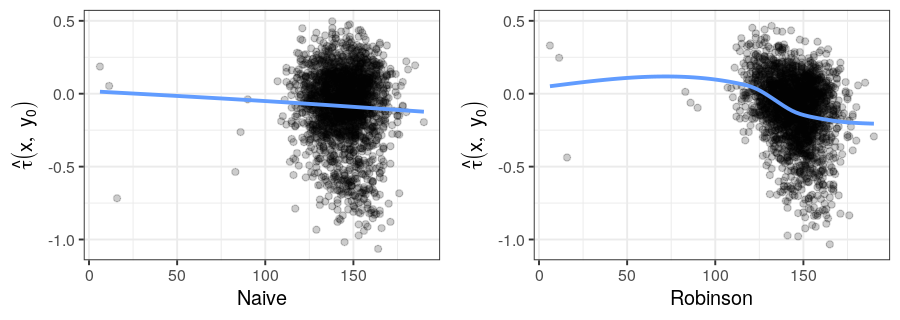}}
	\subfigure[platelets]{
		\includegraphics[width=0.48\textwidth, page=1]{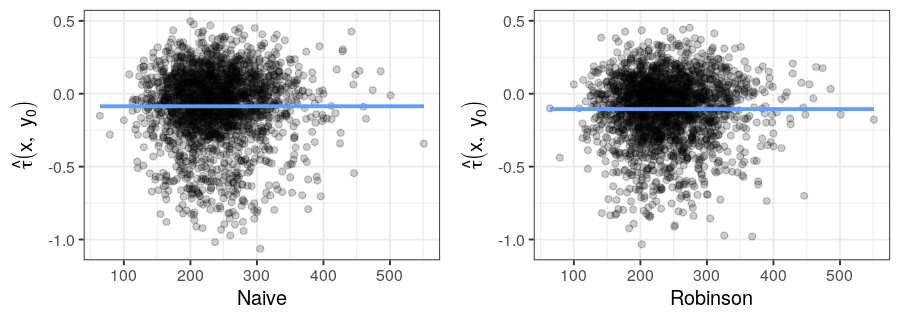}}
	\subfigure[sodium]{
		\includegraphics[width=0.48\textwidth, page=1]{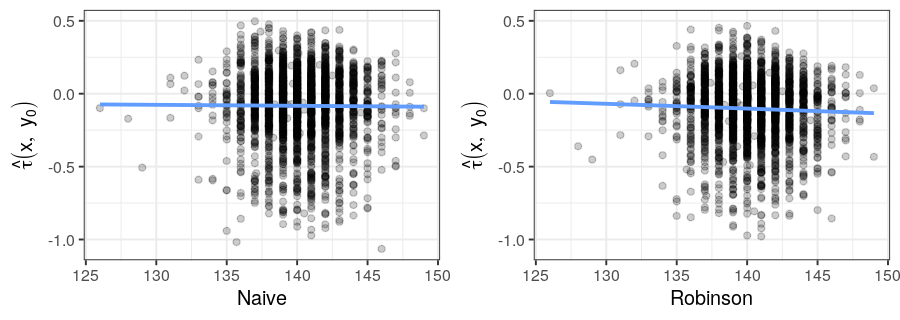}}
	\subfigure[blood urea nitrogen bun]{
		\includegraphics[width=0.48\textwidth, page=1]{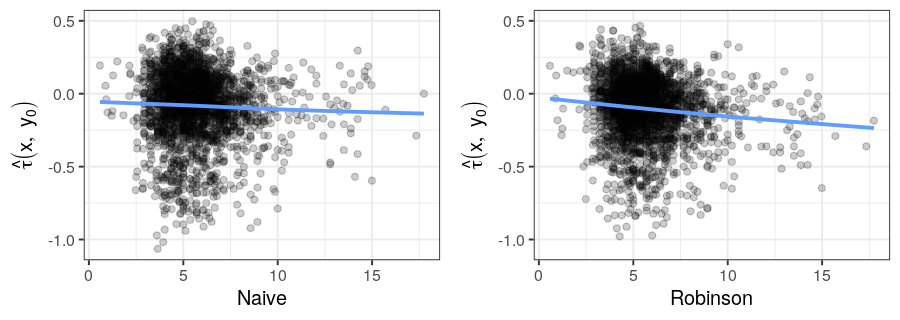}}
	\subfigure[potassium]{
		\includegraphics[width=0.48\textwidth, page=1]{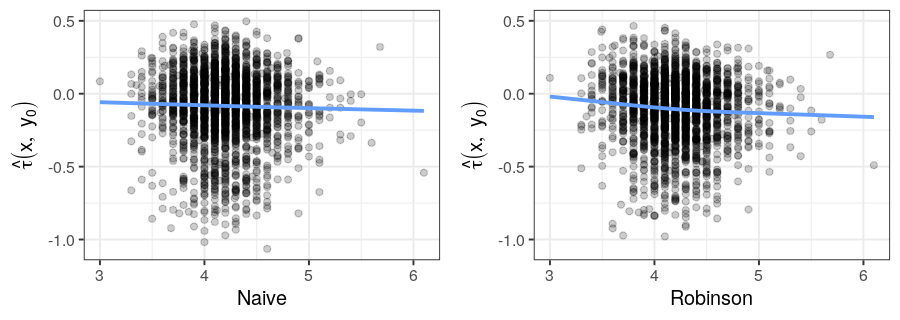}}
	\subfigure[total bilirubin]{
		\includegraphics[width=0.48\textwidth, page=1]{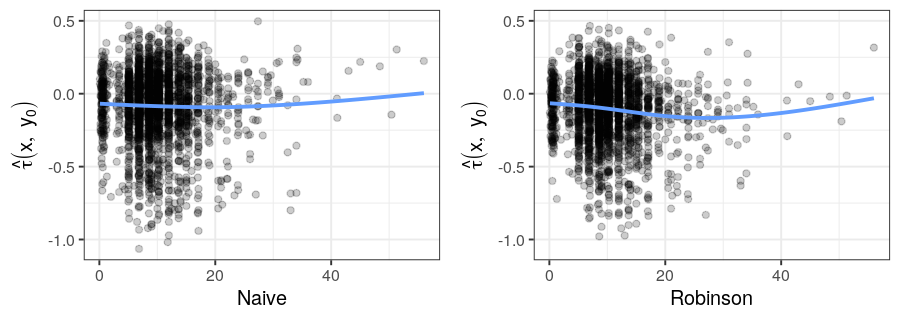}}
	\subfigure[lymphocytes]{
		\includegraphics[width=0.48\textwidth, page=1]{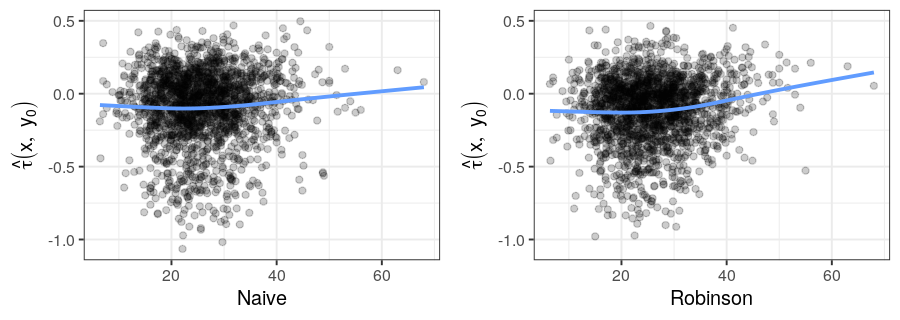}}
	\subfigure[red blood cells]{
		\includegraphics[width=0.48\textwidth, page=1]{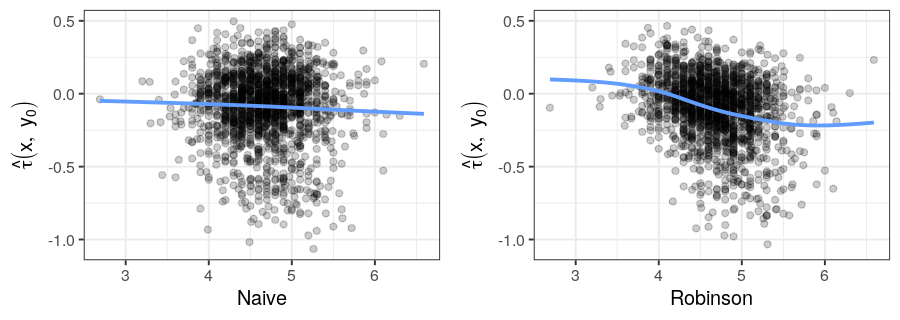}}
	\caption{Handwriting ability score: dependency plot of individual average treatment effects calculated by model-based forest without (left) and with Robinson centering (right). Blue lines and diamond points depict (smooth conditional) mean effects.}
	\label{fig:dependplotalsfrs4}
\end{figure}

\begin{figure}[h!]
	\subfigure[protein]{
		\includegraphics[width=0.48\textwidth, page=1]{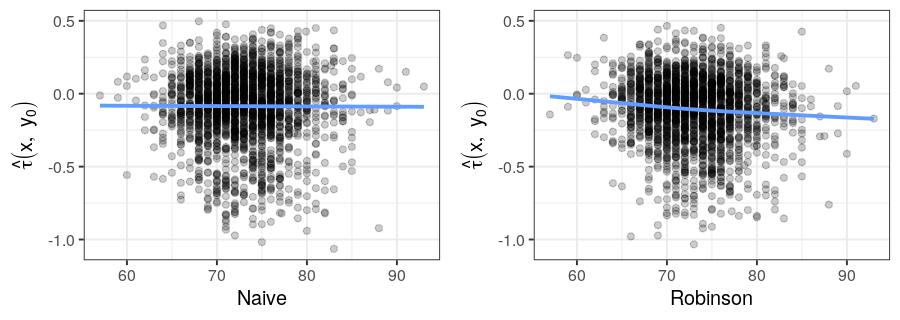}}
	\subfigure[phosphorus]{ 
		\includegraphics[width=0.48\textwidth, page=1]{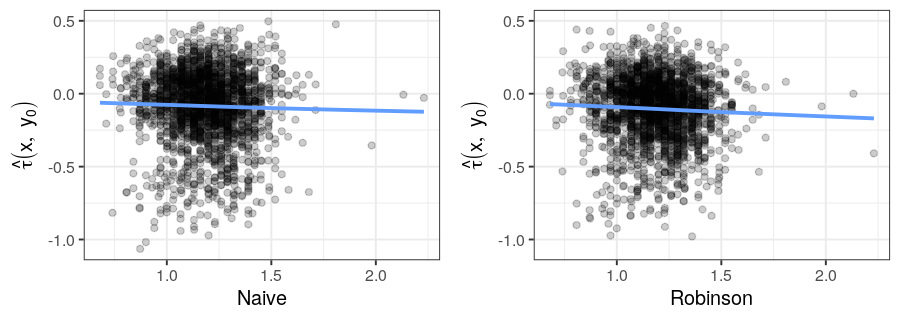}}
	\subfigure[altsgpt]{
		\includegraphics[width=0.48\textwidth, page=1]{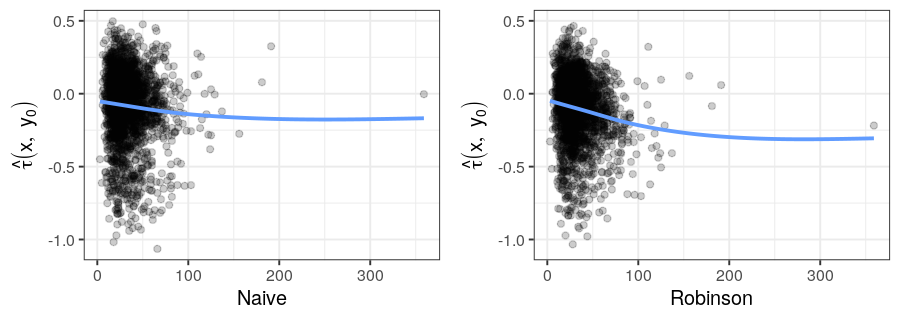}}
	\subfigure[albumin]{
		\includegraphics[width=0.48\textwidth, page=1]{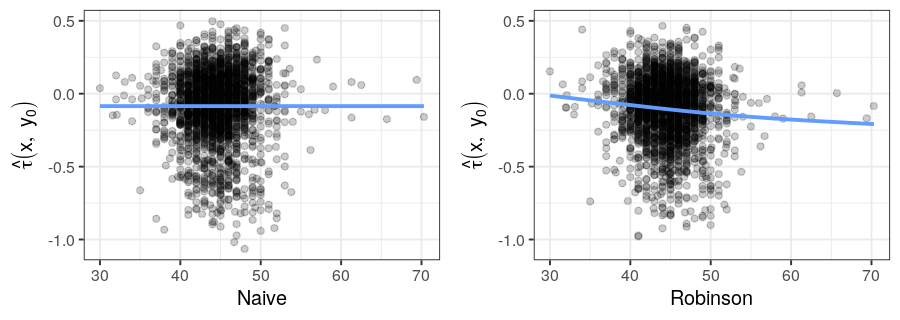}}
	\subfigure[hematocrit]{
		\includegraphics[width=0.48\textwidth, page=1]{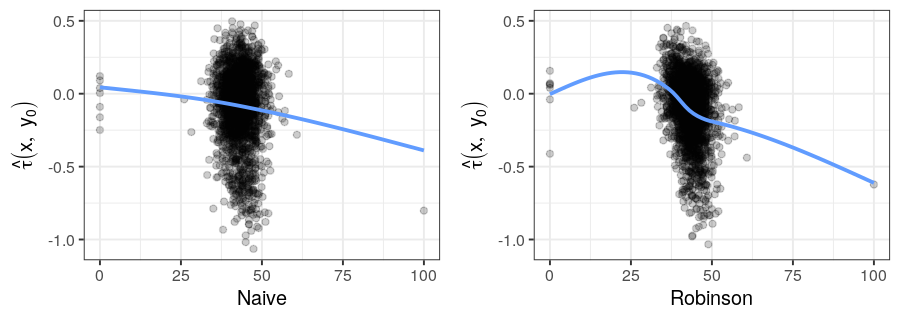}}
	\subfigure[bicarbonate]{
		\includegraphics[width=0.48\textwidth, page=1]{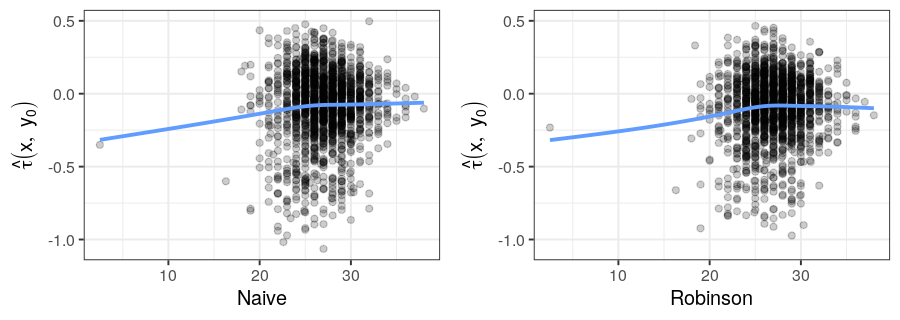}}
	\subfigure[absolute eosinophils count]{
		\includegraphics[width=0.48\textwidth, page=1]{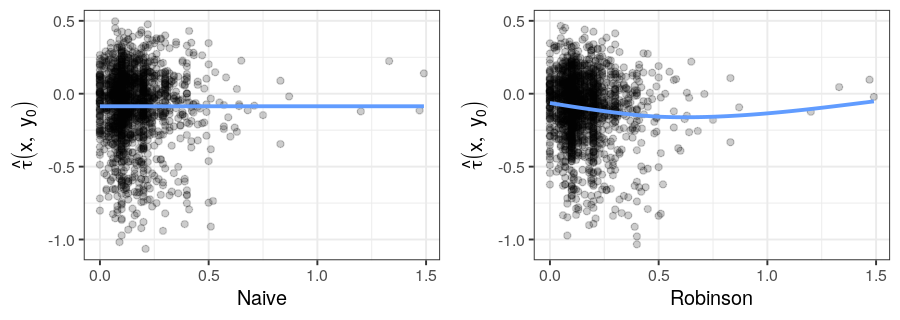}}
	\subfigure[creatinine]{
		\includegraphics[width=0.48\textwidth, page=1]{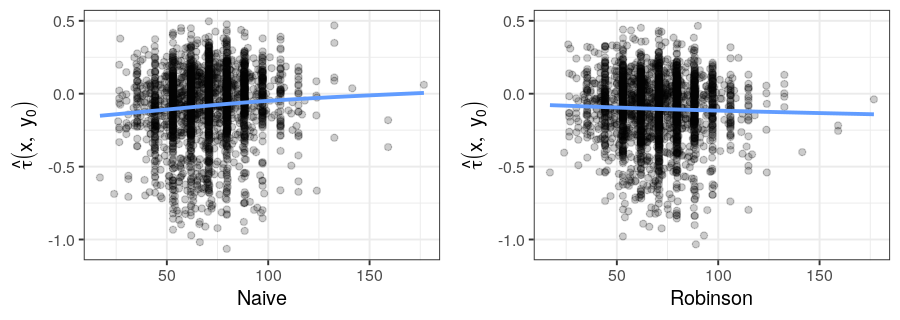}}
	\subfigure[eosinophils]{
		\includegraphics[width=0.48\textwidth, page=1]{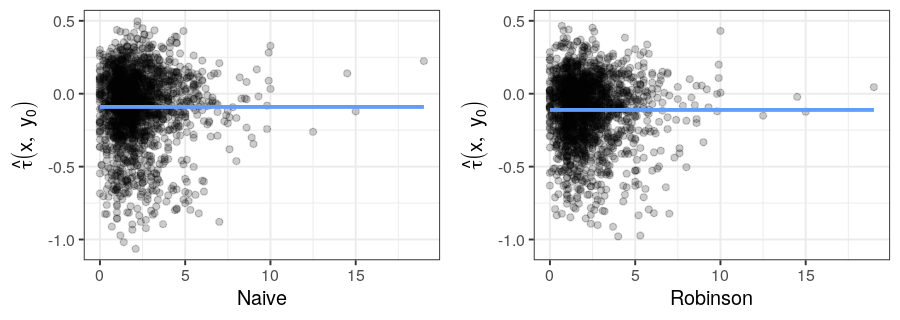}}
	\subfigure[neutrophils]{
		\includegraphics[width=0.48\textwidth, page=1]{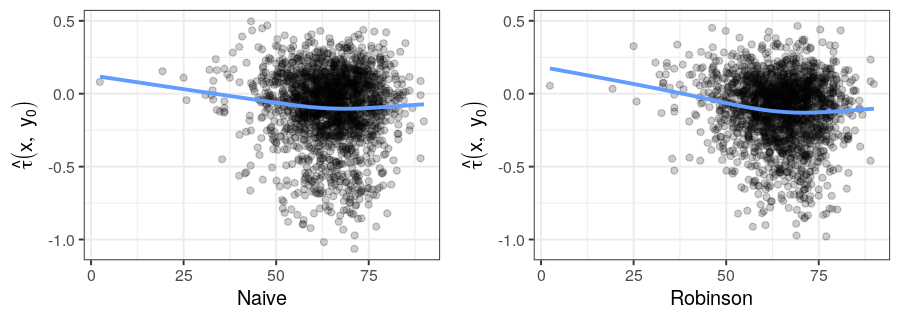}}
	\caption{Handwriting ability score: dependency plot of individual average treatment effects calculated by model-based forest without (left) and with Robinson centering (right). Blue lines and diamond points depict (smooth conditional) mean effects.}
	\label{fig:dependplotalsfrs5}
\end{figure}

\begin{figure}[h!]
	\subfigure[urine ph]{
		\includegraphics[width=0.48\textwidth, page=1]{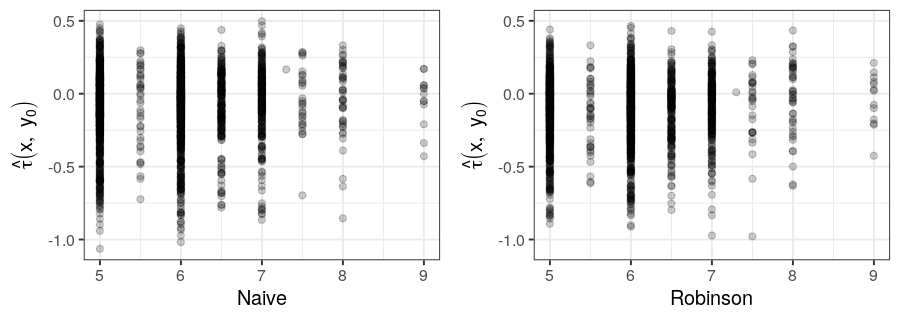}}
	\subfigure[glutamyltransferase]{
		\includegraphics[width=0.48\textwidth, page=1]{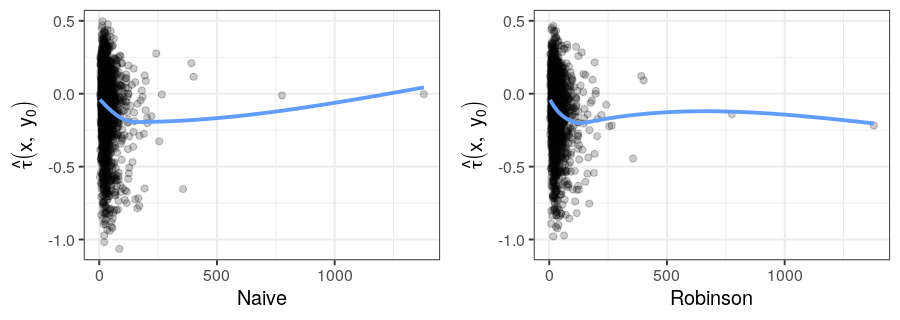}}
	\subfigure[glycated hemoglobin]{
		\includegraphics[width=0.48\textwidth, page=1]{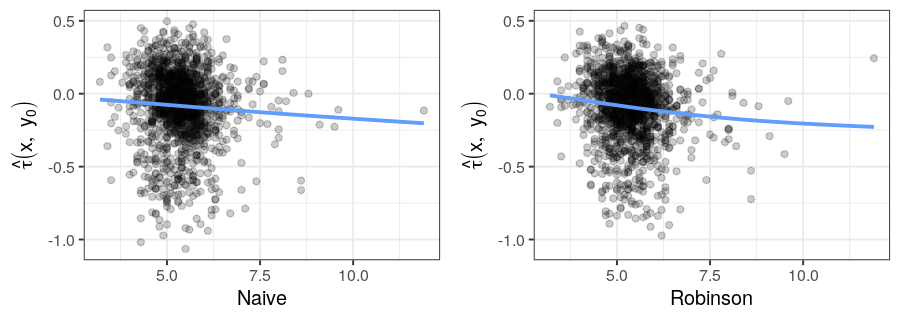}}
	\subfigure[absolute monocyte count]{
		\includegraphics[width=0.48\textwidth, page=1]{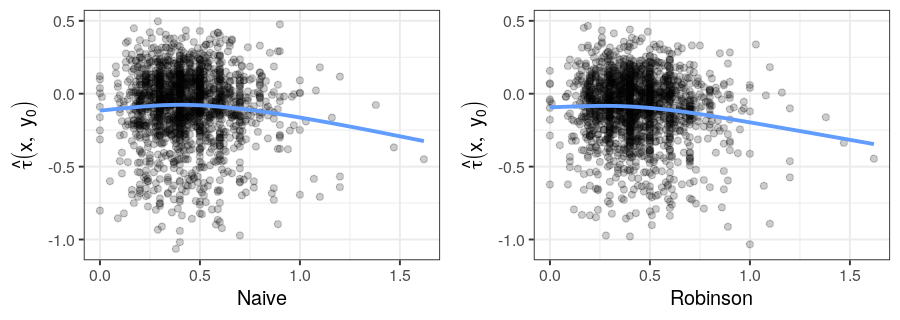}}
	\subfigure[absolute neutrophil count]{
		\includegraphics[width=0.48\textwidth, page=1]{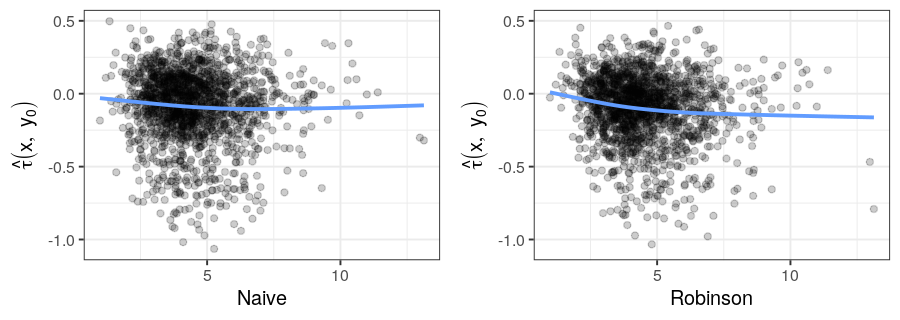}}
	\subfigure[absolute lymphocyte count]{
		\includegraphics[width=0.48\textwidth, page=1]{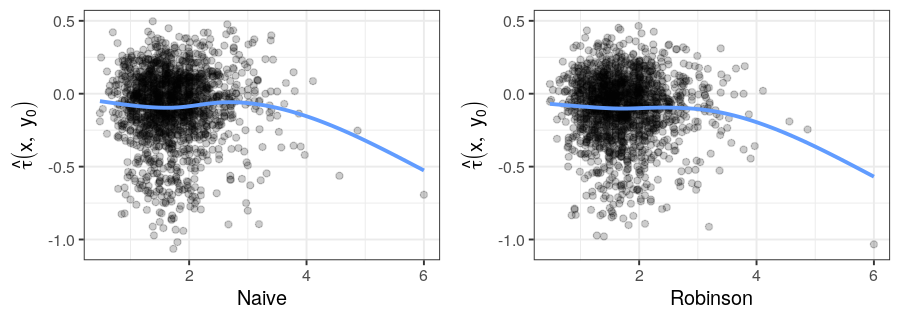}}
	\subfigure[total cholesterol]{
		\includegraphics[width=0.48\textwidth, page=1]{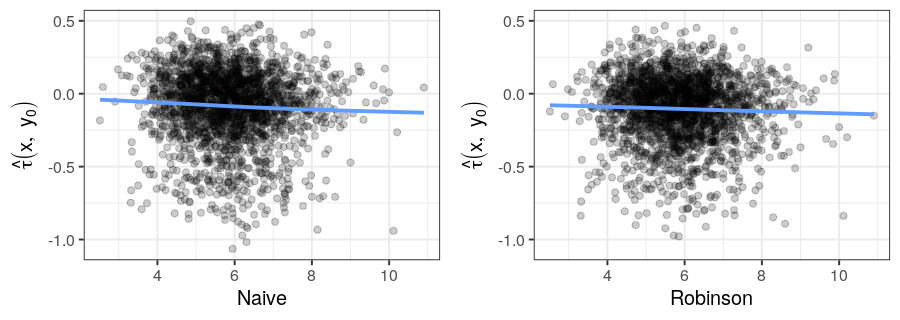}}
		\subfigure[triglycerides]{
		\includegraphics[width=0.48\textwidth, page=1]{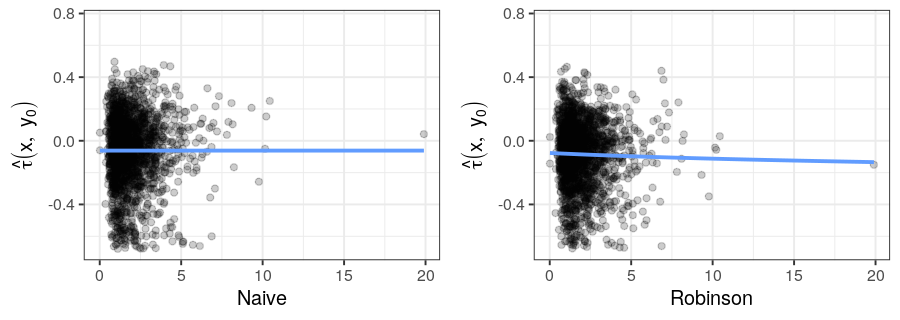}}
	\caption{Handwriting ability score: dependency plot of individual average treatment effects calculated by model-based forest without (left) and with Robinson centering (right). Blue lines and diamond points depict (smooth conditional) mean effects.}
	\label{fig:dependplotalsfrs6}
\end{figure} 
\end{appendix}

\clearpage

\end{document}